\documentclass[review,onefignum,onetabnum]{siamart171218}
\usepackage{booktabs,physics}   
\usepackage{tabularx}  
\usepackage{array,graphicx,float,amsmath} 
\usepackage{xcolor} 
\usepackage{hyperref} 
\usepackage[capitalize]{cleveref} 
\usepackage[section]{placeins}
\usepackage{graphicx}
\usepackage{tikz}
     \usetikzlibrary{positioning,arrows.meta,fit,backgrounds,calc}
\usepackage{subcaption}
\usepackage[
  letterpaper,
  top=1in,
  bottom=1in,
  left=1in,
  right=1in
]{geometry}

\nolinenumbers 

\usepackage{lipsum}
\usepackage{amsfonts}
\usepackage{graphicx,subcaption}
\usepackage{epstopdf}
\usepackage{algorithmic}
\ifpdf
  \DeclareGraphicsExtensions{.eps,.pdf,.png,.jpg}
\else
  \DeclareGraphicsExtensions{.eps}
\fi

\newsiamremark{remark}{Remark}
\newsiamremark{hypothesis}{Hypothesis}
\crefname{hypothesis}{Hypothesis}{Hypotheses}
\newsiamthm{claim}{Claim}

\headers{ NNLCI for solving the 3D Stokes Equations: Novel Techniques \& Advances}{F. Siddiqua, H. Huang, W. Ding, H. Sun, V. Yang \& Y. Liu}

\title{Neural Networks with Local Converging Inputs for Solving the three dimensional Stokes Equations on Unstructured Meshes: Novel Techniques and Advances
}

\author{Farjana Siddiqua\thanks{School of Mathematics, Georgia Institute of
  Technology, Atlanta, GA 30332, USA. Present address: [Department], Iowa State
  University, Ames, IA 50011, USA (\email{siddiqua@iastate.edu}).}
  \and Haoxiang Huang\thanks{Daniel Guggenheim School of Aerospace Engineering, Georgia Institute of Technology, Atlanta, GA, 30332
  (\email{hcwong@gatech.edu}).}
\and Weiming Ding\thanks{School of Mathematics, Georgia Institute of Technology, Atlanta, GA, 30332  
  (\email{wmding@gatech.edu}).}
  \and Haotian Sun\thanks{School of Computational Science and Engineering, Georgia Institute of Technology, Atlanta, GA, 30332
  (\email{htsun@gatech.edu}).}
   \and Vigor Yang\thanks{Daniel Guggenheim School of Aerospace Engineering, Georgia Institute of Technology, Atlanta, GA, 30332
  (\email{vigor@gatech.edu}).}
\and Yingjie Liu\thanks{School of Mathematics, Georgia Institute of Technology, Atlanta, GA, 30332 
  (\email{yingjie@math.gatech.edu})}
  }
\usepackage{amsopn}

\makeatletter
\newcommand*{\addFileDependency}[1]{
  \typeout{(#1)}
  \@addtofilelist{#1}
  \IfFileExists{#1}{}{\typeout{No file #1.}}
}
\makeatother

\ifpdf
\hypersetup{
  pdftitle={Neural Networks with Local Converging Inputs for Solving the Three-Dimensional Stokes Equations on Unstructured Meshes: Novel Techniques and Advances},
  pdfauthor={Authors}
}
\fi
\begin{document}

\maketitle

\begin{abstract}
Neural-network surrogates for fluid-flow partial differential equations usually learn the solution globally, which requires large networks and many high-fidelity training simulations. Neural networks with local converging inputs (NNLCI) instead work locally. They predict the high-fidelity solution at a point from two converging coarse-grid solutions sampled near that point, which keeps the network small, needs little training data, and generalizes across geometries. Here, the two low-fidelity meshes, LF1 and LF2, have on average about 300 and 730 tetrahedra, compared with about 155,000 in the high-fidelity mesh (HF).
The contribution in artificial intelligence is an NNLCI framework for the three-dimensional Stokes equations on unstructured tetrahedral meshes, with two techniques for building local inputs near complex curved boundaries. First, input patches may extend beyond the boundary, with the outside values obtained from a local least-squares fit, so that every high-fidelity point can be predicted, including those next to the wall.
Second, the local mesh spacing is estimated from neighboring centroids, which sets the stencil size consistently on unstructured meshes.
The engineering application is the fast prediction of flow in pipes with local constrictions. For pipes with one to three bumps, including a circumferentially rotated bump, NNLCI reduces the relative $L^2$ error of the finer low-fidelity solution by $86\%$ to $92\%$, evaluated at all high-fidelity points. Both techniques carry over to other NNLCI problems with smooth solutions on point-cloud discretizations.
\end{abstract}

\section{Introduction}
\hypertarget{target:intro}{}
\label{sec:intro}
The steady incompressible Stokes equations describe viscous flow in which inertia is negligible. They arise in creeping-flow models of microfluidic devices, pore-scale transport, lubrication, and particle suspensions \cite{kim1991microhydrodynamics,squires2005microfluidics}, and also serve as a basic component of numerical methods for more general incompressible flows. Although the Stokes momentum equations are linear, pressure and velocity remain coupled through the incompressibility constraint. Their finite-element discretization produces a saddle-point system that requires stable velocity--pressure approximation spaces \cite{boffi2013mixed} or suitable stabilization, together with effective linear solvers \cite{benzi2005saddle}. For three-dimensional domains with localized constrictions and curved
boundaries, resolving how pressure and velocity vary in space can be computationally
expensive. The cost adds up when the simulation must be repeated for many geometries or boundary conditions, as in design studies, parameter estimation, or uncertainty quantification. 

Surrogate and reduced-order models offer complementary ways to reduce the cost
of repeated high-fidelity simulations. Data-fit surrogates, including polynomial
response surfaces, radial-basis-function models, Gaussian-process regression,
and neural networks, approximate the relationship between problem parameters
and selected simulation outputs. Their effectiveness depends on the
approximation model, the sampling strategy, and the balance between predictive
accuracy and data-generation cost \cite{Forrester2007}. Projection-based
reduced-order models instead approximate the governing equations in
low-dimensional spaces constructed, for example, by proper orthogonal
decomposition or reduced-basis methods
\cite{Benner2015,Rowley2004,Taira2020,Yano2014}. These methods retain a
connection to the numerical formulation while reducing the number of unknowns
needed for repeated evaluations. Nonlinear representations extend this idea
beyond fixed linear spaces: Lee and Carlberg \cite{LeeCarlberg2020}, for example,
constructed nonlinear trial manifolds with convolutional autoencoders and
developed projection-based models on those manifolds. Reduced-coordinate
emulators provide another route to predicting parameter-dependent flow fields
\cite{Swischuk2019,Yeh2018}.

Multifidelity methods use relationships among models of different resolutions
or physical complexity to improve predictions while limiting the number of
expensive evaluations. Such relationships can be represented through
correction, information fusion, or coordinated use of low- and high-fidelity
models \cite{Kennedy2000,Peherstorfer2018,Gano2005,Jiang2018,Song2019,Zhou2017}.
Machine learning provides further ways to represent spatial flow structure
\cite{brunton2020machine,vinuesa2022enhancing}. Convolutional networks have
been used to predict steady and aerodynamic flow fields from geometric or
flow inputs \cite{guo2016cnn,bhatnagar2019prediction,thuerey2020deep}.
Autoencoders compress flow snapshots into nonlinear coordinates
\cite{Xu2020}, whereas super-resolution networks infer fine-scale features
from a single coarse field \cite{Fukami19,Liu-etal-20,Fukami2021}.

These approaches demonstrate several ways to combine numerical simulation and
learned representations. Many construct a representation of the full solution
field, which requires decisions about geometry encoding, correspondence
between input and output grids, and the coverage of high-fidelity training
data. The present method instead predicts pressure and velocity at a specified
point from two coarse numerical solutions sampled in its neighborhood. To make
these local inputs comparable, both solutions must be evaluated at the same
physical locations. Constructing such aligned neighborhoods is particularly
challenging on unstructured meshes, where connectivity, local resolution, and
boundary geometry vary between simulations.

Alongside surrogate models built from simulation data, neural-network methods have been developed to approximate PDE solutions by incorporating governing equations and their initial and boundary conditions into the loss function during training \cite{Dissanayake1994,E2018, raissi2019pinn, Sirignano2018}. For incompressible flow, Jin et al.~\cite{jin2021nsfnets} developed NSFnets using velocity--pressure and velocity--vorticity formulations of the Navier--Stokes equations. Geometry-dependent problems have been addressed through physics-informed PointNet architectures \cite{kashefi2022pipn}, which process irregular point clouds and incorporate PDE residuals, boundary conditions, and sparse observations. Closest to the present work, Kashefi  and Mukerji \cite{kashefi2023porous} applied this approach to steady Stokes flow in pore spaces of porous media. These studies show that neural networks can represent incompressible flow with complex boundaries. How well they work, however, depends on the training formulation, the available data, and how varied the training cases are, and they can cost more than conventional solvers.

Operator-learning methods aim to learn mappings between function spaces, typically formulated through basis-function expansions or other operator representations. In \cite{ChenChen95, lu2021deeponet}, neural networks are used to represent both the coefficients and the basis functions of a functional expansion. Fourier neural operators (FNOs) \cite{Li21FNO} learn neural mappings between Fourier-expanded function spaces and can learn parametric PDE solution operators, including two-dimensional Navier--Stokes dynamics. Graph-based neural simulators provide another direction for flow problems on irregular or unstructured meshes. MeshGraphNets \cite{pfaff2021meshgraphnets}, for example, represent computational meshes as graphs and use message passing to learn physical dynamics based on meshes while accommodating nonuniform and adaptive discretizations. 

These methods show that machine learning can speed up flow simulations. However, many existing approaches learn global solution representations, need large amounts of high-fidelity training data, or require costly
optimization of physics-based residuals. This motivates surrogates that use cheap numerical solutions locally and need fewer full-domain high-fidelity simulations.

In Huang et al.~\cite{Huang23_1, Huang23_2}, neural networks with local converging inputs (NNLCI) were introduced for solving Euler systems in one and two dimensions. By using two local, converging low-fidelity solution patches as input, the method predicts the high-fidelity solution at a given point in space and time, for smooth and discontinuous solutions alike, with little training data and good generalization. Using two converging solutions in this way is related to estimating the discretization error from the change between grid levels, as in Richardson extrapolation \cite{richardson1911approximate,roache1998verification}, but NNLCI learns a nonlinear local correction instead of applying a fixed extrapolation formula. It has also been applied to predicting electromagnetic waves scattered from a variety of 2D metal shapes by Cobb et al.~\cite{Cobb23}.  Ding et al.~\cite{Ding24} extended NNLCI to unstructured grids and validated the method using inviscid supersonic flow through a 2D channel with varying bumps. Lee et al.~\cite{Lee24} applied it to a one-dimensional Poisson--Nernst--Planck
ion-channel model and showed that it learns in context across multiple
subdomains for different nonlinear elliptic systems, without further
modification of the method. Lu et al.~\cite{POD-NNLCI-26} combined NNLCI with encoder--decoder reduced-order modeling for 2D combustion problems. A framework for integrating NNLCI with commercial software, despite limited
access to mesh generation and data formatting, was developed in Huang~\cite{huang2025dissertation} for computing compressible Navier--Stokes equations on unstructured grids around varying 2D airfoils, reducing the complexity by orders of magnitude, which shows its potential for engineering use.

This work applies NNLCI to three-dimensional incompressible Stokes flow.  The steady Stokes system is elliptic and globally coupled \cite{boffi2013mixed}: changes in boundary conditions or geometry can affect pressure and velocity throughout the domain. We show that small 3D patches taken from two converging coarse-grid solutions are enough for a trained network to recover the high-fidelity solution at a point. Because the inputs are local, a network trained on some geometries can be applied to others.

Three-dimensional unstructured meshes introduce further challenges in constructing comparable inputs. Mesh connectivity and centroid spacing vary between fidelity levels and between geometric configurations, so corresponding neighborhoods cannot generally be identified by a common list of mesh indices. Near curved walls, a regular stencil centered inside the fluid may also extend outside the computational mesh. Rejecting every such stencil removes training and prediction locations precisely where boundary effects are important.

An NNLCI construction for this setting needs three things: the same physical neighborhood for both coarse solutions, a measure of the local resolution of each, and usable patches next to the boundary built from nearby data only.

The present work develops this construction for the three-dimensional steady Stokes equations using two low-fidelity Taylor--Hood finite-element solutions and a convolutional neural-network surrogate. Both solutions are evaluated on the same physical \(3\times3\times3\) stencil, and the network predicts pressure and all three velocity components at its center. Two new ingredients make this possible. First, a centroid-based spacing estimator uses neighboring point distances to define the stencil scale and supply separate local-resolution features for the two fidelity levels. 
Second, a hybrid evaluation procedure combines direct finite-element
evaluation inside the mesh with weighted quadratic least-squares
reconstruction, using bounded local support, for stencil points outside the
mesh. Earlier NNLCI work restricted the input stencils to the interior of the
domain, so that the solution next to the boundary could not be predicted
\cite{Cobb23}. Filling the outside stencil points with prescribed boundary
values would require such values for every variable on every boundary, which
are not available in general, for example for the pressure at a wall or the
velocity at an outflow boundary; the least-squares extension instead uses only
the local low-fidelity solution. Local refinement, support, conditioning, and
patch-quality checks determine which training samples are retained, whereas
predictions are made and evaluated at all high-fidelity points.
With both, every high-fidelity point gets inputs in the same format, including points whose stencils cross the boundary.
The numerical assessment presented here considers four full-length
three-dimensional pipe configurations: one bump, two aligned bumps, two bumps
with the second bump rotated circumferentially about the streamwise \(x\)-axis, and three bumps. Held-out combinations of bump positions or orientations and inflow amplitudes test interpolation within the sampled parameter ranges, and one additional case tests prediction outside them. Predictions at all high-fidelity points of each test mesh, including those next to the wall, are compared with high-fidelity finite-element reference fields and with the finer low-fidelity solution, using volume-weighted error measures, cross-sectional comparisons, and line profiles. The study asks whether local reconstruction from two coarse solutions improves the pressure and velocity accuracy for a globally coupled problem and which sampling and boundary treatments make it work on 3D unstructured meshes.

The remainder of this paper is organized as follows.
\hyperlink{target:np}{Section~\ref*{sec:np}} introduces the parametric Stokes
problem and preliminary notation.
\hyperlink{target:problem-setup}{Section~\ref*{sec:problem_setup}} describes the
computational setup.
\hyperlink{target:data-generation}{Section~\ref*{sec:data_generation}} presents
the data-generation procedure and the NNLCI surrogate, and
\hyperlink{target:numerical-results}{Section~\ref*{sec:numerical_results}} reports
the numerical results. Finally,
\hyperlink{target:conclusions}{Section~\ref*{sec:conclusions}} concludes the paper.

\section{Parametric Stokes Problem and Preliminaries}
\hypertarget{target:np}{}
\label{sec:np}
We consider a family of steady incompressible Stokes-flow problems in
three-dimensional pipes with localized wall constrictions. The test problems are chosen to see whether NNLCI can recover high-fidelity pressure and
velocity fields from two low-fidelity finite-element solutions on
parameterized domains. This section defines the geometry family, the
governing equations, and the function spaces; the meshes, boundary
conditions, and parameter values are given in \hyperlink{target:problem-setup}{Section~\ref*{sec:problem_setup}}.
\subsection{Parameterized Pipe Geometries}
\hypertarget{target:geometry}{}
\label{sec:geometry}

The undeformed domain is a straight circular pipe of radius $R=1$ and length $L=5$. Each computational domain $\Omega(\boldsymbol{\mu})$ is obtained by deforming this reference pipe with one or more smooth inward wall bumps. In cylindrical coordinates $(x,r,\theta)$, with $\theta=\operatorname{atan2}(z,y)$, the deformed wall is described by
\begin{equation}
 r_w(x,\theta)=R-A\sum_{k=1}^{N_b}
 \exp\!\left[-\frac{(x-x_{b,k})^2}{2\sigma_x^2}\right]
 \exp\!\left[-\frac{\delta(\theta,\theta_{b,k})^2}{2\sigma_\theta^2}\right],
 \label{eq:wall_deformation}
\end{equation}
where $N_b$ is the number of bumps, $x_{b,k}$ and $\theta_{b,k}$ are the axial and circumferential locations of bump $k$, and $\delta(\theta,\theta_{b,k})\in(-\pi,\pi]$ is the wrapped angular difference. We set $A=0.30$ in all cases. The widths are $\sigma_x=0.35$ and $\sigma_\theta=0.45$ for the one-bump and two-aligned-bump configurations, $\sigma_x=0.35$ and $\sigma_\theta=0.32$ for the rotated two-bump configuration, and $\sigma_x=0.25$ and $\sigma_\theta=0.45$ for the three-bump configuration; hence, an isolated bump decreases the local pipe radius by at most $30\%$.
We denote one such domain by
\(\Omega=\Omega(\boldsymbol{\mu})\subset\mathbb{R}^3\), with boundary
\(\partial\Omega\), where \(\boldsymbol{\mu}\) collects the parameters of a
case, namely the bump positions and orientations and the inlet velocity
(\hyperlink{target:problem-setup}{Section~\ref*{sec:problem_setup}}); the dependence on \(\boldsymbol{\mu}\) is omitted
when it is not needed.

The (steady) incompressible Stokes equations for velocity $\mathbf{u}:\Omega\to\mathbb{R}^3$
and pressure $p:\Omega\to\mathbb{R}$ are
\begin{align}
-\nu \Delta \mathbf{u} + \nabla p &= \mathbf{f} \qquad \text{in }\Omega, \label{eq:stokes_momentum}\\
\nabla\cdot \mathbf{u} &= 0 \qquad \text{in }\Omega, \label{eq:stokes_incomp}
\end{align}
where $\nu>0$ is the kinematic viscosity and $\mathbf{f}$ is a prescribed body force \cite{boffi2013mixed}. 
The spaces \(L^2(\Omega)\) and \(H^1(\Omega)\)
are defined by
\[
L^2(\Omega)
=
\left\{
v:\Omega\rightarrow\mathbb{R}:
\int_\Omega |v|^2\,dx<\infty
\right\},
\]
and
\[
H^1(\Omega)
=
\left\{
v\in L^2(\Omega):
\nabla v\in[L^2(\Omega)]^3
\right\}.
\]
Their vector-valued counterparts are denoted by
\([L^2(\Omega)]^3\) and \([H^1(\Omega)]^3\).

For scalar functions \(v,w\in L^2(\Omega)\), the \(L^2(\Omega)\) inner
product and norm are defined by
\[
(v,w)_\Omega
=
\int_\Omega vw\,dx,
\qquad
\|v\|_{0,\Omega}
=
(v,v)_\Omega^{1/2}.
\]
For vector-valued functions, the inner product is defined componentwise:
\[
(\mathbf{v},\mathbf{w})_\Omega
=
\int_\Omega \mathbf{v}\cdot\mathbf{w}\,dx.
\]
For matrix-valued functions, the corresponding Frobenius product is used.
The \(H^1(\Omega)\) seminorm is denoted by
\[
|v|_{1,\Omega}
=
\|\nabla v\|_{0,\Omega}.
\]
When the domain is clear from the context, the subscript \(\Omega\) is
omitted.

Let \(\Gamma_D\subset\partial\Omega\) denote the portion of the boundary on
which the velocity is prescribed. The homogeneous velocity test space is
\[
X_0
=
\left\{
\mathbf{v}\in[H^1(\Omega)]^3:
\mathbf{v}=\mathbf{0}\ \text{on }\Gamma_D
\right\}.
\]
For prescribed boundary data \(\mathbf{g}\), the corresponding affine
velocity trial space is
\[
X_{\mathbf{g}}
=
\left\{
\mathbf{v}\in[H^1(\Omega)]^3:
\mathbf{v}=\mathbf{g}\ \text{on }\Gamma_D
\right\}.
\]
The continuous pressure space is denoted by
\[
Q=L^2(\Omega).
\]
The boundary portion \(\Gamma_D\), the data \(\mathbf{g}\), and the pressure
reference condition are specified in Section~\ref{sec:problem_setup}.

Let \(\mathcal{T}_h\) be a conforming tetrahedral partition of \(\Omega\),
where \(h\) denotes a representative mesh size. The Stokes equations are
discretized using the Taylor--Hood finite-element pair \cite{taylor1973numerical,Fortin93}. The discrete velocity
and pressure spaces are
\[
X_h
=
\left\{
\mathbf{v}_h\in[C^0(\overline{\Omega})]^3:
\left.\mathbf{v}_h\right|_K\in[P_2(K)]^3
\quad\text{for every }K\in\mathcal{T}_h
\right\},
\]
and
\[
Q_h
=
\left\{
q_h\in C^0(\overline{\Omega}):
\left.q_h\right|_K\in P_1(K)
\quad\text{for every }K\in\mathcal{T}_h
\right\},
\]
where \(P_k(K)\) denotes the space of polynomials of degree at most \(k\)
on the tetrahedron \(K\). Thus, the three velocity components are
approximated by continuous piecewise-quadratic functions, whereas the
pressure is approximated by continuous piecewise-linear functions.

Let
\[
X_{h,0}=X_h\cap X_0
\]
denote the discrete velocity test space. In the computations, the pressure
reference is imposed on a designated pressure boundary
\(\Gamma_p\subset\partial\Omega\), leading to the constrained pressure space
\[
Q_{h,0}
=
\left\{
q_h\in Q_h:
q_h=0\ \text{on }\Gamma_p
\right\}.
\]
For the problems considered in this paper, \(\Gamma_p\) is the outlet boundary.

The Taylor--Hood pair is used to provide a stable mixed approximation of
velocity and pressure \cite{boffi2013mixed}. The associated discrete inf-sup constant is
\[
\beta_h
=
\inf_{\substack{q_h\in Q_{h,0}\\q_h\neq0}}
\sup_{\substack{\mathbf{v}_h\in X_{h,0}\\
\mathbf{v}_h\neq\mathbf{0}}}
\frac{
(q_h,\nabla\cdot\mathbf{v}_h)_\Omega
}{
\|q_h\|_{0,\Omega}
\|\nabla\mathbf{v}_h\|_{0,\Omega}
}.
\]
For the shape-regular mesh families employed in this work, the
Taylor--Hood discretization is assumed to satisfy
\[
\beta_h\geq\beta>0,
\]
where \(\beta\) is independent of the mesh size \(h\).
\section{Problem Setup}
\hypertarget{target:problem-setup}{}
\label{sec:problem_setup}

All finite element computations use FEniCS (version 2019.1.0)
\cite{alnaes2015fenics,logg2012fenics}. The domains of \hyperlink{target:geometry}{Section~\ref*{sec:geometry}} are meshed with mshr,
using resolution parameters 8, 11, and 70 for LF1,
LF2, and HF, which give on average about 300, 730, and 155{,}000 tetrahedra.
The meshes are generated for the undeformed pipe, and the wall
displacement~\eqref{eq:wall_deformation} is extended into the interior by a
vector Laplace equation with fixed inlet and outlet cross-sections
\cite{donea1982ale}. The meshes are generated separately for each case;
because mshr is not deterministic, meshes with the same resolution parameter
can differ slightly between cases, for example between 284 and 303 tetrahedra
for LF1.

Four families of geometries are considered (\Cref{fig:config}). A case is
defined by \(\boldsymbol{\mu}=(U_{\max},\mathbf{x}_b,\boldsymbol{\theta}_b)\),
the maximum inlet velocity and the axial and circumferential bump positions,
with \(\theta_{b,k}=\pi/2\) for aligned bumps and
\(\theta_{b,2}=\pi/2+\Delta\theta\) in the rotated configuration; the
parameter values of each family are given in \hyperlink{target:quantitative-results}{Section~\ref*{subsec:quantitative_results}}.

Equations~\eqref{eq:stokes_momentum}--\eqref{eq:stokes_incomp} are solved with
\(\mathbf{f}=\mathbf{0}\) and \(\nu=0.005\), a parabolic inlet profile
\(\mathbf{u}=\bigl(U_{\max}(1-r^2/R^2),0,0\bigr)\), \(r=\sqrt{y^2+z^2}\),
no slip on the wall, and \(p=0\) at the outlet, where the natural condition
\(\nu\,\partial_n\mathbf{u}-p\,\mathbf{n}=\mathbf{0}\) holds. In the notation
of \hyperlink{target:np}{Section~\ref*{sec:np}}, \(\Gamma_D\) is the union of the inlet and the wall and
\(\Gamma_p\) is the outlet. The saddle-point system is solved with MUMPS
\cite{amestoy2001mumps}. Because the problem is linear and the inflow is the
only nonzero boundary data, the solution for a fixed geometry is proportional
to \(U_{\max}\), so the bump positions and orientation constitute the
nontrivial part of the interpolation tests.
\subsection{Computational Configurations}
\hypertarget{target:config}{}
\label{sec:configurations}

\begin{figure*}[!h]
    \centering

    \subfloat[One-bump configuration.\label{fig:onebump}]{
        \includegraphics[
            width=0.48\textwidth,
            height=3.2cm,
        ]{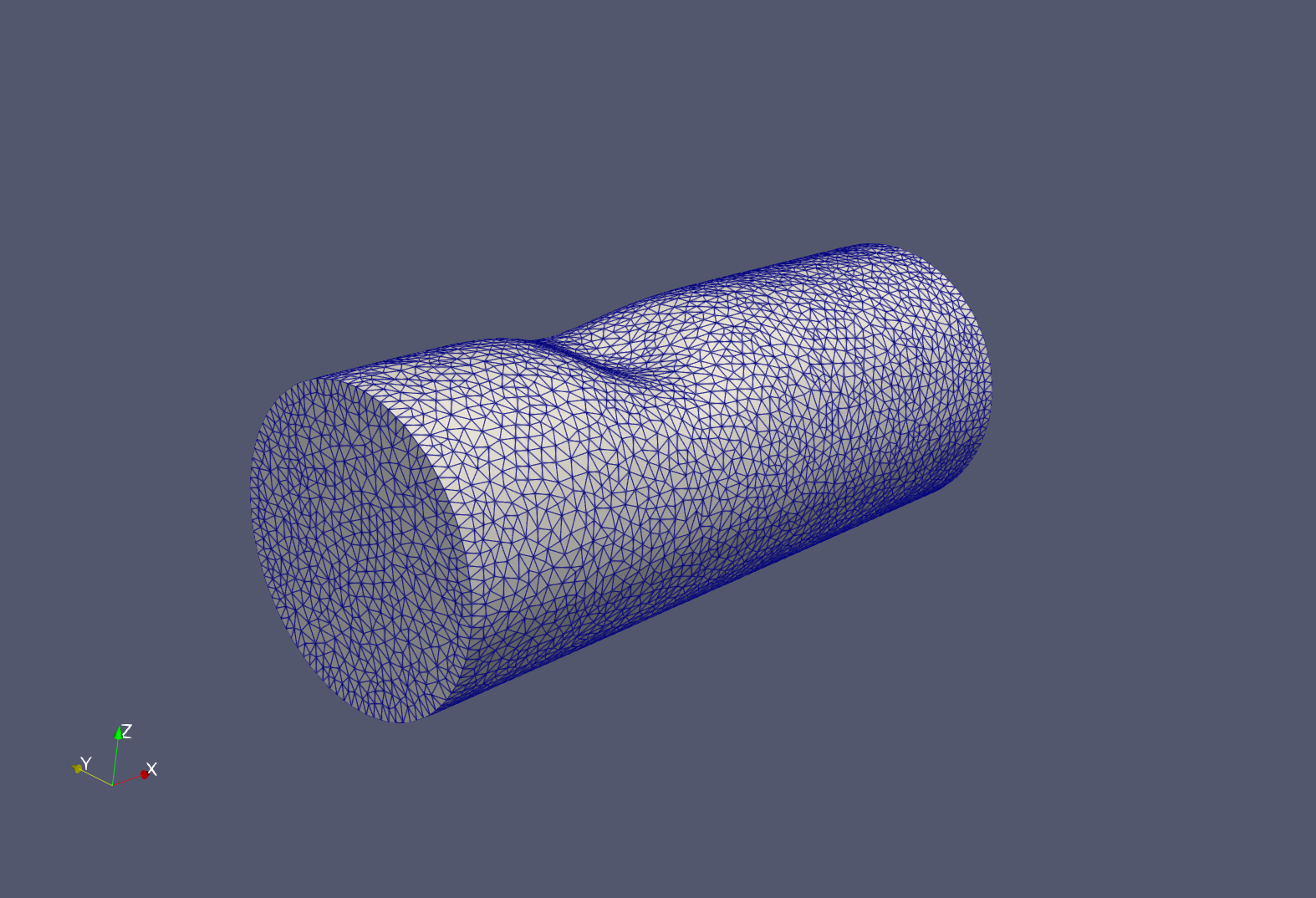}
    }
    \hfill
    \subfloat[Two-bump configuration.\label{fig:twobumps}]{
        \includegraphics[
            width=0.48\textwidth,
            height=3.2cm,
        ]{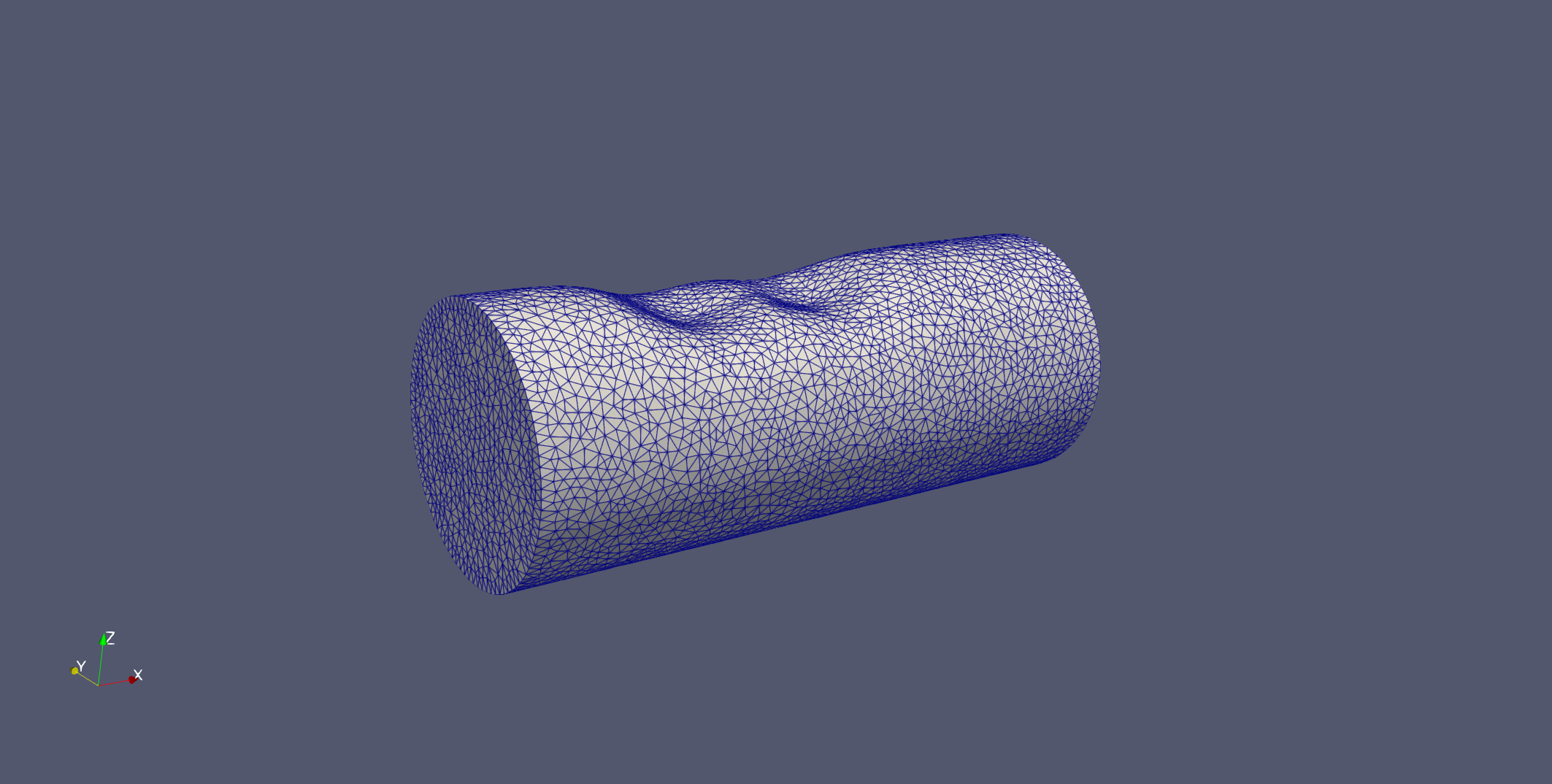}
    }

    \par\medskip

    \subfloat[Rotated two-bump configuration.\label{fig:2b_2ndrotates}]{
        \includegraphics[
            width=0.48\textwidth,
            height=3.2cm,
        ]{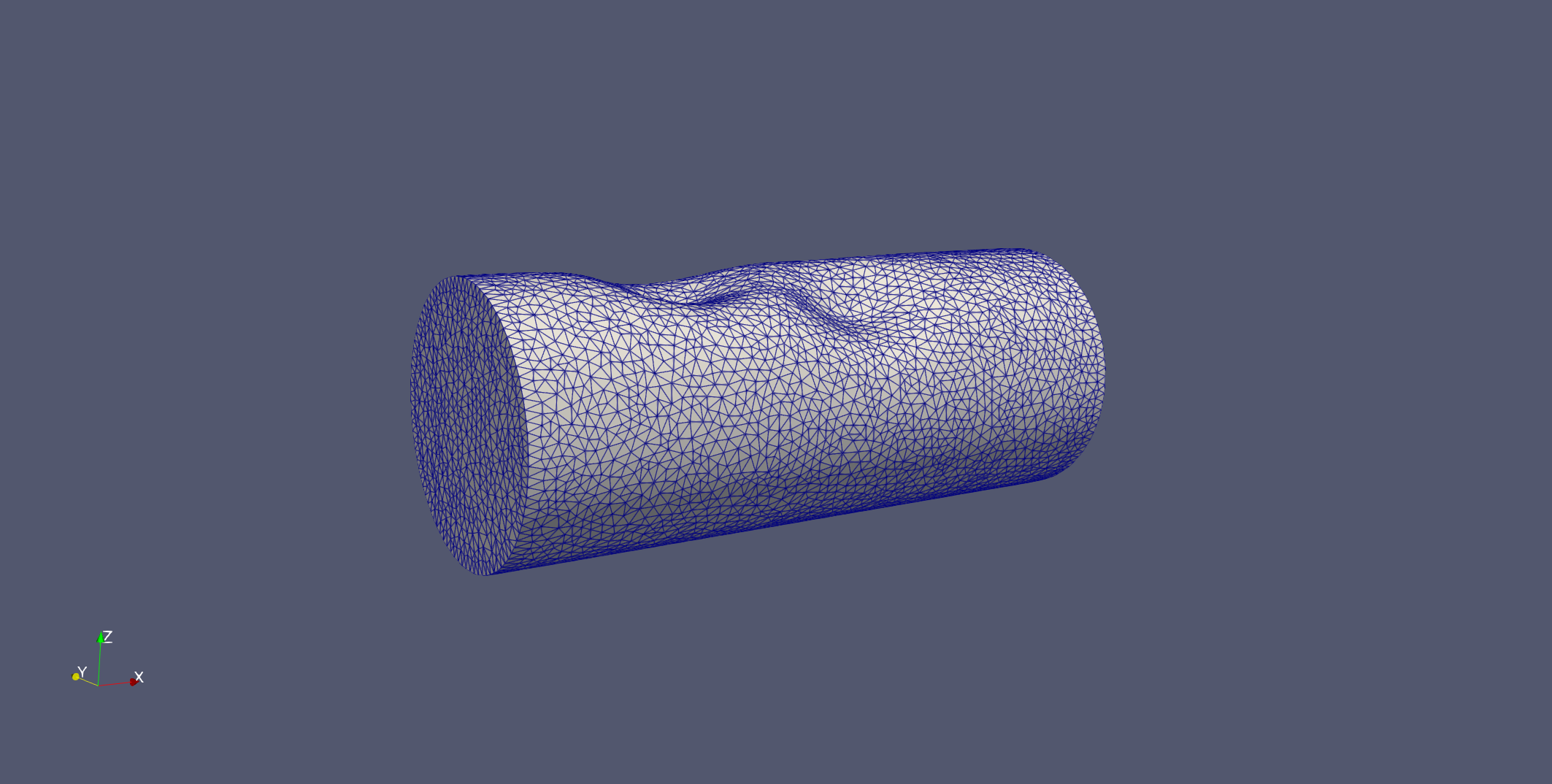}
    }
    \hfill
    \subfloat[Three-bump configuration.\label{fig:threebumps}]{
        \includegraphics[
            width=0.48\textwidth,
            height=3.2cm,
        ]{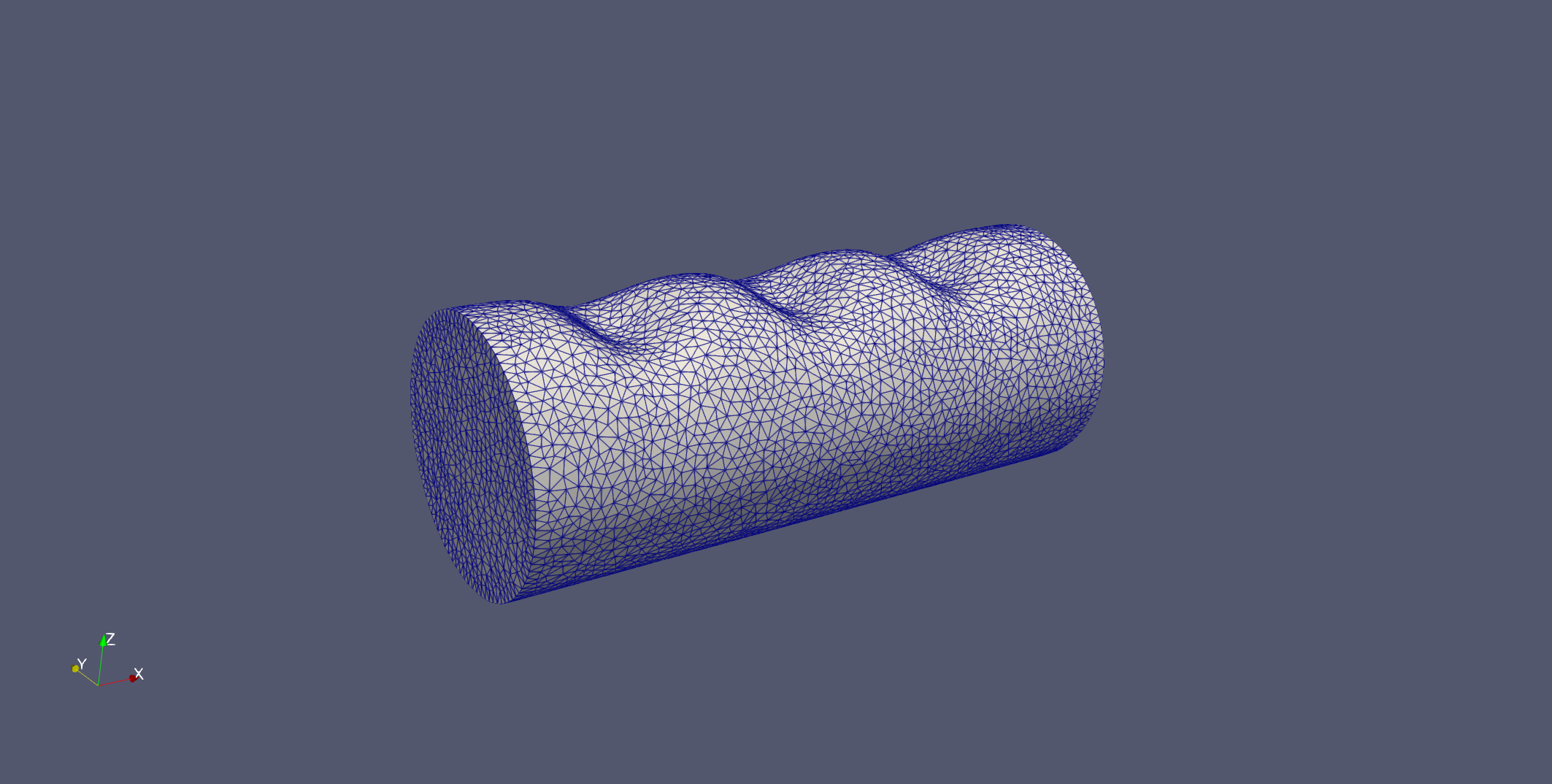}
    }
\FloatBarrier
    \caption{Computational-domain configurations: (a) one bump,
(b) two aligned bumps, (c) two bumps with the second bump rotated
circumferentially, and (d) three bumps.}
    \label{fig:config}
\end{figure*}

The four configurations are shown in \Cref{fig:config}(a)--(d).
Varying the number and circumferential
arrangement of the bumps changes the local geometry while retaining the same
underlying flow problem. These cases test how well NNLCI captures the changes in pressure and velocity caused by the geometry.

\subsection{Velocity-Magnitude Isosurfaces}
\hypertarget{target:Isosurfaces}{}
\label{sec:velocity_isosurfaces}

\begin{figure}[!h]
\centering

\subfloat[$\|u\|=0.06$\label{fig:velmag_006}]{
    \includegraphics[width=0.48\linewidth]{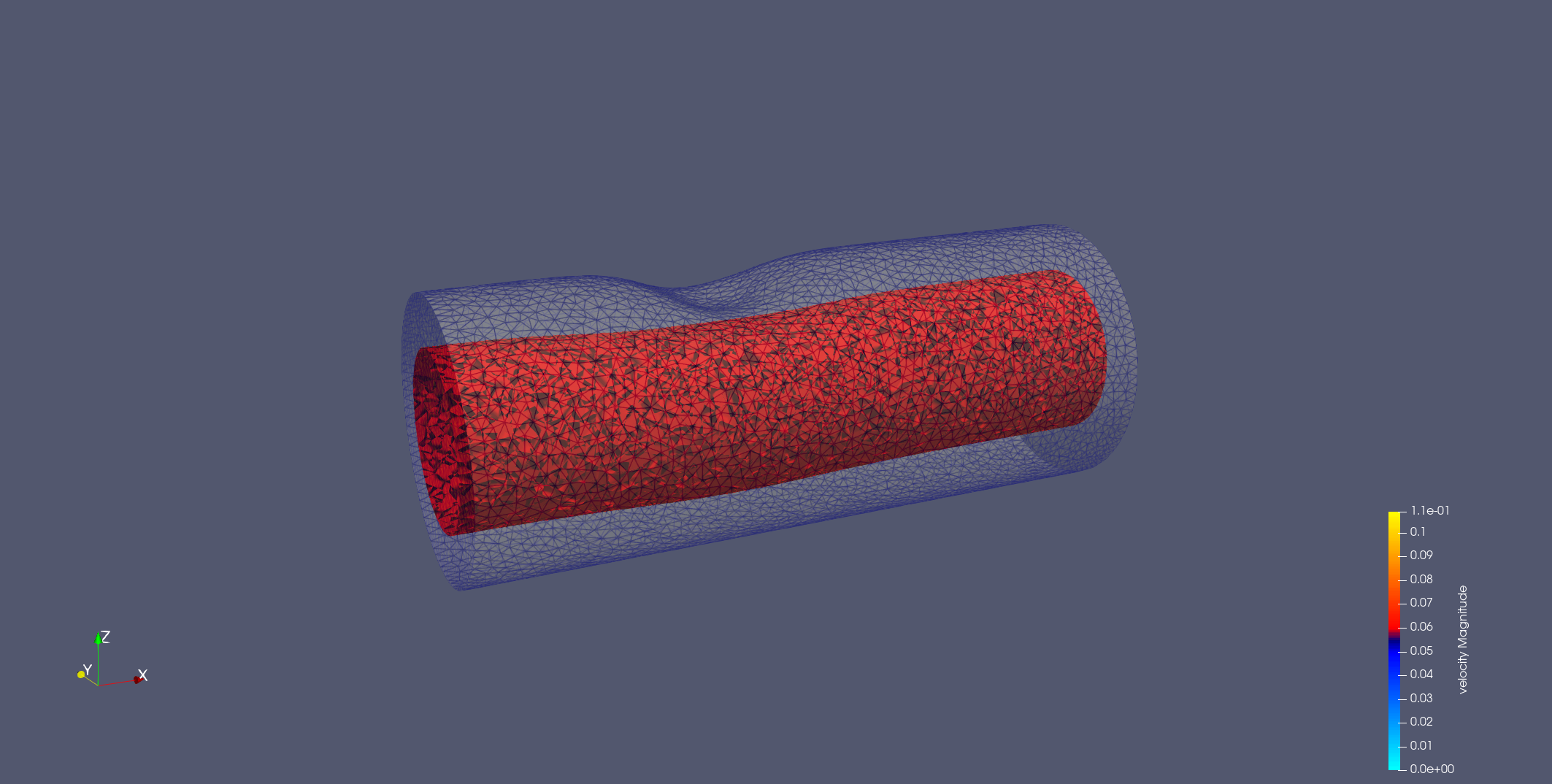}
}
\hfill
    \subfloat[$\|u\|=0.1$\label{fig:velmag_01}]{
    \includegraphics[width=0.48\linewidth]{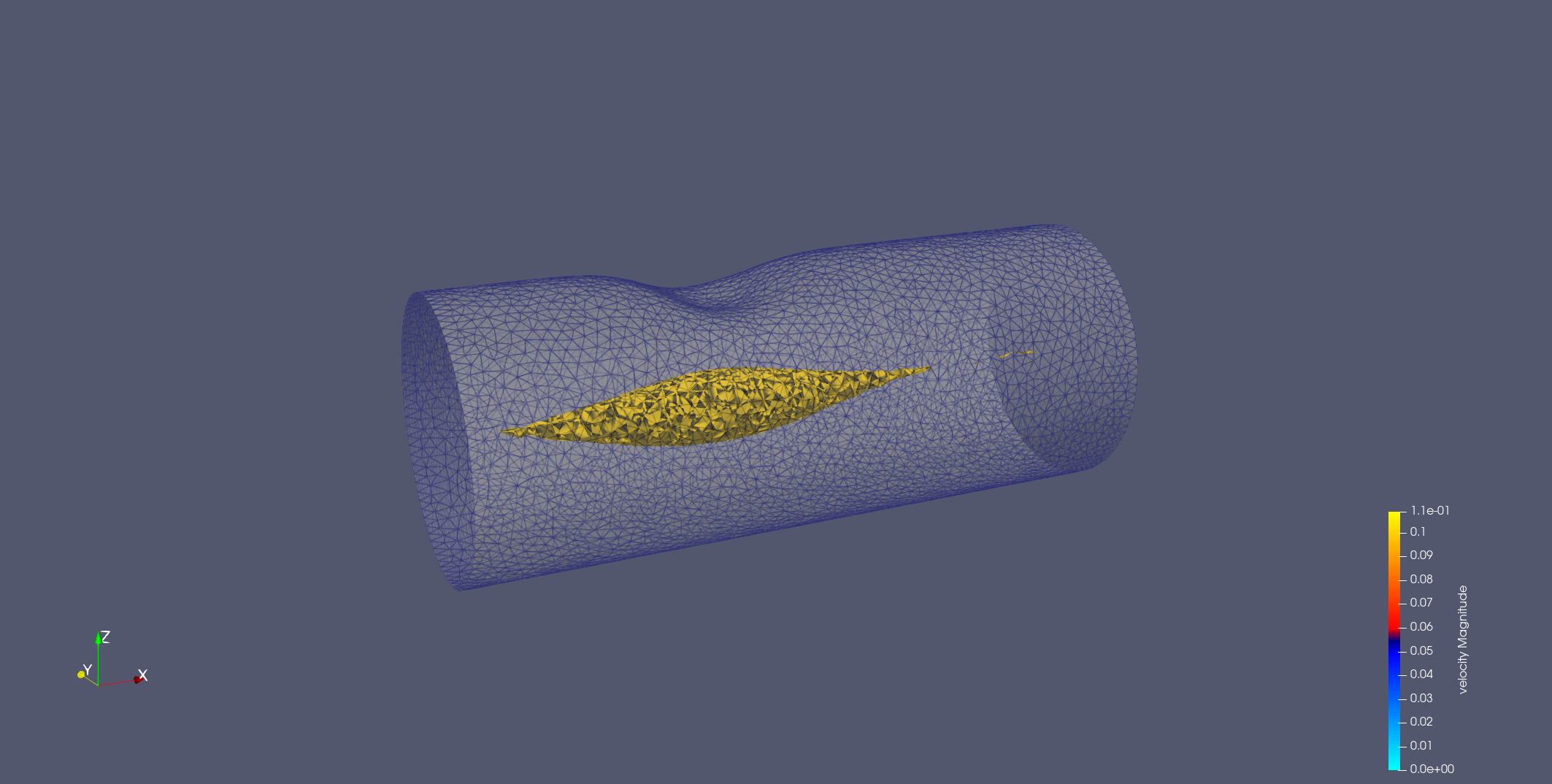}
}
\caption{Velocity-magnitude isosurfaces ($\|u\|$) for the one-bump pipe. The pipe surface is shown as a transparent wireframe; each subfigure shows an isosurface at the indicated $\|u\|$ level, illustrating contraction of the high-speed core and deflection induced by the constriction.}
\label{fig:velmag_onebump_contours}
\end{figure}

\begin{figure}[!h]
\centering

\subfloat[$\|u\|=0.06$\label{fig:velmag_2b_006}]{
    \includegraphics[width=0.48\linewidth]{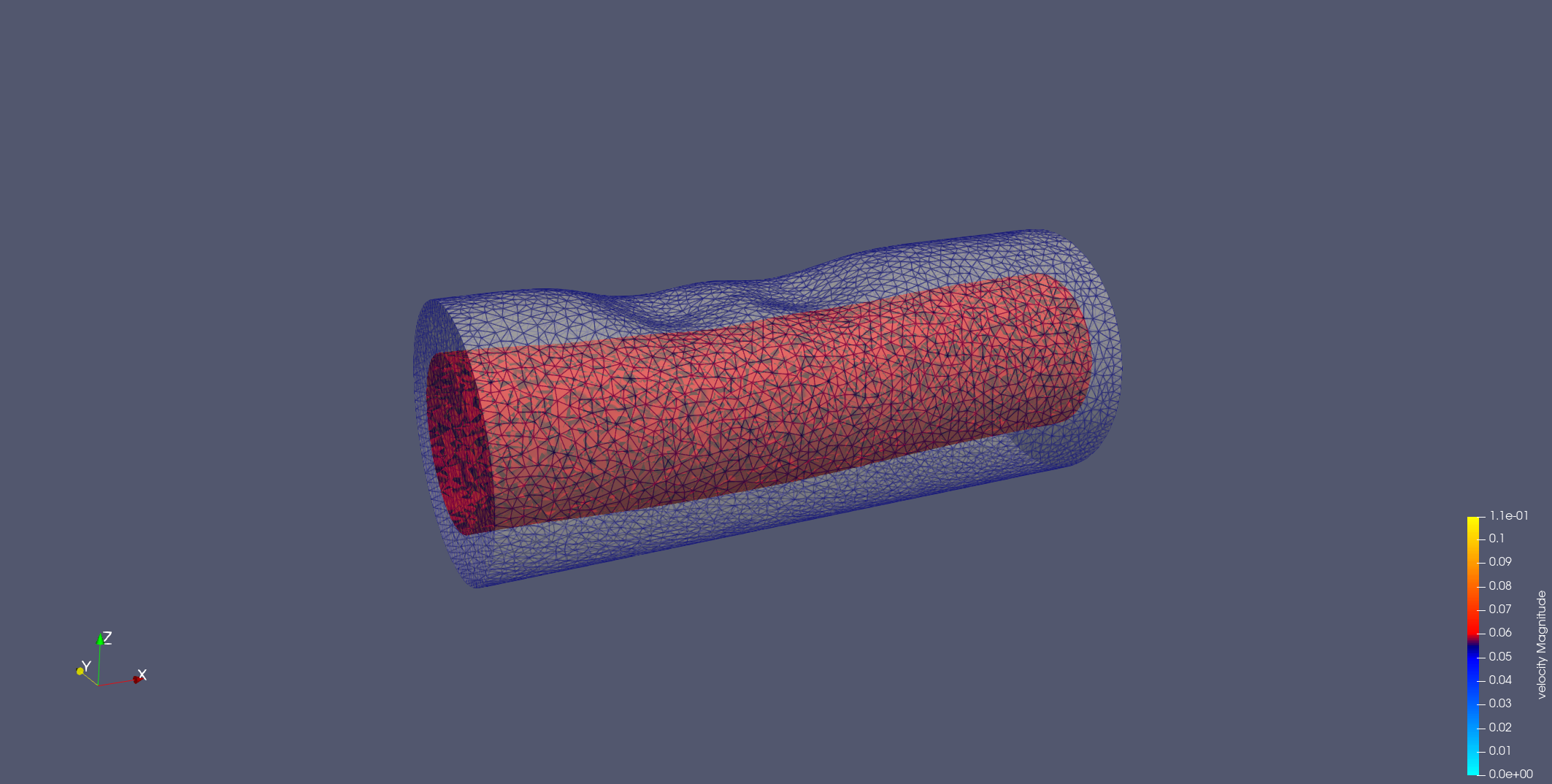}
}
\hfill
\subfloat[$\|u\|=0.1$\label{fig:velmag_2b_01}]{
    \includegraphics[width=0.48\linewidth]{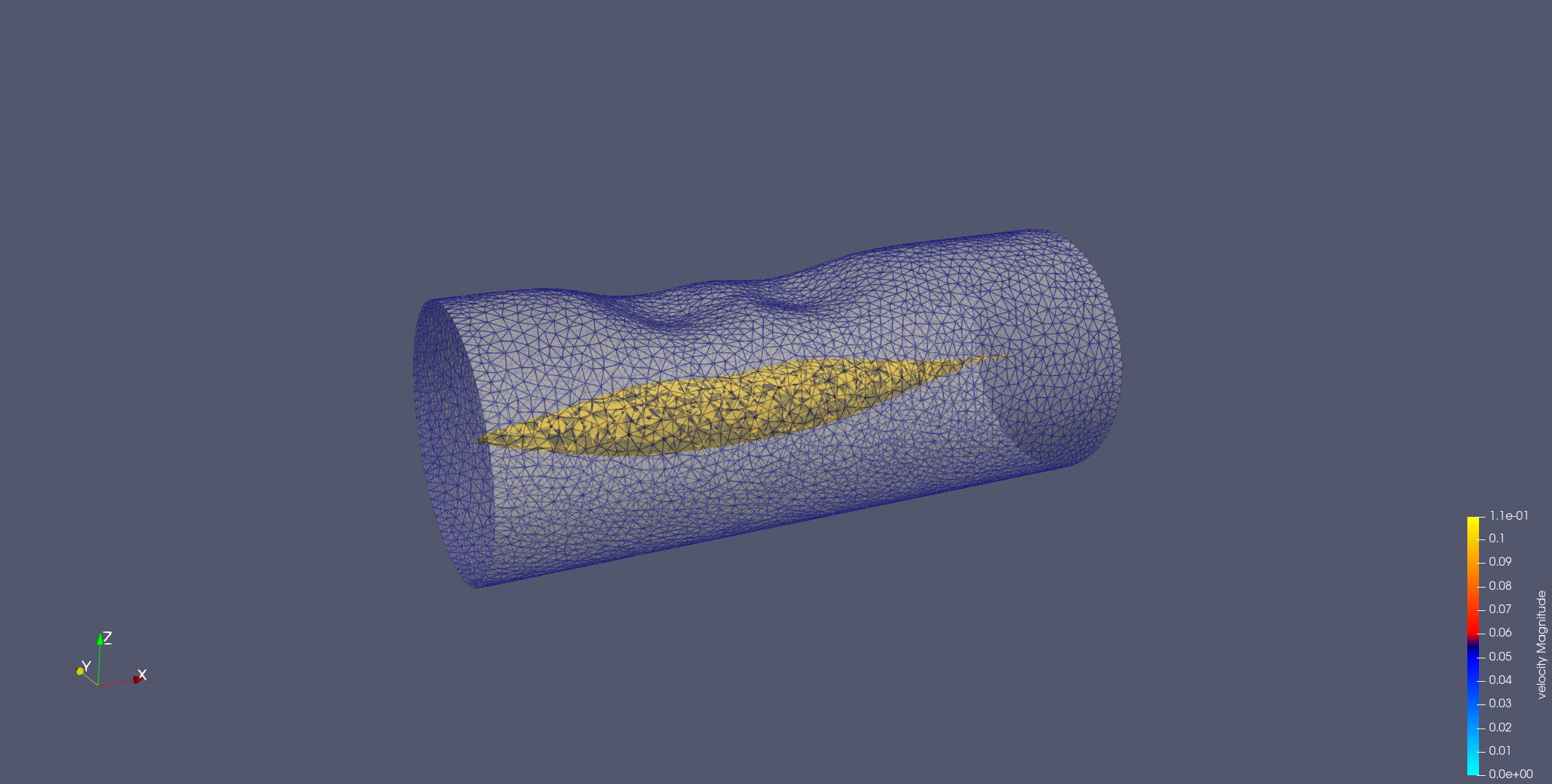}
}
\caption{Velocity-magnitude isosurfaces ($\|u\|$) for the two-bump pipe. The pipe surface is shown as a transparent wireframe; each subfigure shows the $\|u\|=\text{constant}$ isosurface at the indicated level, illustrating how the high-speed core contracts and becomes asymmetric due to the two local constrictions.}
\label{fig:velmag_twobumps_contours}
\end{figure}

\begin{figure}[!h]
\centering

\subfloat[$\|u\|=0.06$\label{fig:velmag_2br_006}]{
    \includegraphics[width=0.48\linewidth]{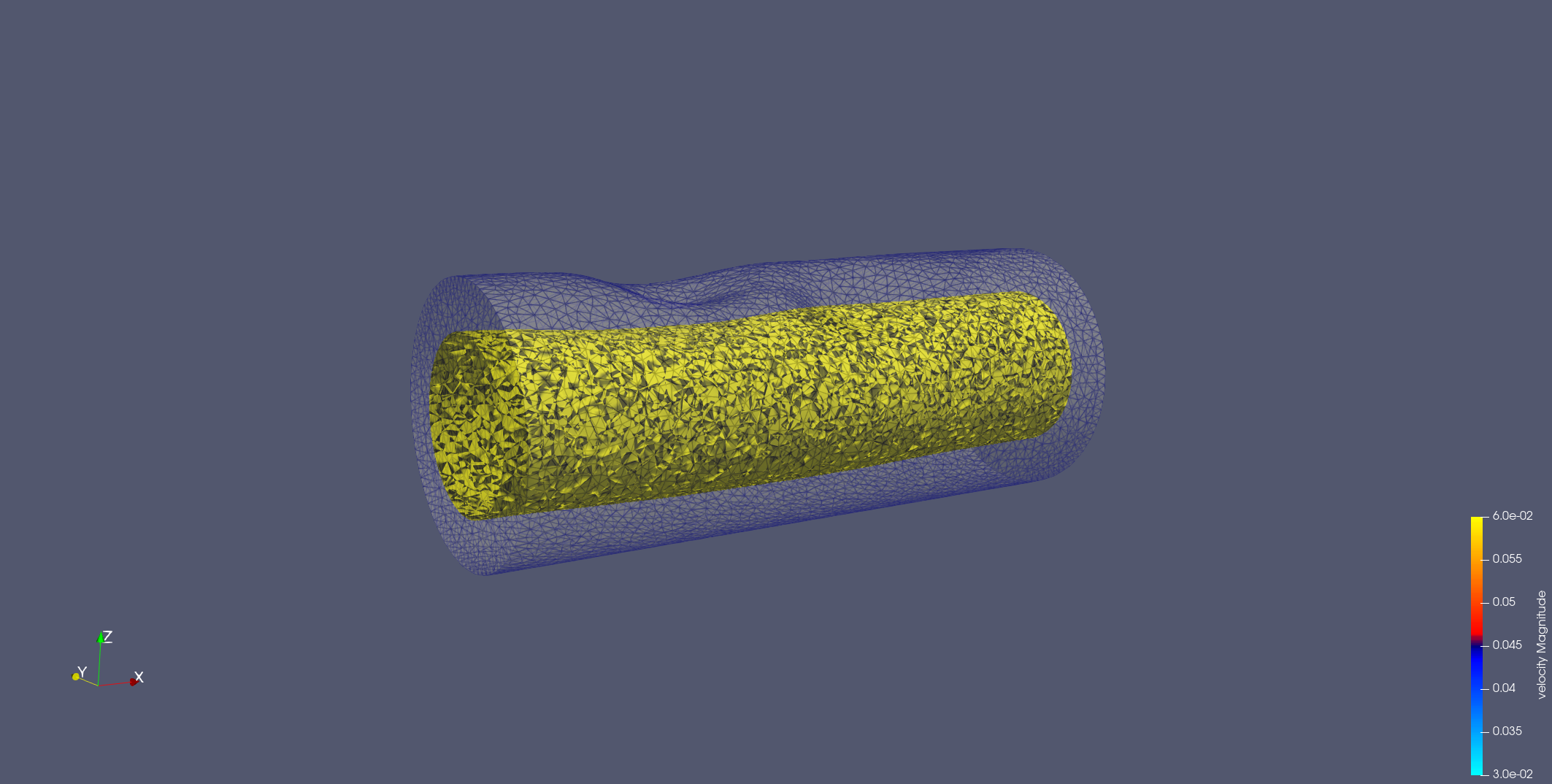}
}
\hfill
\subfloat[$\|u\|=0.1$\label{fig:velmag_2br_01}]{
    \includegraphics[width=0.48\linewidth]{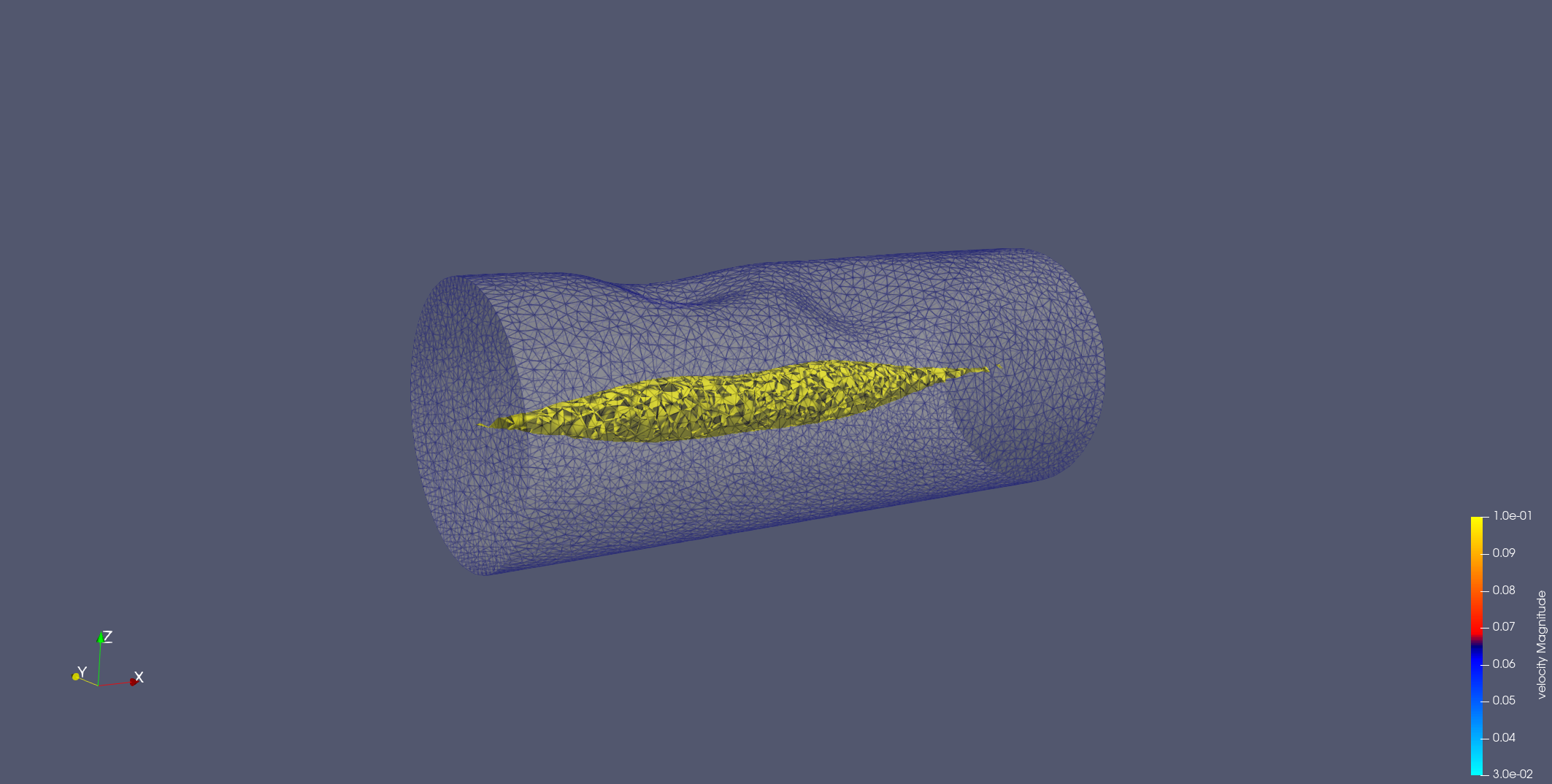}
}
\caption{Velocity-magnitude isosurfaces (\(\lVert\mathbf{u}\rVert\)) for the
two-bump pipe, with the second bump rigidly rotated about the
streamwise \(x\)-axis relative to the first bump. The pipe surface is shown
as a transparent wireframe. Each subfigure displays an
\(\lVert\mathbf{u}\rVert=\text{constant}\) isosurface at the indicated level,
illustrating the contraction and circumferential asymmetry of the high-speed
core induced by the two non-aligned local constrictions.}
\label{fig:velmag_twobumps_rotated_contours}
\end{figure}

\begin{figure}[!h]
\centering

\subfloat[$\|u\|=0.06$\label{fig:velmag_3b_006}]{
    \includegraphics[width=0.48\linewidth]{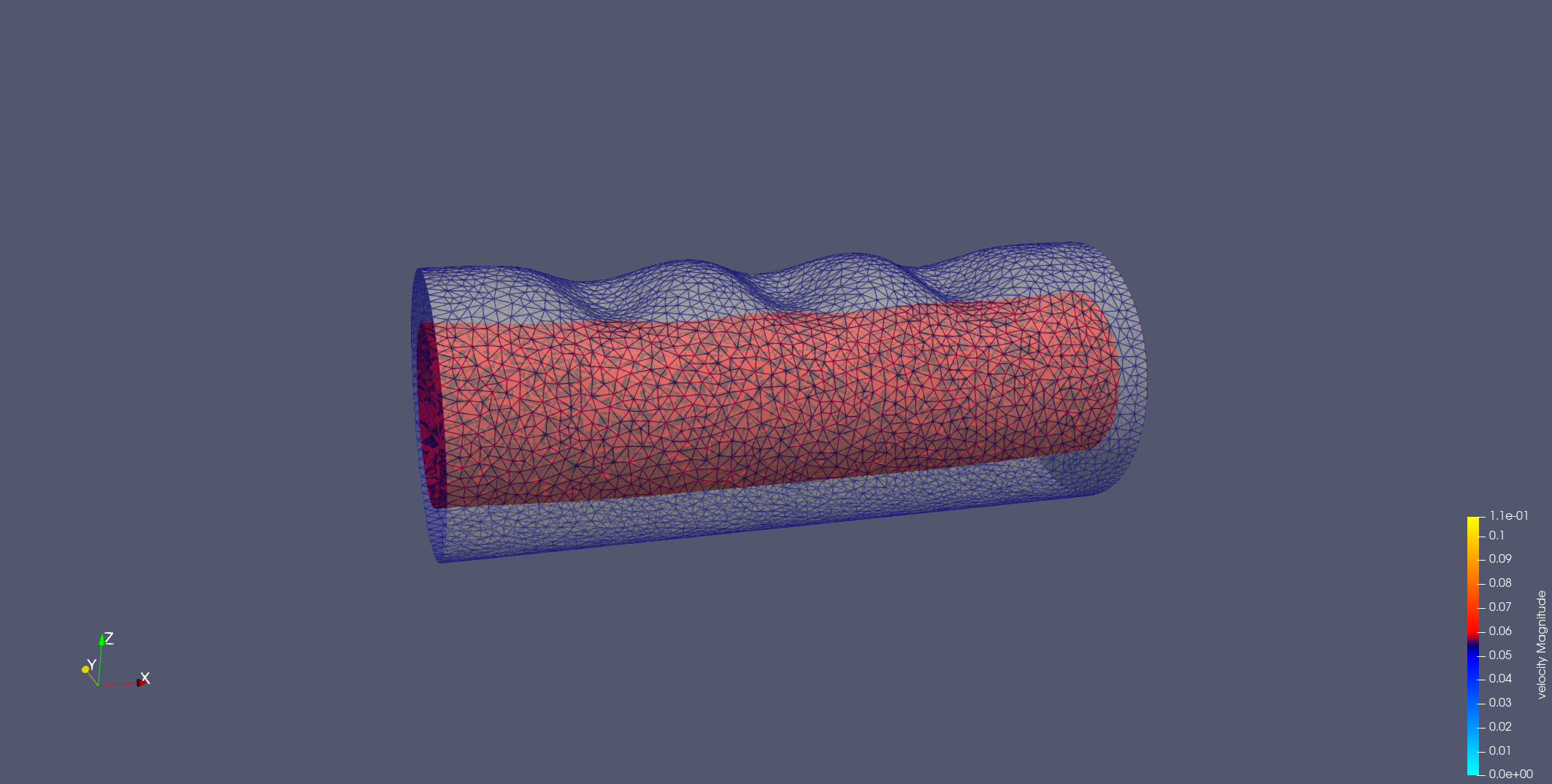}
}
\hfill
\subfloat[$\|u\|=0.1$\label{fig:velmag_3b_01}]{
    \includegraphics[width=0.48\linewidth]{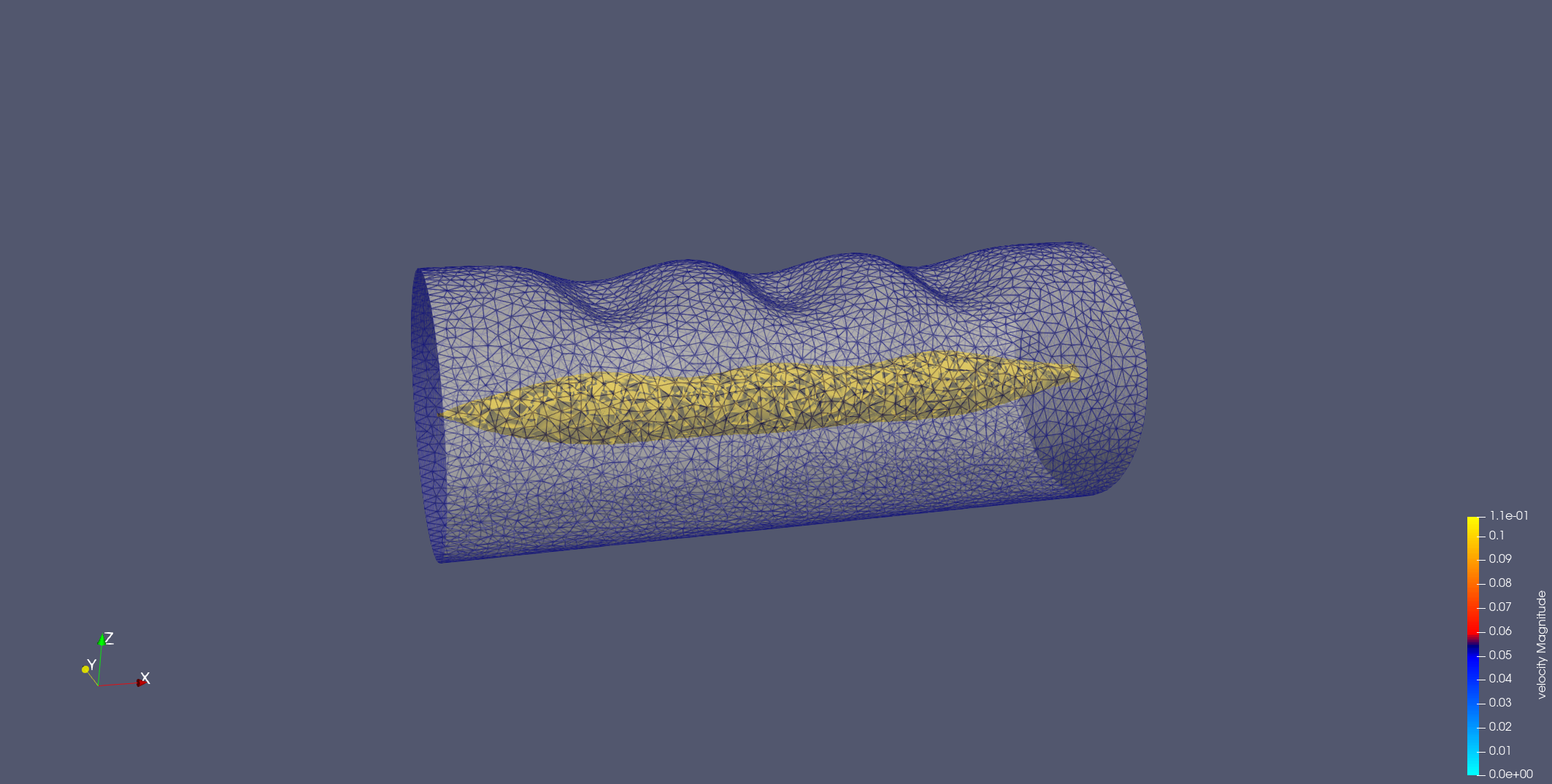}
}

\caption{Velocity-magnitude isosurfaces ($\|u\|$) for the three-bump pipe. The pipe surface is shown as a transparent wireframe; each subfigure shows the $\|u\|=\text{constant}$ isosurface at the indicated level, illustrating how the high-speed core contracts and becomes asymmetric due to the three local constrictions.}
\label{fig:velmag_threebumps_contours}
\end{figure}

To visualize the finite element flow fields, we compute the discrete velocity
magnitude
\[
\lVert\mathbf{u}_h\rVert_2
=
\sqrt{u_h^2+v_h^2+w_h^2}.
\]
The isosurfaces in
\Cref{fig:velmag_onebump_contours,fig:velmag_twobumps_contours,fig:velmag_twobumps_rotated_contours,fig:velmag_threebumps_contours}
provide a qualitative comparison of the four geometries. In every case, lower
isovalues occupy a relatively large portion of the pipe interior, whereas
higher isovalues isolate progressively smaller regions of high-speed flow. The figures show how the fast-flow core contracts and changes position as the magnitude threshold and wall geometry vary; they only illustrate the flow,
and the accuracy of NNLCI is measured with the error measures of \hyperlink{target:cnn-metrics}{Section~\ref*{subsubsec:cnn_metrics}}.

For the one-bump configuration
(\Cref{fig:velmag_onebump_contours}), the lower-level isosurfaces form a broad,
elongated interior region. As the isovalue increases, this region narrows and
becomes localized near the constriction. The displacement of the highest-level
isosurfaces from the undeformed pipe centerline indicates that the bump both
contracts and deflects the high-speed core.

For the two-aligned-bump configuration
(\Cref{fig:velmag_twobumps_contours}), the isosurfaces exhibit stronger axial
variation. At low isovalues, the visible region remains approximately
tube-like, while the higher-level surfaces become thinner and less uniform
along the pipe. Thus, the largest velocity magnitudes are confined to a
smaller interior region shaped by the two constrictions.

For the rotated two-bump configuration
(\Cref{fig:velmag_twobumps_rotated_contours}), the circumferential
misalignment of the second bump introduces an additional three-dimensional
asymmetry. With increasing isovalue, the high-speed region becomes more
localized and shifts circumferentially in response to the non-aligned wall
deformations.

For the three-bump configuration
(\Cref{fig:velmag_threebumps_contours}), the central fast-flow region narrows
over a longer axial portion of the pipe. Compared with the one-bump and two-bump
cases, the higher-level isosurfaces form a thinner and more confined structure,
showing the cumulative influence of the three local constrictions on the
location and extent of the high-speed core.

\section{Data Generation and NNLCI Construction}
\hypertarget{target:data-generation}{}
\label{sec:data_generation}
The workflow proceeds in three steps. For each parameter value, the Stokes
problem is solved on the two low-fidelity meshes and, for training, on the
high-fidelity mesh (Section~\ref{sec:stokes_fem}). The solutions are then sampled on
local stencils centered at the high-fidelity points to form input--output
pairs (Section~\ref{subsec:matrix_generation}), and a convolutional neural network
is trained to map the two low-fidelity patches to the high-fidelity state at
the stencil center (Section~\ref{subsec:cnn_error_evaluation}).

\subsection{Finite Element Formulation of the Stokes Problem}
\hypertarget{target:stokes-fem}{}
\label{sec:stokes_fem}

For each computational domain, the steady incompressible Stokes problem is
discretized with the Taylor--Hood spaces \(X_h\) and \(Q_h\) in
Section~\ref{sec:np} \cite{Fortin93}. The prescribed velocity boundary values define the trial space
\(X_{h,\mathbf{g}}=X_h\cap X_{\mathbf{g}}\), and the homogeneous conditions
define the test space \(X_{h,0}\). We seek
$(\mathbf{u}_h,p_h)\in X_{h,\mathbf{g}}\times Q_{h,0}$, where
$\mathbf{u}_h=(u_h,v_h,w_h)^T$, such that
\begin{align}
\nu(\nabla\mathbf{u}_h,\nabla\mathbf{v}_h)_{\Omega}
-(p_h,\nabla\cdot\mathbf{v}_h)_{\Omega}
&=(\mathbf{f},\mathbf{v}_h)_{\Omega}
&&\forall\,\mathbf{v}_h\in X_{h,0},
\label{eq:discrete_stokes_momentum}\\
(q_h,\nabla\cdot\mathbf{u}_h)_{\Omega}
&=0
&&\forall\,q_h\in Q_{h,0}.
\label{eq:discrete_stokes_incompressibility}
\end{align}
Here, $\nu$ is the kinematic viscosity, $\mathbf{f}$ is the body force, and
$(\cdot,\cdot)_{\Omega}$ denotes the $L^2(\Omega)$ inner product. The condition \(p_h=0\) on the outlet removes the additive pressure ambiguity.

The finite element calculations provide the three solution levels used by the
NNLCI pipeline: a coarse low-fidelity solution, a finer low-fidelity solution,
and a high-fidelity reference solution. At a sampling point
$\mathbf{x}_i=(x_i,y_i,z_i)$, the exported state is
\[
\bigl(
x_i,y_i,z_i,
p_h(\mathbf{x}_i),u_h(\mathbf{x}_i),v_h(\mathbf{x}_i),w_h(\mathbf{x}_i)
\bigr).
\]
Collecting these records over the selected sampling points and over all domain
configurations produces the raw finite element data used to construct the
NNLCI training and testing matrices.

\subsection{Generation of the Training and Testing Matrices}
\hypertarget{target:matrix-generation}{}
\label{subsec:matrix_generation}

The NNLCI matrices are constructed from three finite element solutions: the
coarsest low-fidelity solution LF1, the finer low-fidelity solution LF2, and the
high-fidelity solution HF. Each consists of a tetrahedral mesh, the velocity and
pressure functions, the cell centroids, and the cell volumes. Each input row
contains low-fidelity information on a local three-dimensional stencil, and the
corresponding output row contains the high-fidelity state at the stencil center.

Let \(\mathbf{u}_h=(u_h,v_h,w_h)^T\) and \(p_h\) denote the Taylor--Hood velocity
and pressure of Section~\ref{sec:np}. The state at a point \(x\) is
\[
    S_h(x)=\bigl[p_h(x),\,u_h(x),\,v_h(x),\,w_h(x)\bigr]\in\mathbb{R}^4,
\]
with the component ordering \([p,u,v,w]\) used throughout. When \(x\) lies inside
a tetrahedron \(T\), \(S_h(x)\) is evaluated directly from the four \(P_1\)
pressure and ten \(P_2\) velocity nodal values of \(T\), using the Lagrange basis
functions in barycentric coordinates \cite{boffi2013mixed}.

The samples are centered at the high-fidelity cell centroids
\(x_i^{HF}\), \(i=1,\ldots,N_{HF}\). The cell volumes \(V_i^{HF}\) are not input
features; they are used only in the volume-weighted training loss and error
metrics. For each center, local mesh sizes \(h_{1,i}\) and \(h_{2,i}\) are
estimated from the LF1 and LF2 centroid clouds. For a given level, let
\(d_1,\ldots,d_{26}\) be the distances from \(x_i^{HF}\) to its 26 nearest
non-coincident centroids. In a regular \(3\times3\times3\) cube of spacing \(h\),
the 26 non-center points consist of six face neighbors at distance \(h\), twelve
edge neighbors at \(\sqrt{2}h\), and eight corner neighbors at \(\sqrt{3}h\).
Matching the sum of these distances gives
\[
    h=\frac{\sum_{j=1}^{26}d_j}{6+12\sqrt{2}+8\sqrt{3}} .
\]
The LF1 spacing \(h_{1,i}\) sets the stencil size, and \(h_{2,i}\) is an
additional resolution feature.

The stencil consists of the 27 points
\[
    z_{i,m}=x_i^{HF}+h_{1,i}\alpha_m,\qquad \alpha_m\in\{-1,0,1\}^3,
    \quad m=1,\ldots,27,
\]
with the center offset placed last. The LF1 and LF2 solutions are evaluated on this shared stencil, so that both
describe the same physical neighborhood, with the finer solution providing
higher local resolution. This alignment preserves the converging-input structure of NNLCI.
                         
At a stencil point inside the low-fidelity mesh, the finite element state is
evaluated directly. At a point outside the mesh, either beyond the physical
boundary or between the polygonal low-fidelity boundary and the curved physical
boundary, the value is reconstructed from nearby low-fidelity data by a local
weighted quadratic least-squares fit. Filling these points, rather than
discarding the stencil and thus the corresponding row of the matrix, retains the high-fidelity centers near the inlet, outlet, and
wall: in a representative two-aligned-bump training matrix, only 17.7\% of the
centers have stencils entirely inside both low-fidelity meshes, whereas the
hybrid evaluation provides inputs at all of them.

The reconstruction uses the finite element states at the low-fidelity cell
centroids \(c_j\). Around a stencil point \(z\) outside the mesh, the support is searched within a
radius that starts at \(2h_{\mathrm{supp}}\), with \(h_{\mathrm{supp}}=h_{1,i}\)
for LF1 and \(h_{2,i}\) for LF2, and grows by a factor of 1.2 until 20 centroids
are found, up to \(5h_{\mathrm{supp}}\); if fewer are found, the stencil is rejected.
This bound keeps the reconstruction strictly local. In the shifted coordinates
\(\xi_j=c_j-z\), each component \(q\in\{p,u,v,w\}\) is fitted by
\[
    q(\xi)=a_0+a_1\xi_1+a_2\xi_2+a_3\xi_3+a_4\xi_1^2+a_5\xi_2^2+a_6\xi_3^2
    +a_7\xi_1\xi_2+a_8\xi_1\xi_3+a_9\xi_2\xi_3 ,
\]
so that the value at \(z\) is \(a_0\). With the basis rows \(\Phi(\xi_j)\), the
weights \(\omega_j=\max(\|c_j-z\|,\varepsilon)^{-2}\)
\cite{levin1998mls}, and \(W=\mathrm{diag}(\omega_j)\), the coefficients solve the
regularized system
\[
    \left(\Phi^T W\Phi+\eta D\right)a=\Phi^T W q,
\]
with \(\eta=10^{-10}\) and \(D=\operatorname{diag}(10^{-3},1,\ldots,1)\), so that
the constant coefficient is regularized more weakly. The stencil is rejected if the
condition number of this system exceeds \(10^{12}\).

The resulting patches are
\[
    P_i^{(\ell)}
    =
    \bigl[S^{LF\ell}(z_{i,1});\,\ldots;\,S^{LF\ell}(z_{i,27})\bigr]
    \in\mathbb{R}^{27\times4},
    \qquad \ell=1,2,
\]
the target is the high-fidelity state at the center,
\(Y_i=S^{HF}(x_i^{HF})\in\mathbb{R}^4\), and the input vector is
\[
    X_i
    =
    \left[
    \operatorname{vec}\!\left((P_i^{(1)})^{T}\right),\,
    \operatorname{vec}\!\left((P_i^{(2)})^{T}\right),\,
    h_{1,i},\,h_{2,i}
    \right]\in\mathbb{R}^{218}.
\]
Rows containing non-finite values are rejected. In the matrices of this study, no row was rejected for insufficient support, conditioning, or
non-finite values.

Two quality criteria are applied to the training matrices. First, since LF2 is
finer than LF1, the ratio \(h_{1,i}/h_{2,i}\) is expected to be about
\((730/300)^{1/3}\approx1.34\) for the present meshes; its median in the
two-aligned-bump training matrices is 1.37--1.41, and rows with
\(h_{1,i}/h_{2,i}<1.1\) are rejected. Second, for each patch \(P_i\in\{P_i^{(1)},
P_i^{(2)}\}\), with entries \(P_{i,m,c}\) at stencil point \(m\) and component
\(c\), the patch-jump indicator
\begin{equation}
    J(P_i)
    =
    \max_{m\neq 27}\;\max_{c}\;
    \frac{\lvert P_{i,m,c}-P_{i,27,c}\rvert}{s_c}
    \label{eq:patch_jump}
\end{equation}
is computed, where \(s_p\) is the largest pressure magnitude and
\(s_u=s_v=s_w\) the largest velocity magnitude over the centroid values of the
corresponding low-fidelity solution, and rows with \(J>3\) for either patch are
rejected. Measuring the jumps against fixed field scales, rather than against the
center value, keeps patches next to the no-slip wall, where the velocity is small.
The criterion removes the rare patches whose variation is large compared with the
field scales and would otherwise dominate the min--max scaling of the network
inputs, 0.4\%--0.7\% of the centers in the two-aligned-bump training matrices.

The test matrices are built in the same way but without these two criteria, so
that the trained network is evaluated at all high-fidelity centers, including
those next to the wall. In all matrices, the rows, centers, and cell volumes are
stored in the same order, and all entries are kept in physical variables; the
scaling used by the network is applied afterward.

\subsection{CNN-Based NNLCI Surrogate and Common Error Evaluation}
\hypertarget{target:cnn-error-eval}{}
\label{subsec:cnn_error_evaluation}

The matrices of Section~\ref{subsec:matrix_generation} define a supervised regression
problem: given the two low-fidelity patches and the two local mesh sizes
associated with a high-fidelity center \(x_i^{HF}\), predict the high-fidelity
state \(Y_i\) at that center. We approximate this map by a convolutional neural
network (CNN) \(\mathcal{F}_\theta\), where \(\theta\) denotes all trainable
weights and biases. A convolutional architecture is natural here because each
patch is sampled on a regular \(3\times3\times3\) stencil and can be processed
like a small three-dimensional image, with the four solution components \(p\),
\(u\), \(v\), and \(w\) playing the role of color channels.

The same methodology is applied to the one-bump, two-aligned-bump, rotated
two-bump, and three-bump configurations. For each configuration, a separate
network is trained on the training matrices of that configuration and
evaluated on its held-out test matrix, which plays no role in training,
validation, or normalization. The input representation, network architecture,
normalization, optimization settings, and error measures are identical across
the four studies; only the number of training matrices differs, as listed in
Section~\ref{subsec:quantitative_results}.

\subsubsection{Input and Target Representation}
\hypertarget{target:cnn-input}{}
\label{subsubsec:cnn_input}

The network receives the input vector \(X_i\in\mathbb{R}^{218}\) and the target
\(Y_i=[p_i^{HF},u_i^{HF},v_i^{HF},w_i^{HF}]\) of Section~\ref{subsec:matrix_generation}.
In \(X_i\), each patch is flattened point by point: the four components
\([p,u,v,w]\) of stencil point \(z_{i,m}\) occupy four consecutive entries and
are followed by those of \(z_{i,m+1}\), with the 27 stencil points in the order
of the offsets and the center last.

\paragraph{Arranging each patch as a three-dimensional grid}
This storage order does not reflect which stencil points are neighbors in
space. Before entering the network, each patch is rearranged into an
array \(\mathcal{P}_i^{(\ell)}\in\mathbb{R}^{4\times3\times3\times3}\),
\(\ell=1,2\), defined by
\begin{equation}
    \mathcal{P}_i^{(\ell)}[c,a,b,d]
    =
    S_c^{LF\ell}\!\left(
    x_i^{HF}+h_{1,i}\,(a-2,\;b-2,\;d-2)
    \right),
    \qquad
    c=1,\ldots,4,\quad a,b,d\in\{1,2,3\},
    \label{eq:cnn_patch_grid}
\end{equation}
where \(S_c^{LF\ell}\) is the \(c\)-th component of \(S^{LF\ell}=[p,u,v,w]\).
The indices \(a\), \(b\), and \(d\) run along the \(x\)-, \(y\)-, and
\(z\)-directions; grid point \((2,2,2)\) is the stencil center; and entries that
are adjacent in the array are adjacent in physical space. The first index \(c\)
selects the solution component and is referred to as a \emph{channel}. The two
mesh sizes are collected
in \(e_i=[h_{1,i},h_{2,i}]\in\mathbb{R}^{2}\).

\paragraph{Normalization}
The entries of \(X_i\) and \(Y_i\) differ in magnitude; in particular, the
transverse velocity components are much smaller than the streamwise component.
Each of the 218 input features and each of the four target components is mapped to \([0,1]\) by min--max normalization,
\begin{equation}
    A^{s}
    =
    \frac{A-A_{\min}}{A_{\max}-A_{\min}},
    \label{eq:cnn_minmax_scaling}
\end{equation}
where \(A_{\min}\) and \(A_{\max}\) are computed separately for each feature
over all rows of the training matrices; if \(A_{\max}-A_{\min}<10^{-12}\), the
denominator is replaced by one. Because the scaling is applied to \(X_i\) before
the rearrangement~\eqref{eq:cnn_patch_grid}, each combination of stencil point
and solution component has its own constants. The same constants are applied
unchanged to the test data, whose scaled values may fall slightly
outside \([0,1]\). We denote the normalized patches and mesh sizes by
\(\mathcal{P}_i^{(\ell),s}\) and \(e_i^{s}\). The network returns normalized
outputs \(\widehat{Y}_i^{s}\in\mathbb{R}^4\), which are mapped back to the
solver variables by
\begin{equation}
    \widehat{Y}_i
    =
    \widehat{Y}_i^{s}\odot(Y_{\max}-Y_{\min})+Y_{\min},
    \label{eq:cnn_inverse_scaling}
\end{equation}
where \(\odot\) denotes componentwise multiplication.

\subsubsection{Network Architecture}
\hypertarget{target:cnn-network}{}
\label{subsubsec:cnn_architecture}

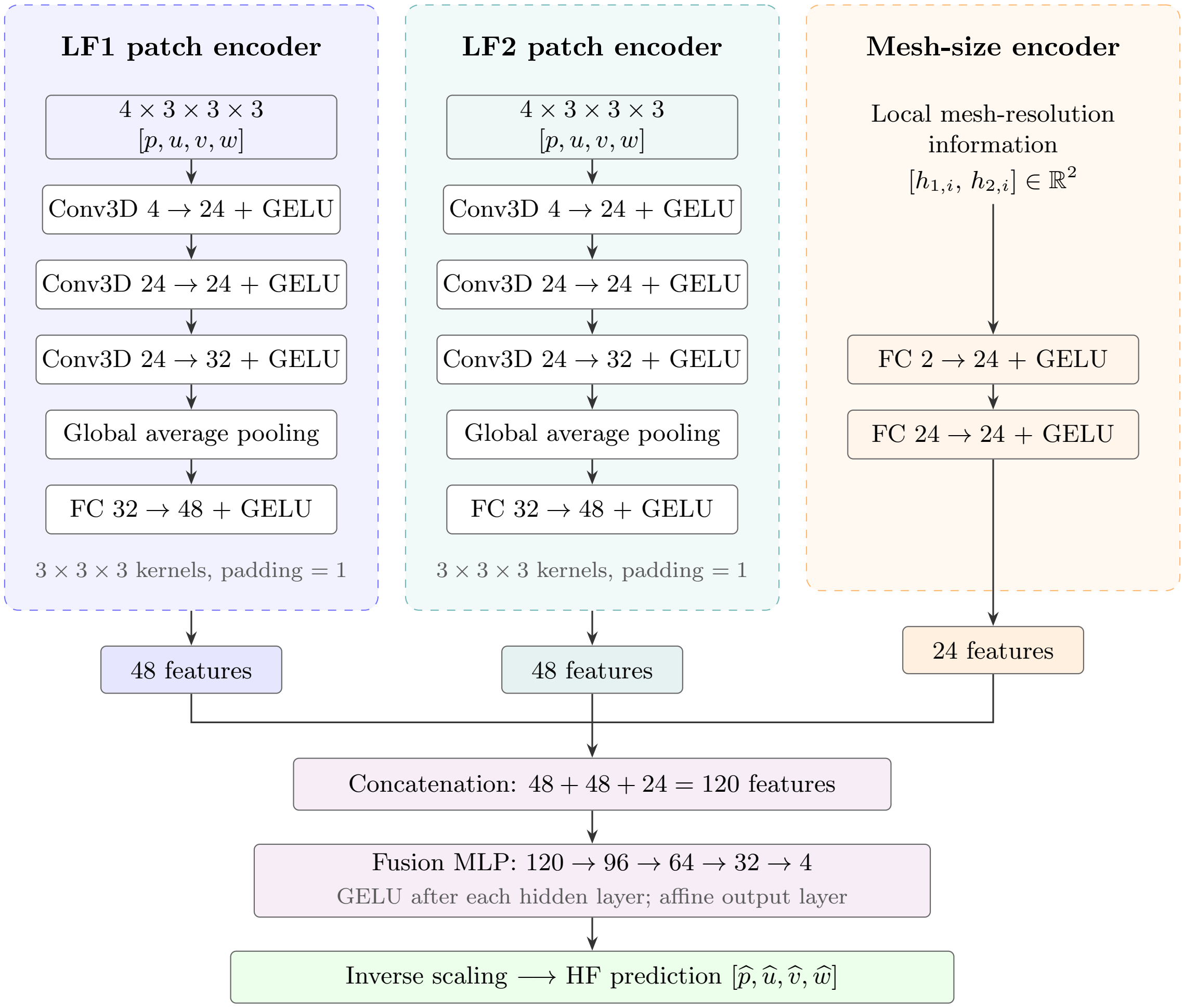
\begin{figure}[htbp]
\centering
\begin{tikzpicture}[
    font=\small,
    node distance=0.32cm,
    >={Stealth[length=1.8mm,width=1.4mm]},
    layer/.style={draw=black!60, rounded corners=2pt, fill=white,
                  minimum width=3.7cm, minimum height=0.62cm, align=center,
                  inner sep=2pt},
    feat/.style={draw=black!60, rounded corners=2pt, minimum width=2.3cm,
                 minimum height=0.6cm, align=center},
    wide/.style={draw=black!60, rounded corners=2pt, minimum height=0.66cm,
                 align=center, inner xsep=10pt},
    note/.style={font=\footnotesize, text=black!65, align=center},
    title/.style={font=\normalsize\bfseries, align=center, text height=1.8ex, text depth=0.5ex},
    box/.style={draw=#1!55, dashed, rounded corners=5pt, fill=#1!5, minimum width=4.75cm,
                inner sep=7pt},
    flow/.style={->, black!80, semithick},
  ]

  \node[title] (t1) at (0,0) {LF1 patch encoder};
  \node[layer, fill=blue!6, below=0.30cm of t1] (in1)
       {$4\times3\times3\times3$\\$[p,u,v,w]$};
  \node[layer, below=of in1] (c11) {Conv3D $4\to24$ + GELU};
  \node[layer, below=of c11] (c12) {Conv3D $24\to24$ + GELU};
  \node[layer, below=of c12] (c13) {Conv3D $24\to32$ + GELU};
  \node[layer, below=of c13] (p1)  {Global average pooling};
  \node[layer, below=of p1]  (f1)  {FC $32\to48$ + GELU};
  \node[note,  below=0.22cm of f1] (k1) {$3\times3\times3$ kernels, padding $=1$};
  \foreach \a/\b in {in1/c11, c11/c12, c12/c13, c13/p1, p1/f1}
    \draw[flow] (\a) -- (\b);

  \node[title] (t2) at (5.1,0) {LF2 patch encoder};
  \node[layer, fill=teal!6, below=0.30cm of t2] (in2)
       {$4\times3\times3\times3$\\$[p,u,v,w]$};
  \node[layer, below=of in2] (c21) {Conv3D $4\to24$ + GELU};
  \node[layer, below=of c21] (c22) {Conv3D $24\to24$ + GELU};
  \node[layer, below=of c22] (c23) {Conv3D $24\to32$ + GELU};
  \node[layer, below=of c23] (p2)  {Global average pooling};
  \node[layer, below=of p2]  (f2)  {FC $32\to48$ + GELU};
  \node[note,  below=0.22cm of f2] (k2) {$3\times3\times3$ kernels, padding $=1$};
  \foreach \a/\b in {in2/c21, c21/c22, c22/c23, c23/p2, p2/f2}
    \draw[flow] (\a) -- (\b);

  \node[title] (t3) at (10.2,0) {Mesh-size encoder};
  \node[align=center, below=0.30cm of t3] (in3)
       {Local mesh-resolution\\information\\[2pt]
        $[h_{1,i},\,h_{2,i}]\in\mathbb{R}^2$};
  \node[layer, fill=orange!8] (m1) at (in3 |- c13) {FC $2\to24$ + GELU};
  \node[layer, fill=orange!8] (m2) at (in3 |- p1)  {FC $24\to24$ + GELU};
  \draw[flow] (in3) -- (m1);
  \draw[flow] (m1) -- (m2);

  \begin{scope}[on background layer]
    \node[box=blue,   fit=(t1)(in1)(f1)(k1)] (B1) {};
    \node[box=teal,   fit=(t2)(in2)(f2)(k2)] (B2) {};
    \node[box=orange, fit=(t3)(in3)(m2) (in3 |- k2)] (B3) {};
  \end{scope}

  \node[feat, fill=blue!10]   (o1) at ($(B1.south)+(0,-0.75)$) {48 features};
  \node[feat, fill=teal!10]   (o2) at ($(B2.south)+(0,-0.75)$) {48 features};
  \node[feat, fill=orange!12] (o3) at ($(B3.south)+(0,-0.75)$) {24 features};
  \draw[flow] (f1.south |- B1.south) -- (o1);
  \draw[flow] (f2.south |- B2.south) -- (o2);
  \draw[flow] (m2) -- (m2 |- B3.south) -- (o3);

  \node[wide, fill=violet!7, minimum width=7.6cm] (cat) at ($(o2.south)+(0,-1.15)$)
       {Concatenation: $48+48+24=120$ features};
  \coordinate (merge) at ($(cat.north)+(0,0.45)$);
  \draw[black!80, semithick] (o1.south) |- (merge);
  \draw[black!80, semithick] (o3.south) |- (merge);
  \draw[flow] (o2.south) -- (cat.north);

  \node[wide, fill=violet!7, minimum width=8.6cm, below=0.42cm of cat] (mlp)
       {Fusion MLP: $120\to96\to64\to32\to4$\\[1pt]
        {\footnotesize\color{black!65}GELU after each hidden layer; affine output layer}};
  \node[wide, fill=green!8, minimum width=9.2cm, below=0.42cm of mlp] (out)
       {Inverse scaling $\longrightarrow$ HF prediction
        $[\widehat{p},\widehat{u},\widehat{v},\widehat{w}]$};
  \draw[flow] (cat) -- (mlp);
  \draw[flow] (mlp) -- (out);
\end{tikzpicture}
\caption{CNN architecture for local high-fidelity Stokes reconstruction. The two
low-fidelity solutions are evaluated at the same physical $3\times3\times3$
stencil. Independent 3D convolutional encoders extract 48 features per
low-fidelity level, while a separate fully connected branch extracts 24
features from the local grid spacings. The combined 120 features are mapped
to pressure and the three velocity components at the stencil center.}
\label{fig:cnn_architecture}
\end{figure}

The network consists of three parallel branches followed by a fusion network,
as shown in \Cref{fig:cnn_architecture} and listed layer by layer in
\Cref{tab:cnn_architecture}. The LF1 branch \(E_1\) and the LF2 branch \(E_2\)
each convert one low-fidelity patch into 48 features, the mesh-size branch
\(M\) converts \(e_i^{s}\) into 24 features, and the fusion network \(G\) maps
the combined 120 features to the four outputs:
\begin{equation}
    \widehat{Y}_i^{s}
    =
    \mathcal{F}_\theta\!\left(
    \mathcal{P}_i^{(1),s},\mathcal{P}_i^{(2),s},e_i^{s}
    \right)
    =
    G\!\left(
    \bigl[
    E_1(\mathcal{P}_i^{(1),s}),\,
    E_2(\mathcal{P}_i^{(2),s}),\,
    M(e_i^{s})
    \bigr]
    \right).
    \label{eq:cnn_composite_map}
\end{equation}

\paragraph{Low-fidelity patch encoders}
Each patch encoder contains three three-dimensional convolutional layers with
channel dimensions \(4\to24\to24\to32\). All convolutions use a
\(3\times3\times3\) kernel, unit stride, and one-cell zero padding, so the
spatial dimensions of the patch are preserved, and the Gaussian error linear
unit (GELU) \cite{hendrycks2016gelu} is applied after each convolution. Global
average pooling then converts the \(32\times3\times3\times3\) feature array
into a 32-component vector, which a fully connected layer with GELU maps to 48
features. The two encoders have identical architectures but independent
parameters, so the two resolutions are processed separately before fusion. The
network does not explicitly evaluate a truncation error or impose the Stokes
equations.

\paragraph{Mesh-spacing encoder and feature fusion}
The two mesh sizes are processed by a fully connected encoder with dimensions
\(2\to24\to24\) and GELU after each layer. The concatenated features are mapped
to the four normalized outputs by a fully connected module with dimensions
\(120\to96\to64\to32\to4\), with GELU after the first three layers and an
affine output layer.

Because the network receives a fixed-size local stencil, its input dimension
and number of parameters do not depend on the number of finite elements in
the domain. All weights are initialized from the Kaiming normal distribution
\cite{he2015delving} and all biases to zero; the network has \(101{,}796\)
trainable parameters (\Cref{tab:cnn_architecture}).

\begin{table}[htbp]
\centering
\caption{Layer-by-layer structure of the NNLCI network for a single sample.
Each convolution uses a \(3\times3\times3\) kernel with zero padding. FC
denotes a fully connected layer, and the activation column indicates the
function applied after each layer.}
\label{tab:cnn_architecture}
\small
\renewcommand{\arraystretch}{1.1}
\begin{tabular}{@{}lllr r@{}}
\toprule
Branch & Layer & Activation & Output size & Parameters \\
\midrule
LF1 patch \(E_1\)
  & Input                      & ---  & \(4\times3\times3\times3\)  & --- \\
  & Convolution \(4\to24\)     & GELU & \(24\times3\times3\times3\) & 2{,}616 \\
  & Convolution \(24\to24\)    & GELU & \(24\times3\times3\times3\) & 15{,}576 \\
  & Convolution \(24\to32\)    & GELU & \(32\times3\times3\times3\) & 20{,}768 \\
  & Global average pooling     & ---  & \(32\)                      & --- \\
  & FC \(32\to48\)             & GELU & \(48\)                      & 1{,}584 \\
\addlinespace
LF2 patch \(E_2\)
  & \multicolumn{2}{l}{Same as \(E_1\), separate weights} & \(48\) & 40{,}544 \\
\addlinespace
Mesh size \(M\)
  & Input                      & ---  & \(2\)  & --- \\
  & FC \(2\to24\)              & GELU & \(24\) & 72 \\
  & FC \(24\to24\)             & GELU & \(24\) & 600 \\
\addlinespace
Fusion \(G\)
  & Concatenation              & ---  & \(120\) & --- \\
  & FC \(120\to96\)            & GELU & \(96\)  & 11{,}616 \\
  & FC \(96\to64\)             & GELU & \(64\)  & 6{,}208 \\
  & FC \(64\to32\)             & GELU & \(32\)  & 2{,}080 \\
  & FC \(32\to4\)              & none & \(4\)   & 132 \\
\midrule
Total & & & & 101{,}796 \\
\bottomrule
\end{tabular}
\end{table}

\subsubsection{Training Procedure}
\hypertarget{target:cnn-training}{}
\label{subsubsec:cnn_training}

\paragraph{Training and validation subsets}
For each configuration, the rows of all training matrices are pooled, randomly
permuted, and divided into a training subset \(\mathcal{D}_{\mathrm{train}}\)
of 90\% of the rows and a validation subset \(\mathcal{D}_{\mathrm{val}}\) of
the remaining 10\%. Only \(\mathcal{D}_{\mathrm{train}}\) is used to update
\(\theta\); \(\mathcal{D}_{\mathrm{val}}\) controls the learning-rate schedule,
early stopping, and the selection of the final parameters. Because the division
is made row by row, the validation rows come from the same geometries and
inflow conditions as the training rows and do not measure accuracy for unseen
parameter values.

\paragraph{Loss function}
The training loss is evaluated in the solver variables: each normalized output
is mapped back by~\eqref{eq:cnn_inverse_scaling} and compared with the
high-fidelity target. For a set of rows \(\mathcal{B}\), the loss is the
volume-weighted, component-weighted mean-square error
\begin{equation}
    \mathcal{L}(\theta;\mathcal{B})
    =
    \frac{
    \displaystyle
    \sum_{i\in\mathcal{B}}V_i
    \sum_{c=1}^{4}\gamma_c
    \bigl(\widehat{Y}_{i,c}-Y_{i,c}\bigr)^2
    }{
    \displaystyle
    \sum_{i\in\mathcal{B}}V_i+\varepsilon
    },
    \qquad
    (\gamma_p,\gamma_u,\gamma_v,\gamma_w)=(1,1,3,5),
    \label{eq:cnn_training_score}
\end{equation}
where \(V_i\) is the volume of the high-fidelity cell of row \(i\) and
\(\varepsilon=10^{-12}\). Weighting by \(V_i\) makes~\eqref{eq:cnn_training_score}
a discrete analogue of a mean-square error integrated over the flow domain, so
that regions with many small cells do not dominate merely because they contain
more samples. The weights \(\gamma_v=3\) and \(\gamma_w=5\) increase the
contribution of the transverse velocity components, whose magnitudes are much
smaller than those of \(p\) and \(u\).

\paragraph{Optimization}
The network is implemented in PyTorch \cite{paszke2019pytorch}. At the start
of each epoch, the rows of \(\mathcal{D}_{\mathrm{train}}\) are
shuffled and divided into mini-batches \(\mathcal{B}\) of 2048 rows. For each
mini-batch, the gradient \(\nabla_\theta\mathcal{L}(\theta;\mathcal{B})\) is
computed by backpropagation and the parameters are updated with the AdamW
optimizer \cite{loshchilov2019adamw}, which adapts the step size of each
parameter to the history of its gradients and applies a decoupled weight decay.
We use an initial learning rate of \(5\times10^{-4}\) and a weight-decay
coefficient of \(10^{-6}\), and the gradient is rescaled before each update so
that its Euclidean norm does not exceed one.

After every epoch, the loss~\eqref{eq:cnn_training_score} is evaluated on the
validation subset as the average of its mini-batch losses, weighted by the
number of rows in each mini-batch. The learning rate is halved whenever the
validation loss has not decreased for more than six consecutive epochs.
Training terminates when the validation loss has not decreased for 20
consecutive epochs or after at most 200 epochs, and the parameters with the
lowest validation loss are retained. All random number generators, including
the one for the training--validation split, are initialized with seed 1234.
For the two-aligned-bump configuration, the training is repeated with seeds
2024 and 7 to assess the variation between training runs (Sections~\ref{subsec:near_wall}--\ref{subsec:ablation_degree}).

\subsubsection{Prediction for a New Configuration}
\hypertarget{target:cnn-pred}{}
\label{subsubsec:cnn_prediction}

Once trained, the network requires only low-fidelity information. For a new
parameter value \(\boldsymbol{\mu}\), the LF1 and LF2 problems are solved, the
inputs are formed at each target point as described in Section~\ref{subsec:matrix_generation},
and the network is evaluated with the normalization of the training matrices.
In the present study, the target points are
all centroids of the high-fidelity test mesh, and the high-fidelity test
solution is used solely to measure the accuracy of the prediction.

\subsubsection{Aggregate Test Metrics}
\hypertarget{target:cnn-metrics}{}
\label{subsubsec:cnn_metrics}

The network is compared with the high-fidelity target and with an LF2-center
baseline, \(Y_i^{\mathrm{LF2}}=S^{LF2}(x_i^{HF})\), the last row of
\(P_i^{(2)}\); when \(x_i^{HF}\) lies outside the LF2 mesh, this value is the
local least-squares reconstruction of Section~\ref{subsec:matrix_generation}. The
baseline represents the accuracy obtained by evaluating the finer low-fidelity
solution at the target point. All metrics are computed in the solver variables.

For a test matrix with \(N\) rows, define the pointwise errors
\(E_{i,c}=\widehat{Y}_{i,c}-Y_{i,c}\), where \(c=1,2,3,4\) corresponds to
\(p,u,v,w\). The volume-weighted \(L^2\) error and the corresponding norm of
the reference solution are
\begin{equation}
    E_{L^2}
    =
    \left(
    \sum_{i=1}^{N}V_i
    \sum_{c=1}^{4}E_{i,c}^{2}
    \right)^{1/2},
    \qquad
    Y_{L^2}
    =
    \left(
    \sum_{i=1}^{N}V_i
    \sum_{c=1}^{4}Y_{i,c}^{2}
    \right)^{1/2},
    \label{eq:implemented_weighted_l2}
\end{equation}
and the relative \(L^2\) error is
\begin{equation}
    E_{\mathrm{rel}}
    =
    \frac{E_{L^2}}{Y_{L^2}+\varepsilon}.
    \label{eq:implemented_relative_l2_stable}
\end{equation}
Because this composite measure is dominated by the components of largest
magnitude, the relative error of each component,
\begin{equation}
    E_{\mathrm{rel},c}
    =
    \frac{
    \left(\sum_{i=1}^{N}V_iE_{i,c}^{2}\right)^{1/2}
    }{
    \left(\sum_{i=1}^{N}V_iY_{i,c}^{2}\right)^{1/2}+\varepsilon
    },
    \qquad
    c=1,\ldots,4,
    \label{eq:per_component_relative_l2}
\end{equation}
is also reported (\Cref{tab:per_component_all}). The volume-weighted mean
absolute error is
\begin{equation}
    E_{\mathrm{MAE}}
    =
    \frac{
    \displaystyle
    \sum_{i=1}^{N}V_i
    \sum_{c=1}^{4}\lvert E_{i,c}\rvert
    }{
    \displaystyle
    4\left(\sum_{i=1}^{N}V_i+\varepsilon\right)
    },
    \label{eq:implemented_weighted_mae}
\end{equation}
and, with the volume-weighted component means
\(\overline{Y}_c=\sum_iV_iY_{i,c}/(\sum_iV_i+\varepsilon)\) of the reference
solution, the volume-weighted coefficient of determination is
\begin{equation}
    R^2
    =
    1-
    \frac{
    \displaystyle
    \sum_{i=1}^{N}V_i\sum_{c=1}^{4}E_{i,c}^{2}
    }{
    \displaystyle
    \sum_{i=1}^{N}V_i
    \sum_{c=1}^{4}(Y_{i,c}-\overline{Y}_c)^2+\varepsilon
    },
    \label{eq:implemented_weighted_r2}
\end{equation}
where \(R^2=1\) corresponds to a perfect prediction and \(R^2=0\) to a
prediction no better than the componentwise mean.

For each measure \(M\in\{E_{\mathrm{rel}},E_{\mathrm{MAE}}\}\), the percentage
reduction relative to the LF2-center baseline and the baseline-to-model ratio
are
\begin{equation}
    I_M
    =
    100\,
    \frac{M_{\mathrm{LF2}}-M_{\mathrm{NNLCI}}}
         {\lvert M_{\mathrm{LF2}}\rvert+\varepsilon},
    \qquad
    T_M
    =
    \frac{M_{\mathrm{LF2}}}
         {\max(M_{\mathrm{NNLCI}},\varepsilon)},
    \label{eq:implemented_percent_improvement}
\end{equation}
so that \(I_M>0\) indicates a reduction compared with LF2, and \(T_M=3\) means
that the NNLCI error is three times smaller.

The composite measures~\eqref{eq:implemented_weighted_l2},
\eqref{eq:implemented_weighted_mae}, and~\eqref{eq:implemented_weighted_r2}
combine pressure and velocity components. They are dimensionally consistent
only for nondimensionalized variables; otherwise, they should be read as
composite scores that allow a direct comparison of NNLCI and LF2 on the
same test data rather than as physical norms with a single unit.

\section{Numerical Results and Discussion}
\hypertarget{target:numerical-results}{}
\label{sec:numerical_results}

This section evaluates NNLCI on four held-out tests: one bump, two aligned
bumps, two bumps with a circumferentially rotated second bump, and three
bumps, at all high-fidelity points of each test mesh. We first report the
aggregate test scores relative to the LF2-center baseline and then examine the
predicted fields through cross-sectional error maps, isocurve comparisons, and
one-dimensional profiles. The two-aligned-bump configuration is then studied
in more detail: near the wall, outside the training range, and with a linear
least-squares extension. The section closes with the computational cost. The common network,
training procedure, baseline, and metric definitions are given in
Section~\ref{subsec:cnn_error_evaluation} and are not repeated here.
\subsection{Quantitative Performance on Held-Out Cases}
\hypertarget{target:quantitative-results}{}
\label{subsec:quantitative_results}

Each experiment evaluates one parameter combination excluded from its training
and validation data. The training and test parameters of the four
configurations are given in Sections~\ref{subsec:one_bump_base_test}--\ref{subsec:three_bump_interpolation_test},
and the resulting errors are summarized in Section~\ref{subsec:test_error_summary}.

\subsubsection{One Bump: Base-Geometry and Base-Inflow Interpolation}
\hypertarget{target:one-bump-test}{}
\label{subsec:one_bump_base_test}

The held-out one-bump case uses
\[
    x_b^0=2.50,
    \qquad
    U_{\max}^0=0.10.
\]
The training positions are defined as percentage changes relative to the
reference bump position $x_b^0$:
\[
    x_b^{(\Delta)}
    =
    x_b^0
    \left(
    1+\frac{\Delta x_b}{100}
    \right).
\]
The six training positions are
\[
    x_b\in\{1.25,1.75,2.25,2.75,3.25,3.75\},
\]
corresponding to relative changes of
\[
    \Delta x_b\in\{-50\%,-30\%,-10\%,10\%,30\%,50\%\}.
\]
Each position is paired with $U_{\max}\in\{0.09,0.11\}$, producing 12
training matrices. The held-out pair $(x_b,U_{\max})=(2.50,0.10)$ lies between
the $-10\%$ and $+10\%$ geometric perturbations and between the two training
inflow values.

\subsubsection{Two Aligned Bumps: Base-Geometry and Base-Inflow Interpolation}
\hypertarget{target:two_bumps-base}{}
\label{subsec:two_bump_base_test}

The first bump is fixed at $x_{b,1}=1.60$. Training uses the Cartesian product
\[
    x_{b,2}\in\{2.56,2.77\},
    \qquad
    U_{\max}\in\{0.09,0.11\},
\]
giving four matrices. The held-out case
\[
    (x_{b,1},x_{b,2})=(1.60,2.66),
    \qquad
    U_{\max}=0.10
\]
lies between the training values in both the second-bump position and the
inflow.

\subsubsection{Rotated Two Bumps: Rotation--Inflow Interpolation}
\hypertarget{target:two_bumps-rotation}{}
\label{subsec:two_bump_rotation_test}

Both bumps keep the same axial positions, $x_{b,1}=1.60$ and $x_{b,2}=2.88$,
in all cases. The first bump stays at the top of the pipe
($\theta_{b,1}=\pi/2$), and only the second bump is rotated about the pipe
axis, to $\theta_{b,2}=\pi/2+\Delta\theta$. Training uses
\[
    \Delta\theta\in\{-10^\circ,0^\circ,10^\circ,20^\circ\},
    \qquad
    U_{\max}\in\{0.09,0.10,0.11,0.12\},
\]
giving 16 rotation--inflow combinations. The test case
$(\Delta\theta,U_{\max})=(5^\circ,0.105)$ lies midway between neighboring
training values in both parameters.

\subsubsection{Three Bumps: Geometry--Inflow Interpolation}
\hypertarget{target:three_bumps-base}{}
\label{subsec:three_bump_interpolation_test}

The middle bump is fixed at $x_{b,2}=2.50$, while the first and third bumps are
displaced symmetrically about it. Their distance from the middle bump is
\[
    d = d_0\left(1+\frac{\Delta d}{100}\right),
    \qquad d_0 = 1.35,
\]
so that $x_{b,1}=2.50-d$ and $x_{b,3}=2.50+d$; positive $\Delta d$ moves the
outer bumps away from the middle bump. The six training geometries are
\[
\begin{aligned}
\mathcal{G}_{\mathrm{train}}=\{&
(1.56,2.50,3.44),
(1.42,2.50,3.58),
(1.29,2.50,3.71),\\
&
(1.15,2.50,3.85),
(1.02,2.50,3.98),
(0.88,2.50,4.12)\},
\end{aligned}
\]
corresponding to relative changes of
\[
    \Delta d\in\{-30\%,-20\%,-10\%,0\%,10\%,20\%\}.
\]
Each geometry is paired with $U_{\max}\in\{0.09,0.11\}$, giving 12 training
matrices. The held-out configuration
\[
    (x_{b,1},x_{b,2},x_{b,3})=(1.22,2.50,3.78),
    \qquad U_{\max}=0.10
\]
corresponds to $\Delta d=-5\%$ and lies midway between the $-10\%$ and $0\%$
training geometries and between the two training inflows.

\subsubsection{Summary of Test Errors}
\hypertarget{target:error_summary}{}
\label{subsec:test_error_summary}

All errors are evaluated at all centroids of the high-fidelity test mesh,
including those next to the wall, and are obtained with training seed 1234;
the variation between training runs is assessed for the two-aligned-bump
configuration in Section~\ref{subsec:near_wall}.
\Cref{tab:composite_all} summarizes the composite errors for the four held-out
cases. Across the four cases, NNLCI reduces the relative \(L^2\) error by
86.4\%--91.7\%, a factor of 7.4--12.1 compared with the LF2-center baseline, and
\(R^2\) exceeds 0.9996 in every case. Measured by the squared error summed over
the four components, NNLCI is more accurate than LF2 at 96.3\%--98.8\% of the
test points.

\begin{table}[htbp]
\centering
\caption{Composite test errors for the four held-out cases, evaluated at all
test points, in solver variables. \(T\) is the LF2-to-NNLCI error
ratio~\eqref{eq:implemented_percent_improvement}.}
\label{tab:composite_all}
\small
\setlength{\tabcolsep}{4pt}
\begin{tabular}{@{}lccccccc@{}}
\toprule
 & \multicolumn{3}{c}{Relative \(L^2\) error \(E_{\mathrm{rel}}\)}
 & \multicolumn{3}{c}{Mean absolute error \(E_{\mathrm{MAE}}\)} & \\
\cmidrule(lr){2-4}\cmidrule(lr){5-7}
Configuration & NNLCI & LF2 & \(T\) & NNLCI & LF2 & \(T\) & \(R^2\) \\
\midrule
One bump          & \(9.85\times10^{-3}\) & \(7.24\times10^{-2}\) &  7.4 & \(1.59\times10^{-4}\) & \(1.09\times10^{-3}\) &  6.8 & 0.99962 \\
Two aligned bumps & \(7.92\times10^{-3}\) & \(7.90\times10^{-2}\) & 10.0 & \(1.20\times10^{-4}\) & \(1.18\times10^{-3}\) &  9.8 & 0.99975 \\
Rotated two bumps & \(6.99\times10^{-3}\) & \(8.46\times10^{-2}\) & 12.1 & \(1.07\times10^{-4}\) & \(1.32\times10^{-3}\) & 12.3 & 0.99981 \\
Three bumps       & \(7.58\times10^{-3}\) & \(7.30\times10^{-2}\) &  9.6 & \(1.21\times10^{-4}\) & \(1.18\times10^{-3}\) &  9.8 & 0.99978 \\
\bottomrule
\end{tabular}
\end{table}
Because the composite measures are dominated by \(u\), which is the largest
component, \Cref{tab:per_component_all} lists the relative error of each
component separately. NNLCI reduces the error of every component in every
configuration. The pressure and streamwise-velocity errors are below
1.7\%,
corresponding to reductions by factors of 3.1--4.4 for \(p\) and 6.3--10.3
for \(u\). The largest reductions occur for the transverse components, by
factors of 20--34 for \(v\) and 10--24 for \(w\). For these components the
LF2 errors are comparable to or larger than the components themselves: the
LF2 error in \(v\) is three to five times the magnitude of \(v\). NNLCI
reduces the relative error of \(v\), a component about two orders of
magnitude smaller than \(u\), to 12\%--17\%, and that of \(w\) to 5\%--8\%.

\begin{table}[htbp]
\centering
\caption{Relative \(L^2\) error of each component,
\(E_{\mathrm{rel},c}\) in~\eqref{eq:per_component_relative_l2}, in percent,
for NNLCI and, in parentheses, the LF2-center baseline, evaluated at all test
points.}
\label{tab:per_component_all}
\small
\begin{tabular}{@{}lcccc@{}}
\toprule
Configuration & \(p\) & \(u\) & \(v\) & \(w\) \\
\midrule
One bump          & 1.63 (5.1) & 0.90 (5.7) & 17.2 (465) & 8.4 (82) \\
Two aligned bumps & 1.11 (3.9) & 0.75 (6.3) & 15.4 (425) & 4.9 (82) \\
Rotated two bumps & 0.85 (3.7) & 0.66 (6.8) & 15.5 (315) & 4.7 (114) \\
Three bumps       & 1.07 (4.1) & 0.70 (5.7) & 12.1 (415) & 5.6 (68) \\  
\bottomrule
\end{tabular}
\end{table}

Because every test parameter is bracketed by values represented in training,
these experiments test interpolation within the sampled parameter
ranges; one case outside the training range is examined in
Section~\ref{subsec:extrapolation}.
\subsection{Cross-Sectional Field and Error Maps}
\hypertarget{target:cross_sectional_maps}{}
\label{subsec:cross_sectional_maps}

\begin{figure}[htbp]
\centering
\subfloat{\includegraphics[width=0.92\linewidth]{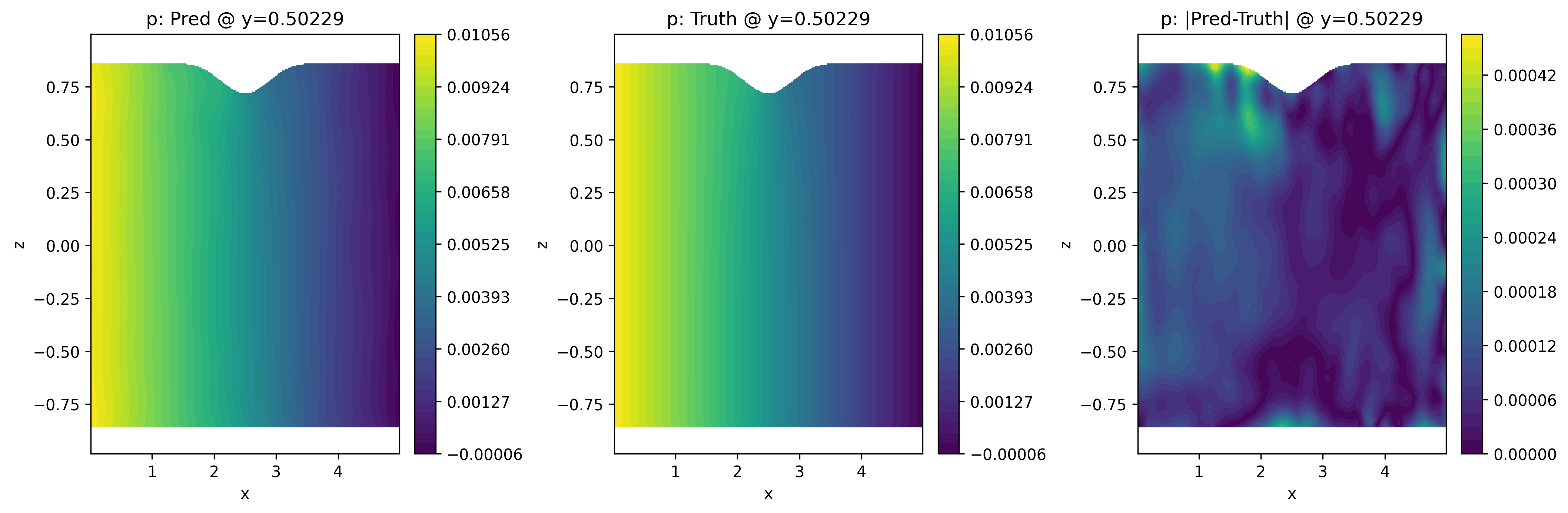}}
\par\medskip
\subfloat{\includegraphics[width=0.92\linewidth]{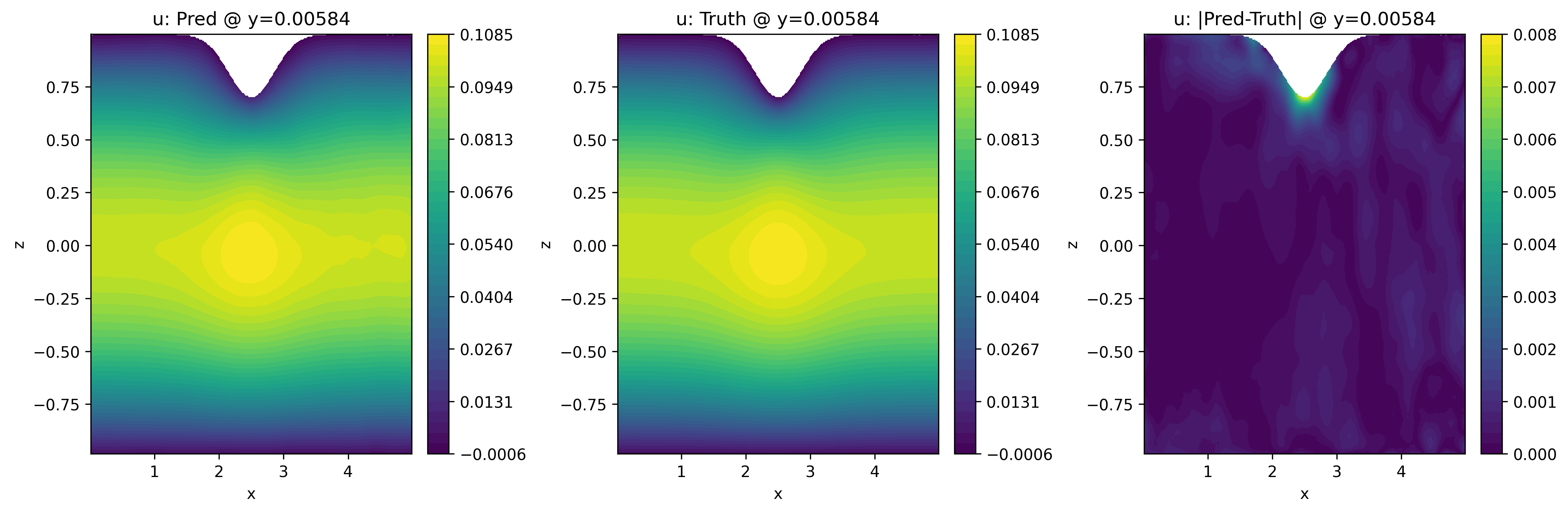}}
\par\medskip
\subfloat{\includegraphics[width=0.92\linewidth]{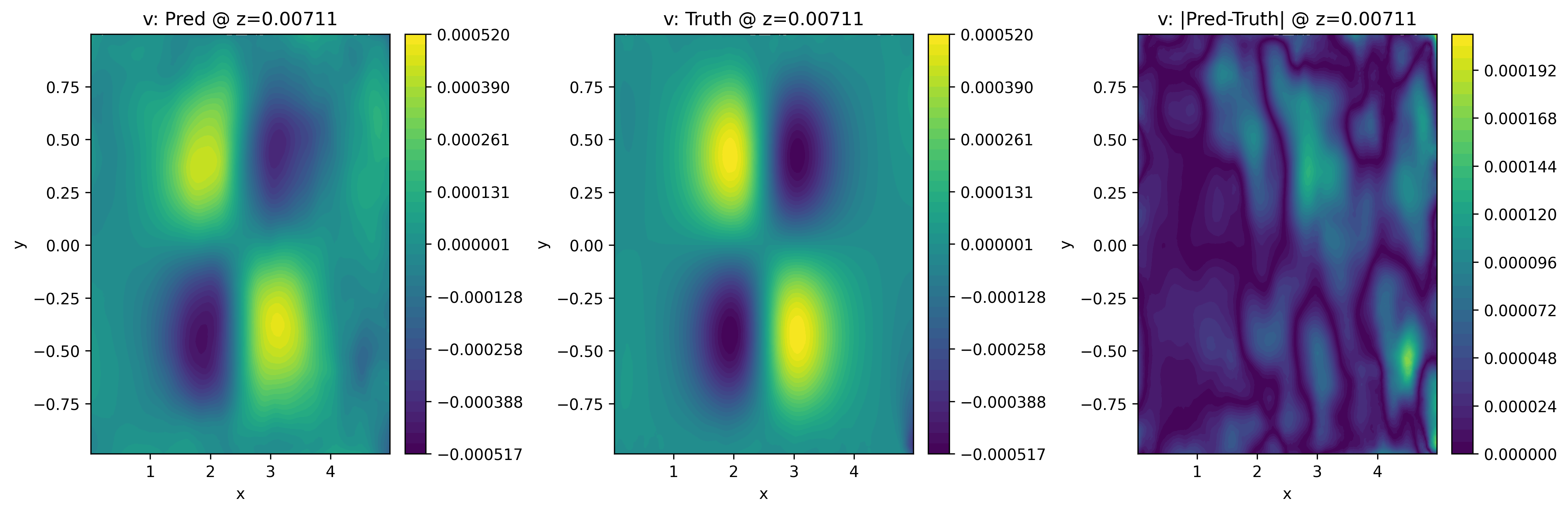}}
\par\medskip
\subfloat{\includegraphics[width=0.92\linewidth]{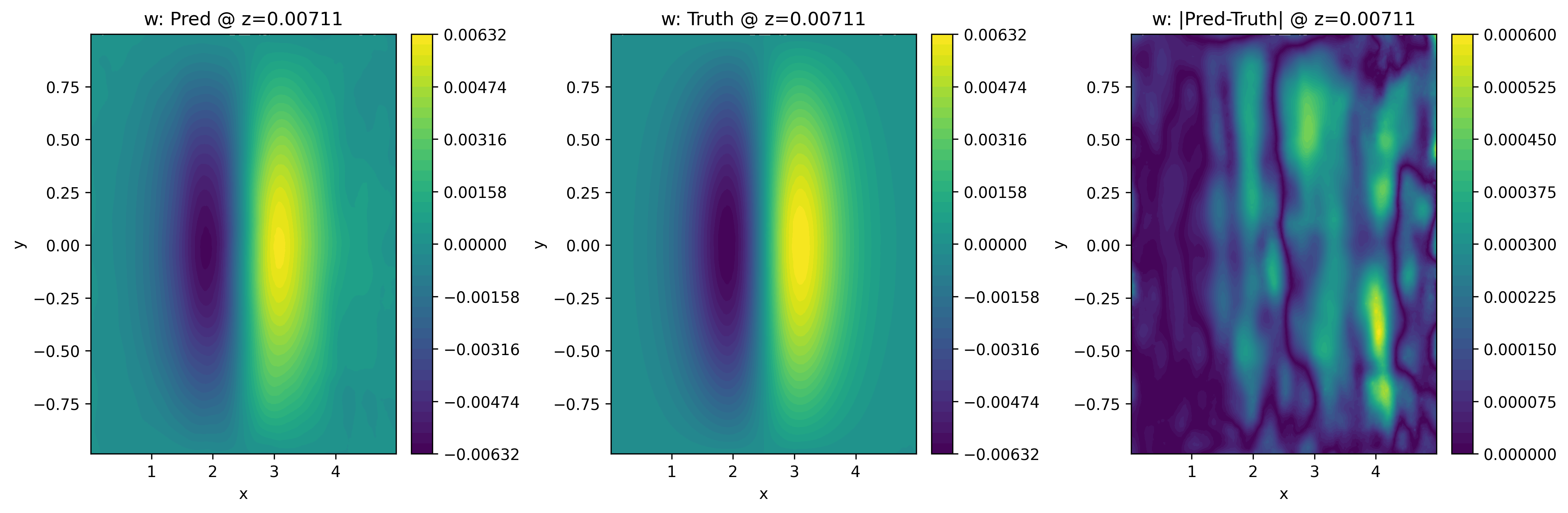}}
\caption{Cross-sectional NNLCI predictions, high-fidelity references, and
absolute-error maps for the one-bump test case: $p$ on the plane $y=0.50$, $u$
on the plane $y=0.01$, through the bump, and $v$ and $w$ on the plane
$z=0.01$.}
\label{fig:1bcross_sectional_flow_comparison}
\end{figure}

\begin{figure}[htbp]
\centering
\subfloat{\includegraphics[width=0.92\linewidth]{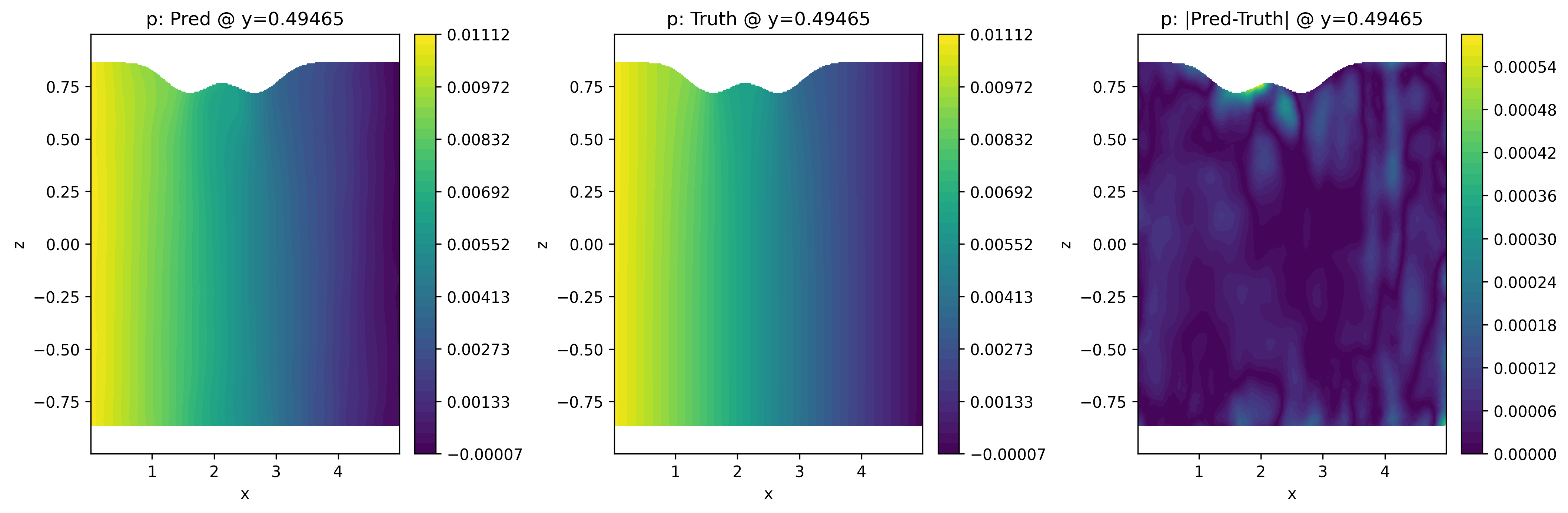}}
\par\medskip
\subfloat{\includegraphics[width=0.92\linewidth]{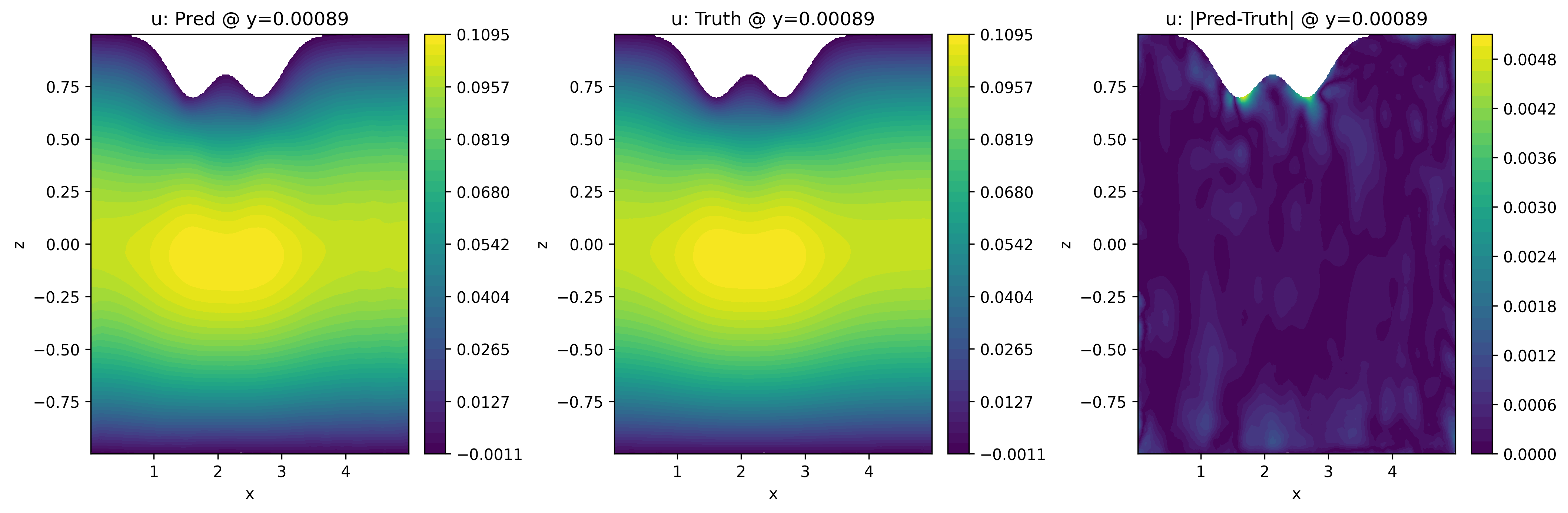}}
\par\medskip
\subfloat{\includegraphics[width=0.92\linewidth]{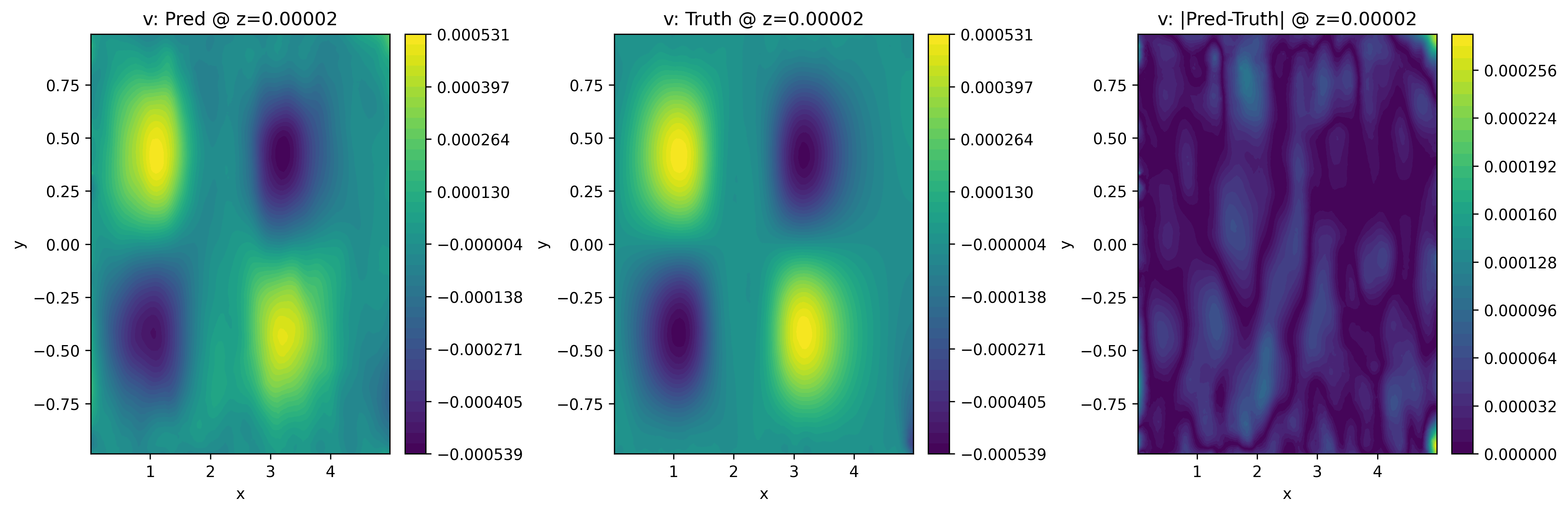}}
\par\medskip
\subfloat{\includegraphics[width=0.92\linewidth]{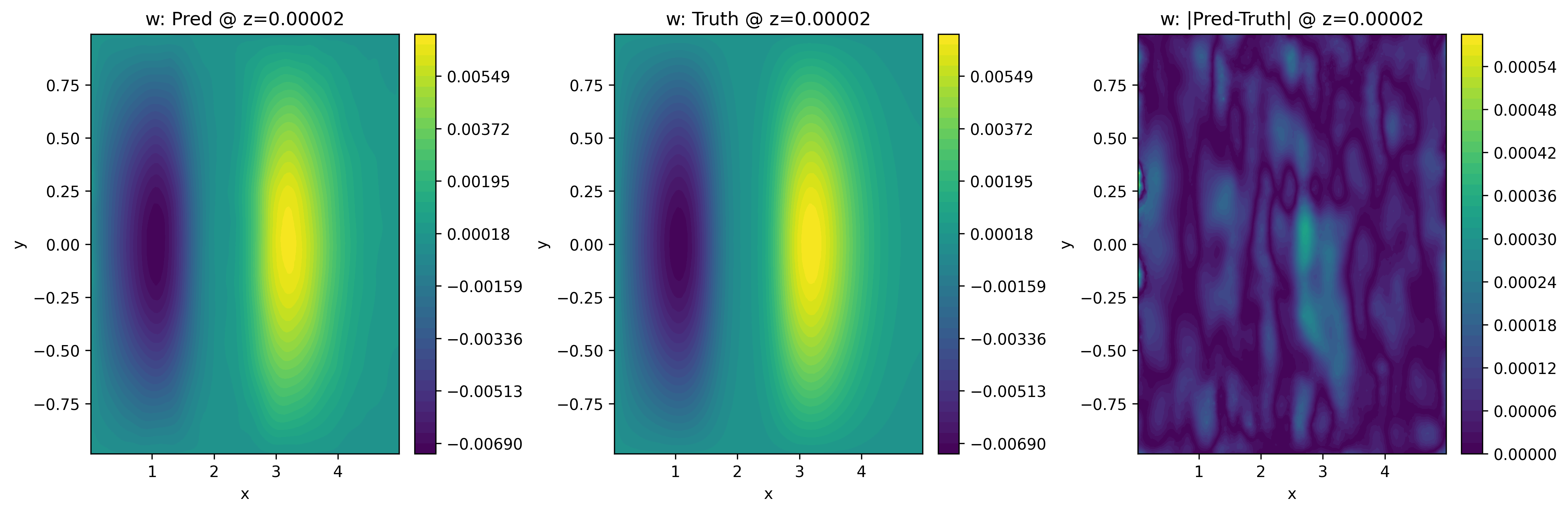}}
\caption{Cross-sectional NNLCI predictions, high-fidelity references, and
absolute-error maps for the two-aligned-bump test case: $p$ on the plane
$y=0.49$, $u$ on the plane $y=0$, through both bumps, and $v$ and $w$ on the
plane $z=0$.}
\label{fig:2bcross_sectional_flow_comparison}
\end{figure}

\begin{figure}[htbp]
\centering
\subfloat{\includegraphics[width=0.92\linewidth]{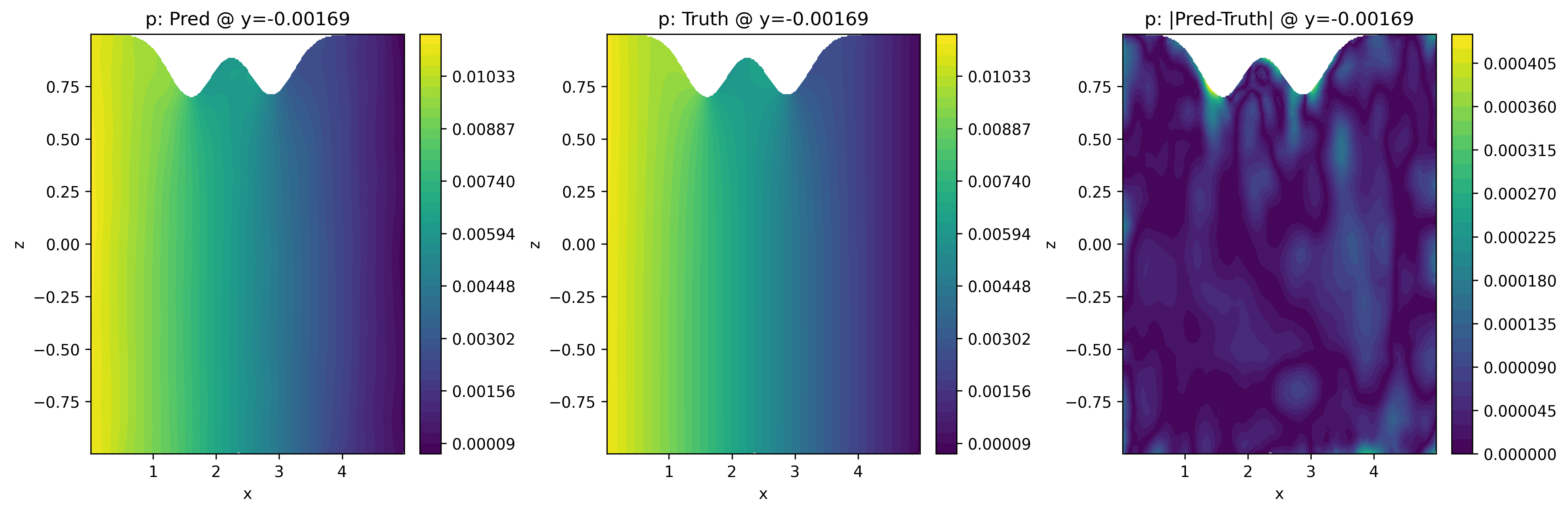}}
\par\medskip
\subfloat{\includegraphics[width=0.92\linewidth]{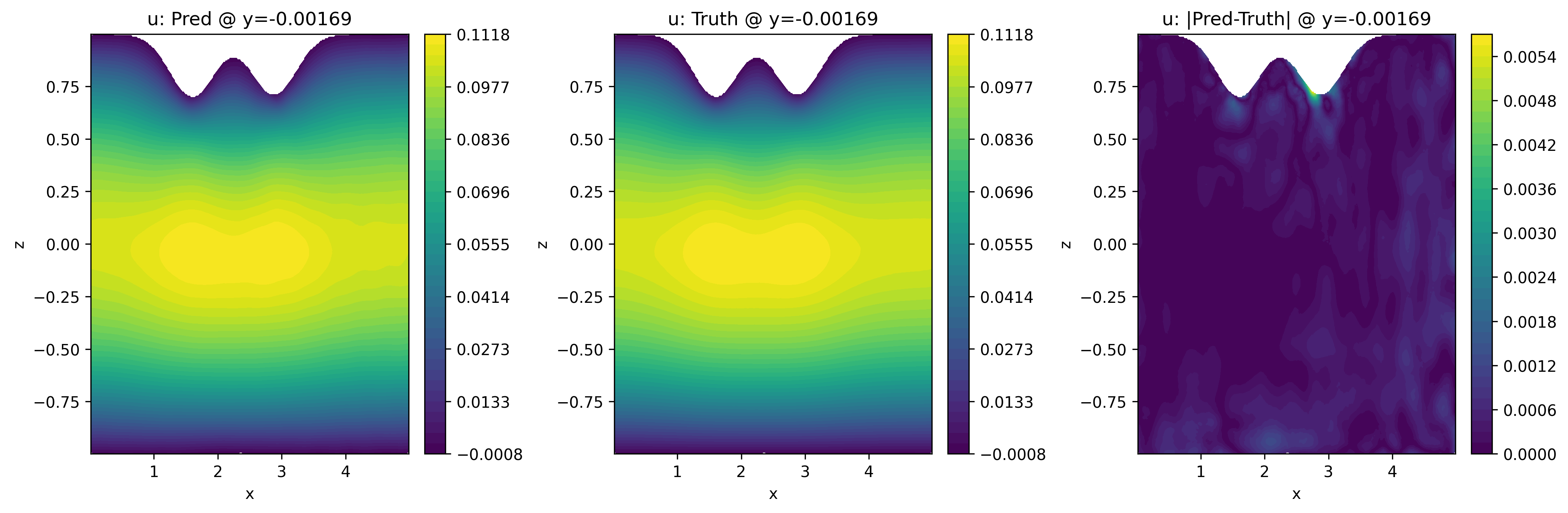}}
\par\medskip
\subfloat{\includegraphics[width=0.92\linewidth]{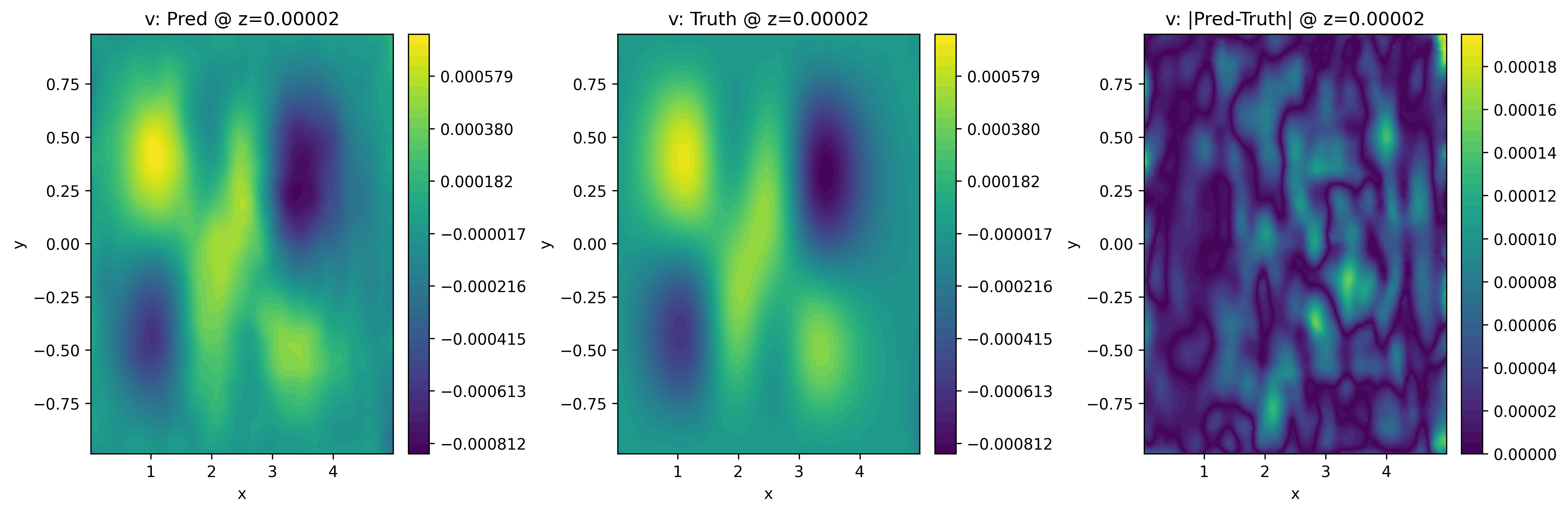}}
\par\medskip
\subfloat{\includegraphics[width=0.92\linewidth]{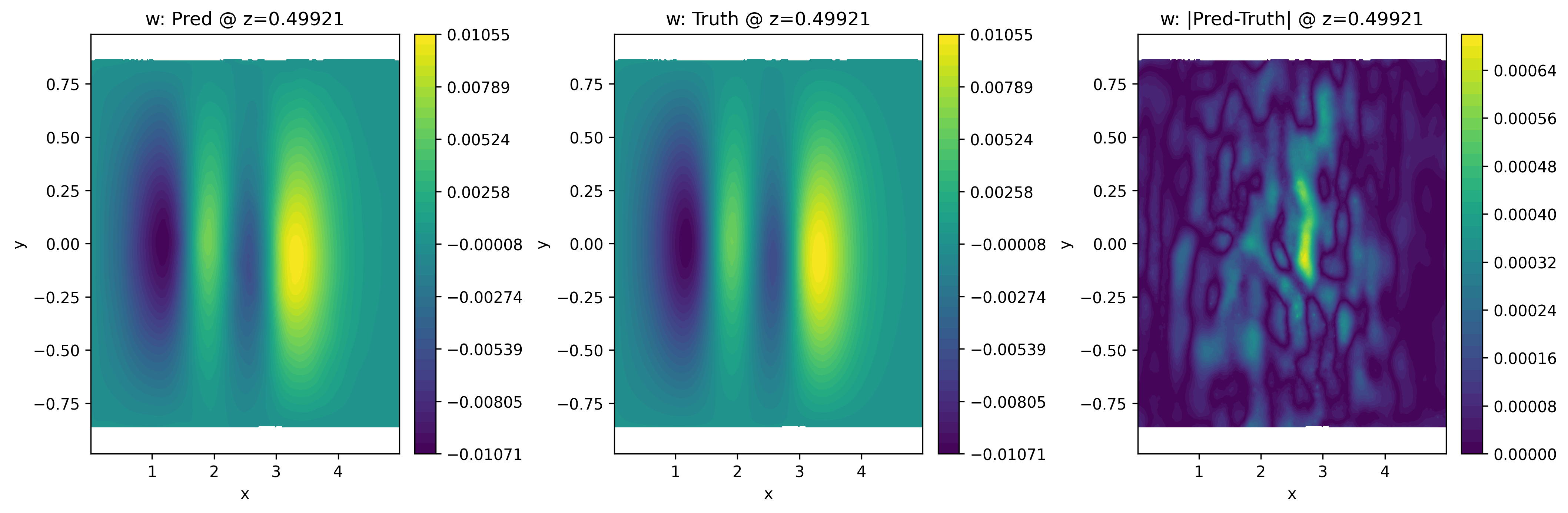}}
\caption{Cross-sectional NNLCI predictions, high-fidelity references, and
absolute-error maps for the rotated two-bump test case $(\Delta\theta=5^\circ)$:
$p$ and $u$ on the plane $y=0$, through both bumps, $v$ on the plane $z=0$,
and $w$ on the plane $z=0.50$, just below the bump crests.}
\label{fig:2br_cross_sectional_flow_comparison}
\end{figure}

\begin{figure}[htbp]
\centering
\subfloat{\includegraphics[width=0.92\linewidth]{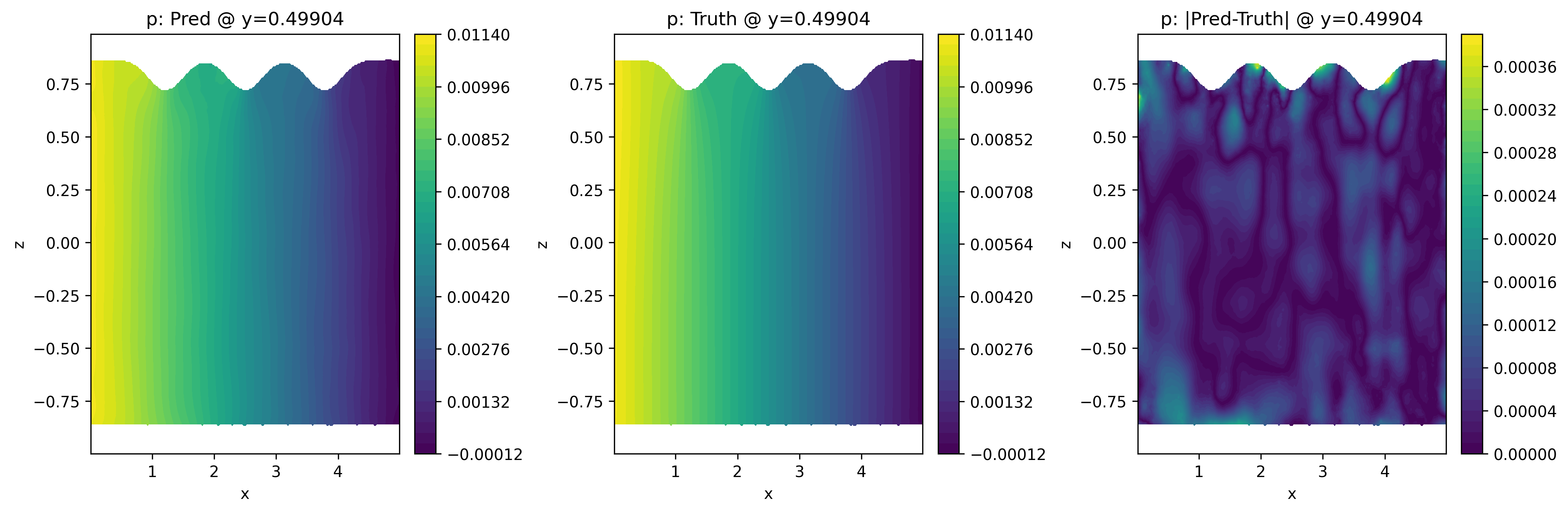}}
\par\medskip
\subfloat{\includegraphics[width=0.92\linewidth]{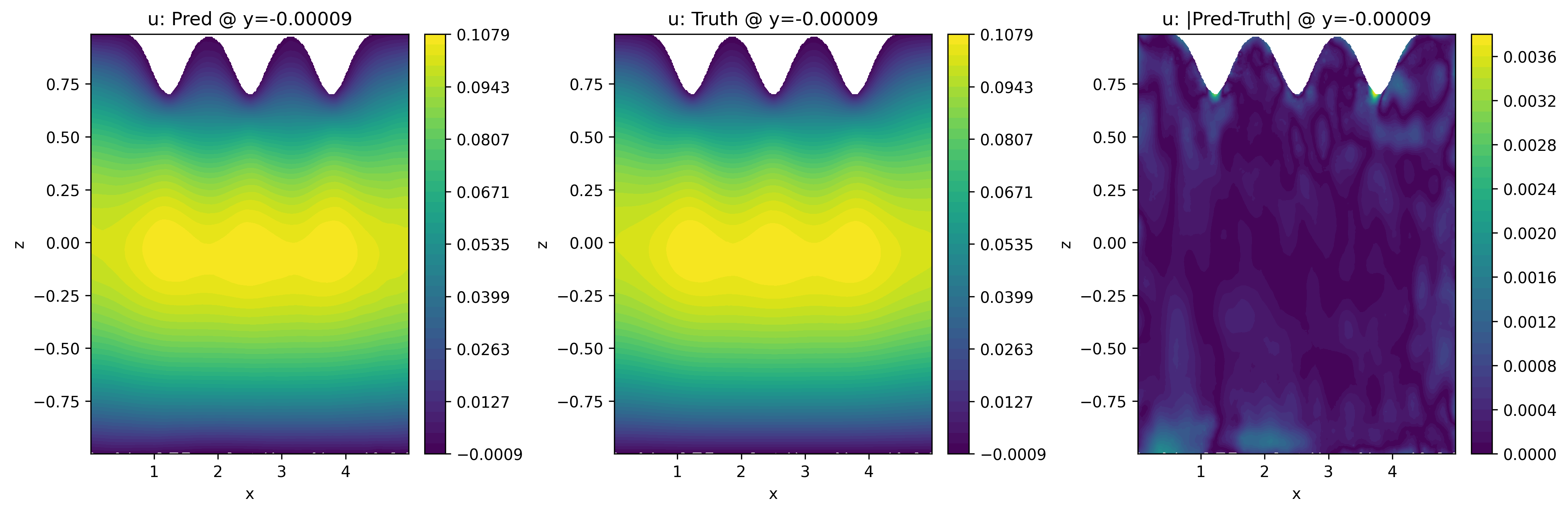}}
\par\medskip
\subfloat{\includegraphics[width=0.92\linewidth]{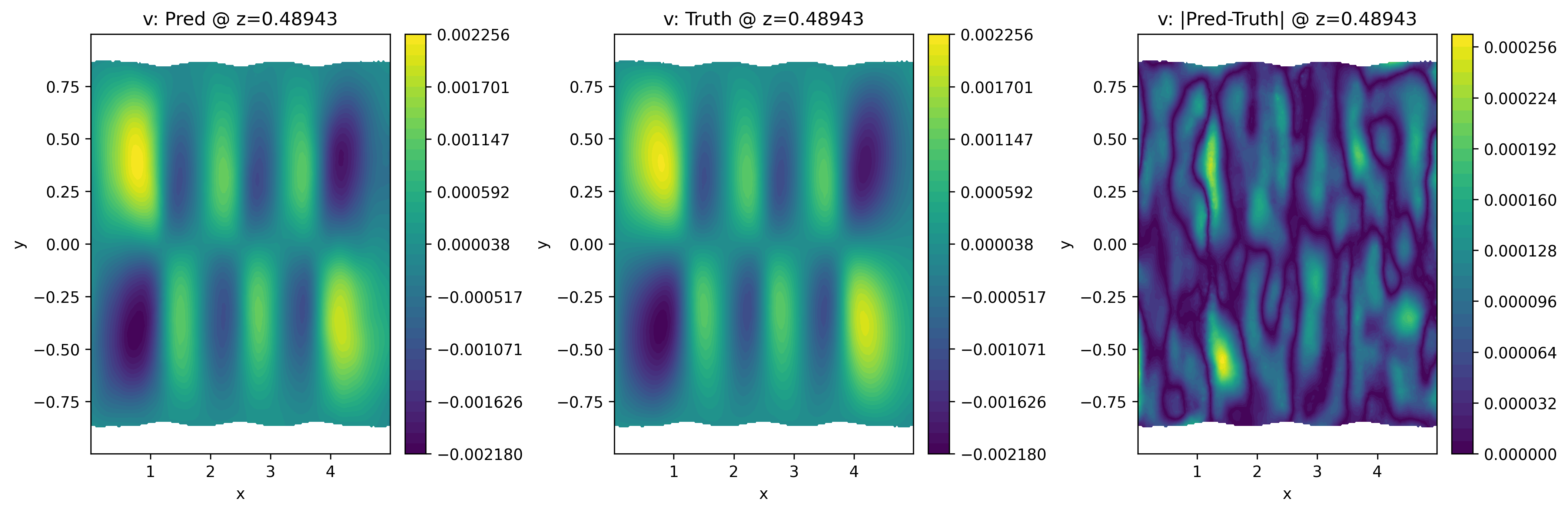}}
\par\medskip
\subfloat{\includegraphics[width=0.92\linewidth]{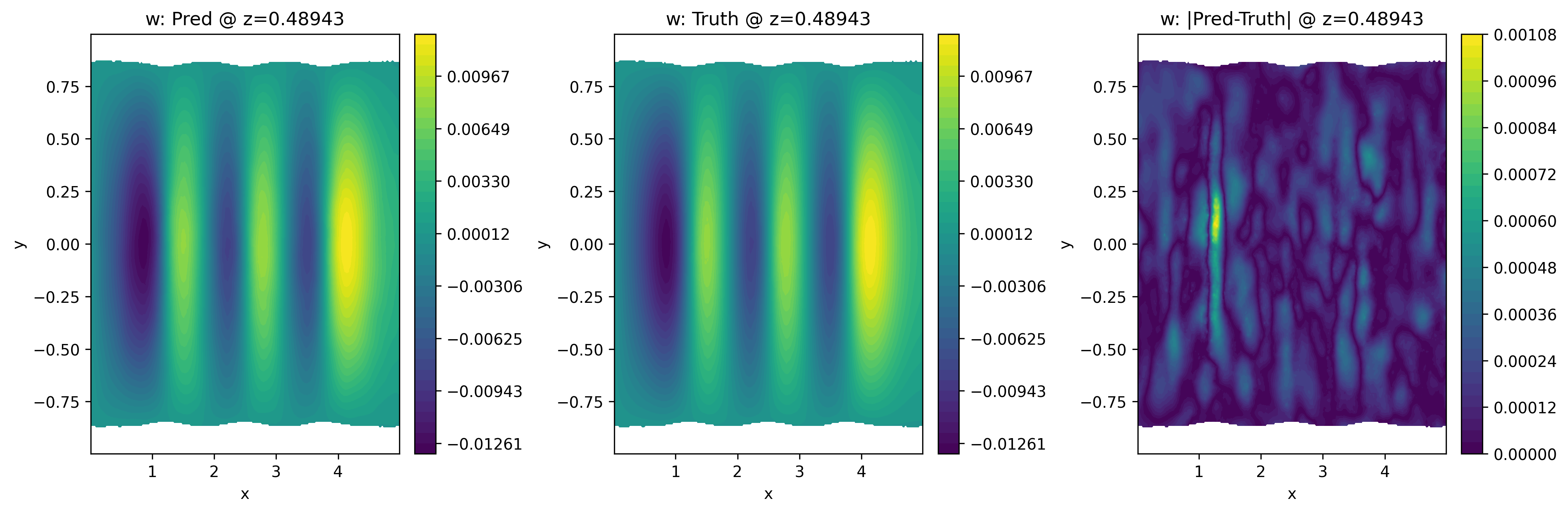}}
\caption{Cross-sectional NNLCI predictions, high-fidelity references, and
absolute-error maps for the three-bump test case: $p$ on the plane $y=0.50$,
$u$ on the plane $y=0$, through all three bumps, and $v$ and $w$ on the plane
$z=0.49$, between the pipe axis and the bump crests.}
\label{fig:3bcross_sectional_flow_comparison}
\end{figure}

Figures~\ref{fig:1bcross_sectional_flow_comparison},
\ref{fig:2bcross_sectional_flow_comparison},
\ref{fig:2br_cross_sectional_flow_comparison}, and
\ref{fig:3bcross_sectional_flow_comparison} compare NNLCI predictions with the
high-fidelity fields on representative two-dimensional slices. For each
variable, the prediction and reference use a common color scale, while the
absolute-error map uses a separate scale. Scattered finite-element data are
reconstructed on the displayed grid using the same local weighted
least-squares procedure for both fields. Grid points outside the pipe are left blank.

\FloatBarrier

\subsection{Cross-Sectional Isocurve Comparisons}
\hypertarget{target:isocurve-comparisons}{}
\label{subsec:isocurve_comparisons}

The isocurves give a second, qualitative view of the predicted fields. For each selected slice and variable, the prediction and
high-fidelity reference use identical contour levels and the same reconstructed
grid, with points outside the pipe left blank. Where the predicted and reference contours coincide, the network reproduces the main gradients and flow features.

\begin{figure}[htbp]
\centering
\subfloat{\includegraphics[width=0.92\linewidth]{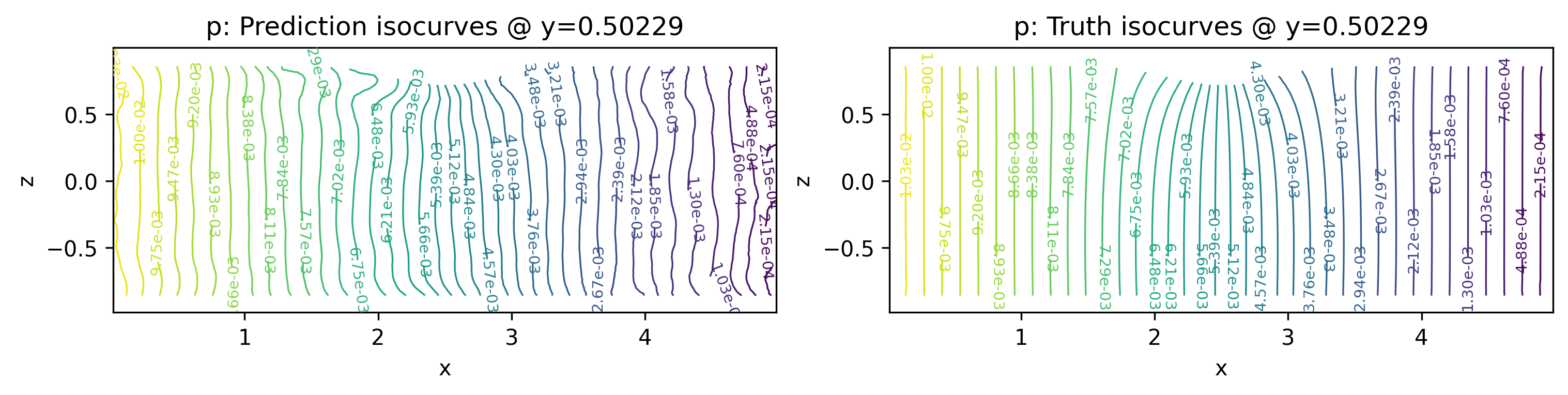}}
\par\medskip
\subfloat{\includegraphics[width=0.92\linewidth]{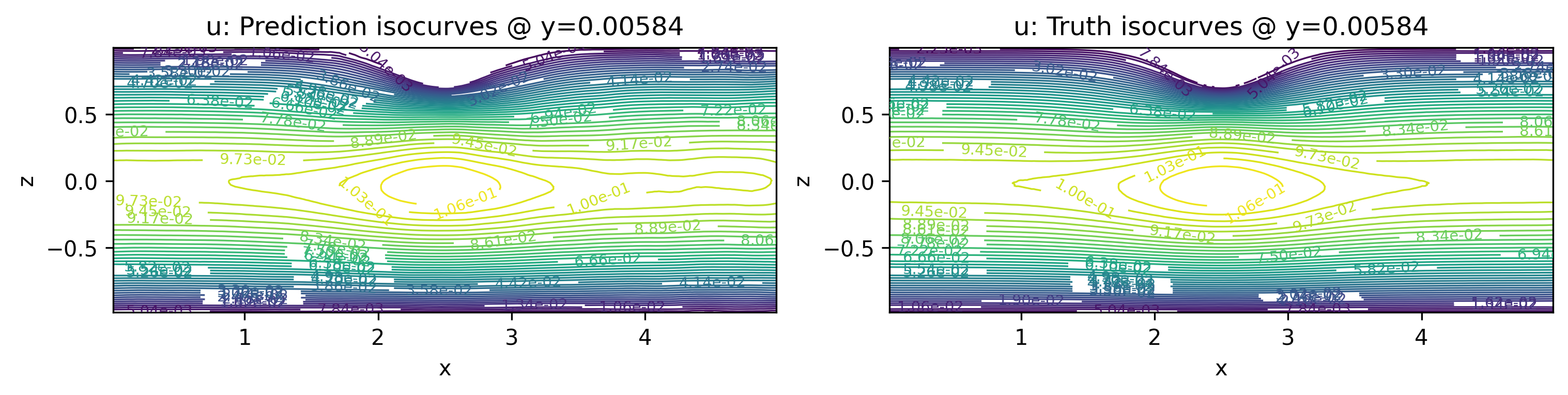}}
\par\medskip
\subfloat{\includegraphics[width=0.92\linewidth]{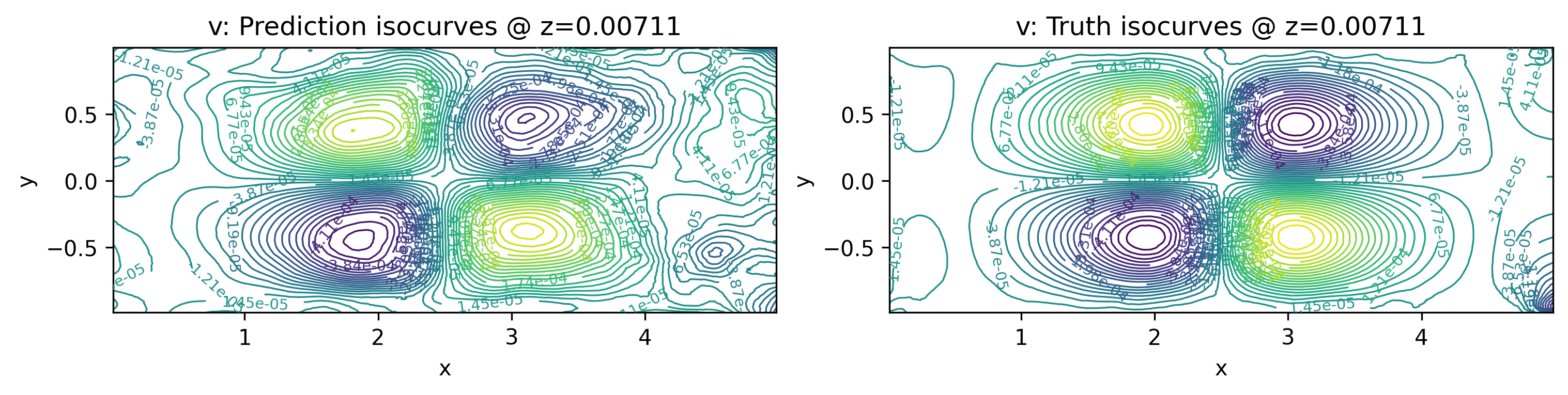}}
\par\medskip
\subfloat{\includegraphics[width=0.92\linewidth]{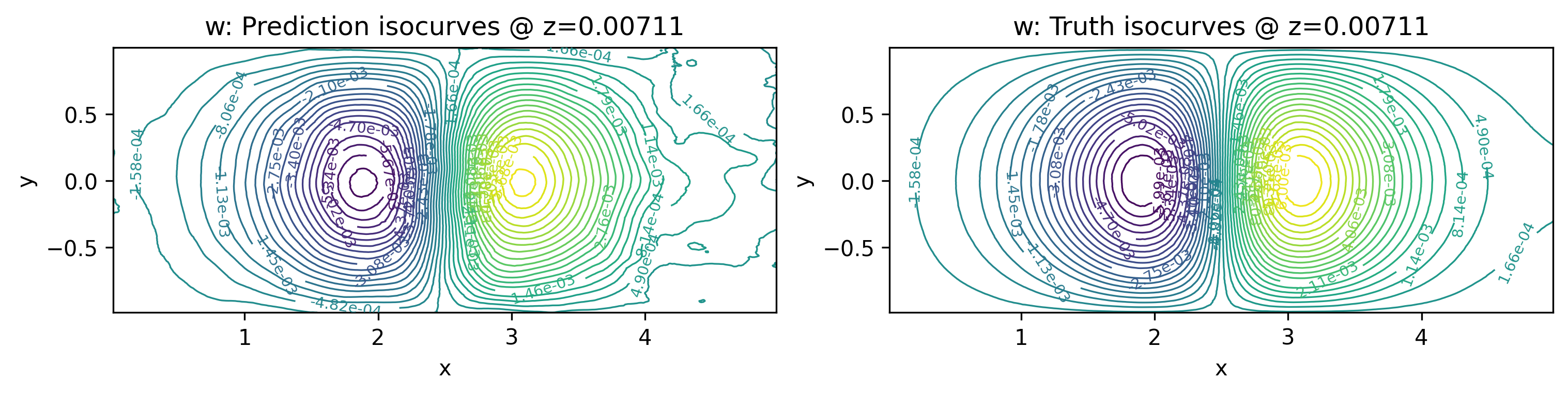}}
\caption{Prediction--reference isocurve comparisons for the one-bump test case,
on the same planes as \Cref{fig:1bcross_sectional_flow_comparison}: $p$ on
$y=0.50$, $u$ on $y=0.01$, through the bump, and $v$ and $w$ on $z=0.01$.}
\label{fig:1biso_cross_sectional_flow_comparison}
\end{figure}

\begin{figure}[htbp]
\centering
\subfloat{\includegraphics[width=0.92\linewidth]{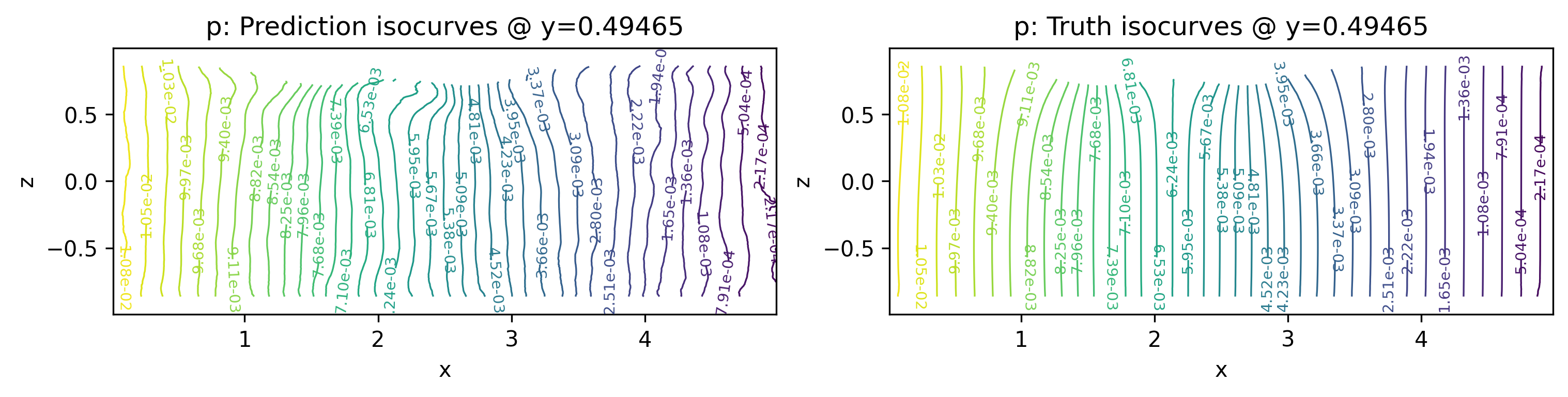}}
\par\medskip
\subfloat{\includegraphics[width=0.92\linewidth]{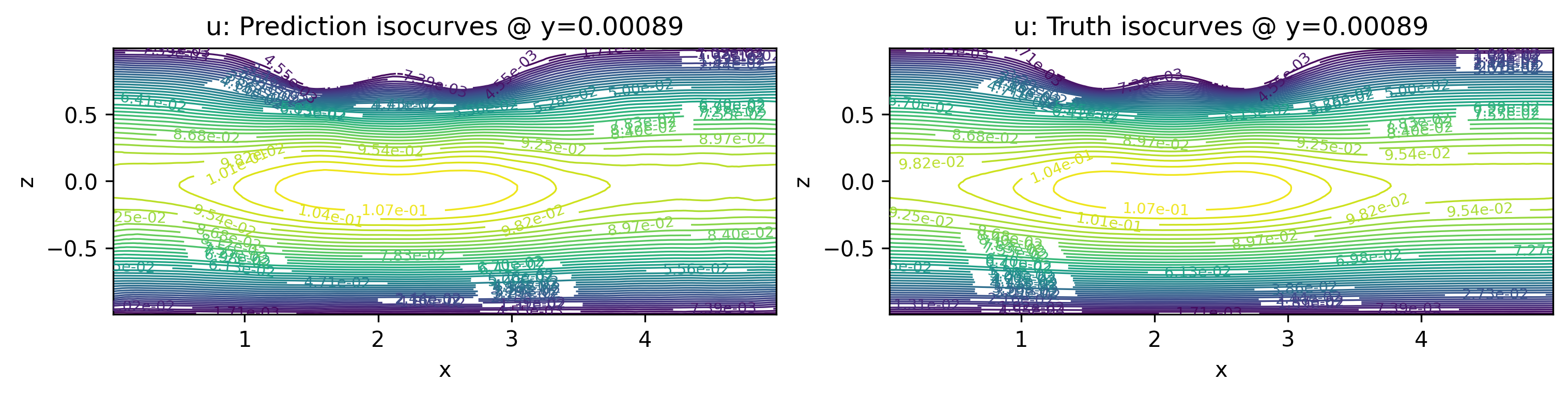}}
\par\medskip
\subfloat{\includegraphics[width=0.92\linewidth]{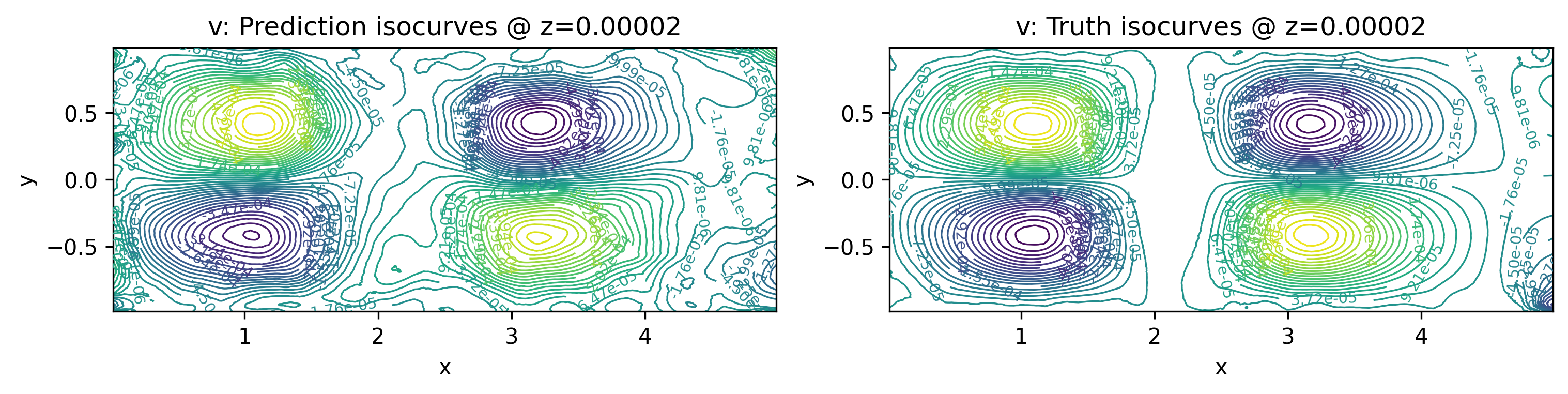}}
\par\medskip
\subfloat{\includegraphics[width=0.92\linewidth]{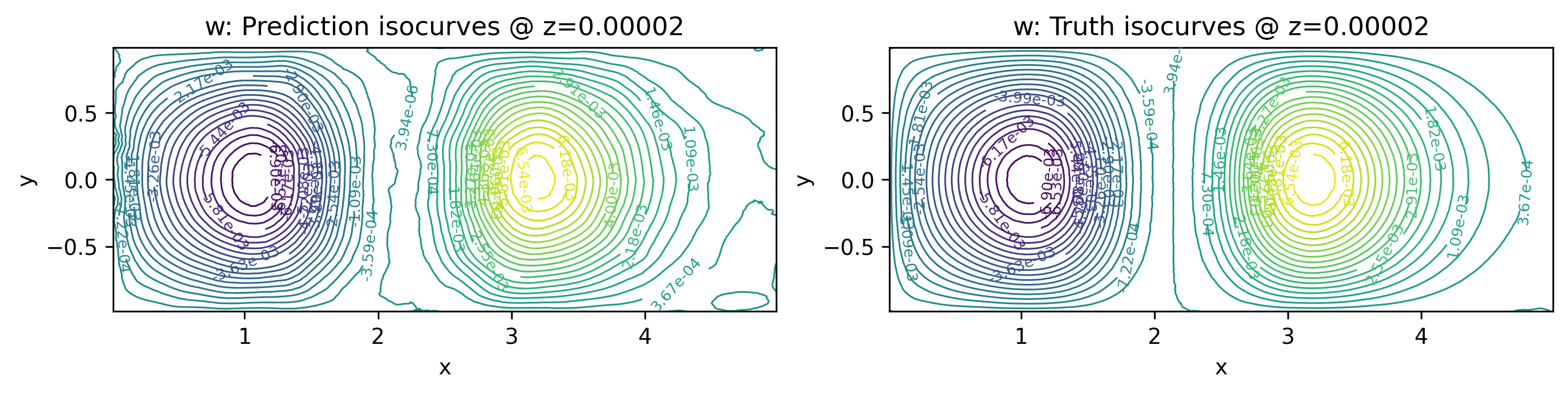}}
\caption{Prediction--reference isocurve comparisons for the two-aligned-bump
test case, on the same planes as \Cref{fig:2bcross_sectional_flow_comparison}:
$p$ on $y=0.49$, $u$ on $y=0$, through both bumps, and $v$ and $w$ on $z=0$.}
\label{fig:2biso_cross_sectional_flow_comparison}
\end{figure}

\begin{figure}[htbp]
\centering
\subfloat{\includegraphics[width=0.92\linewidth]{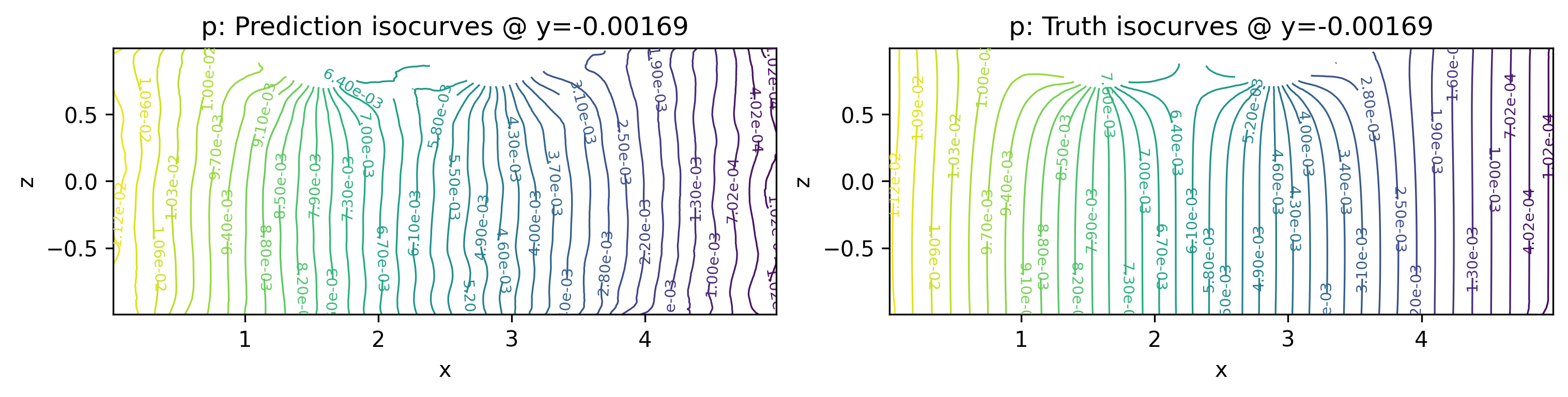}}
\par\medskip
\subfloat{\includegraphics[width=0.92\linewidth]{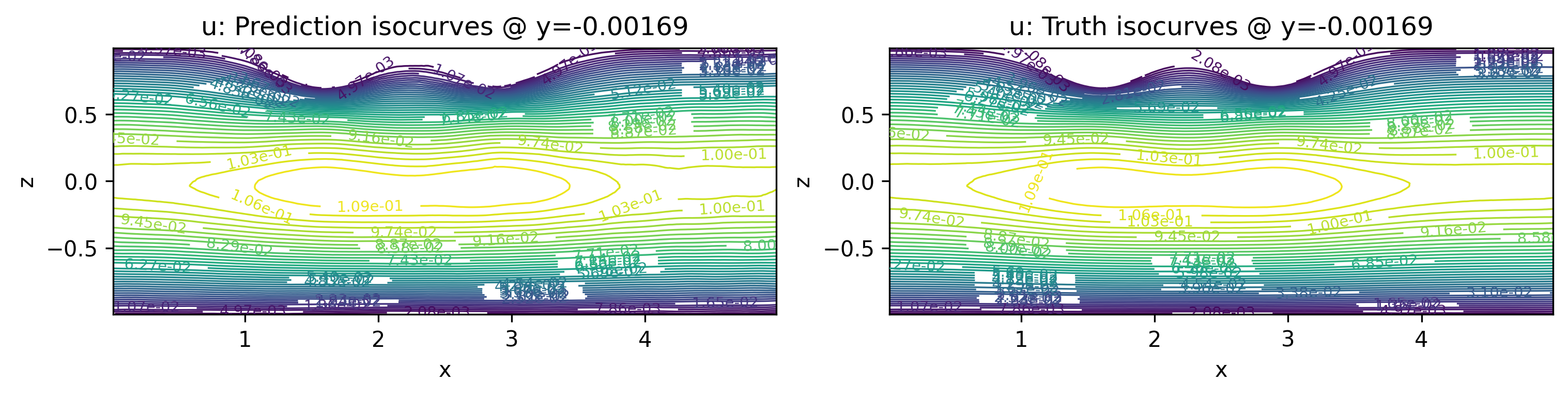}}
\par\medskip
\subfloat{\includegraphics[width=0.92\linewidth]{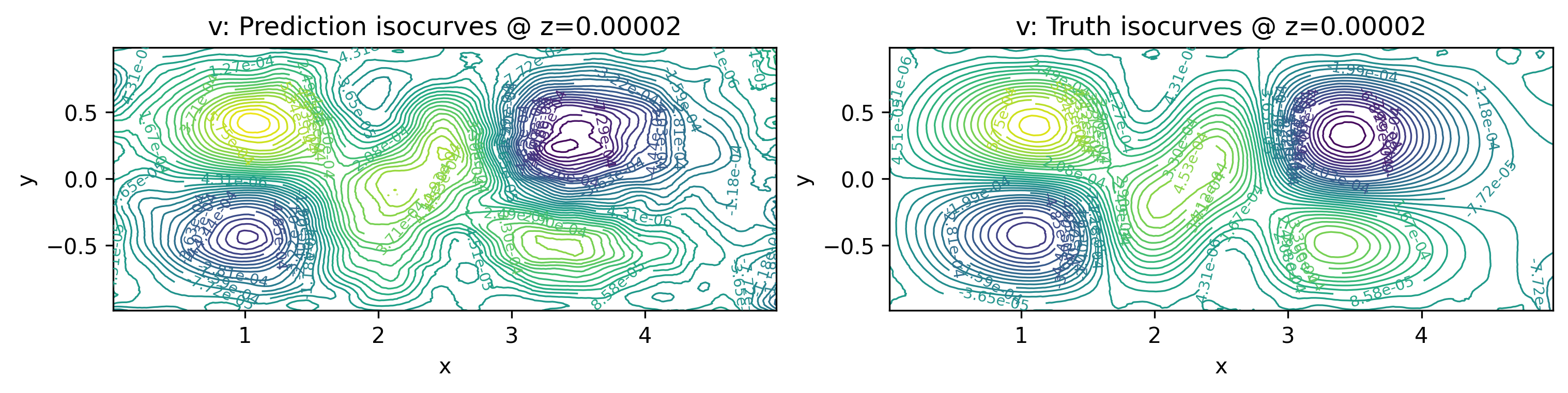}}
\par\medskip
\subfloat{\includegraphics[width=0.92\linewidth]{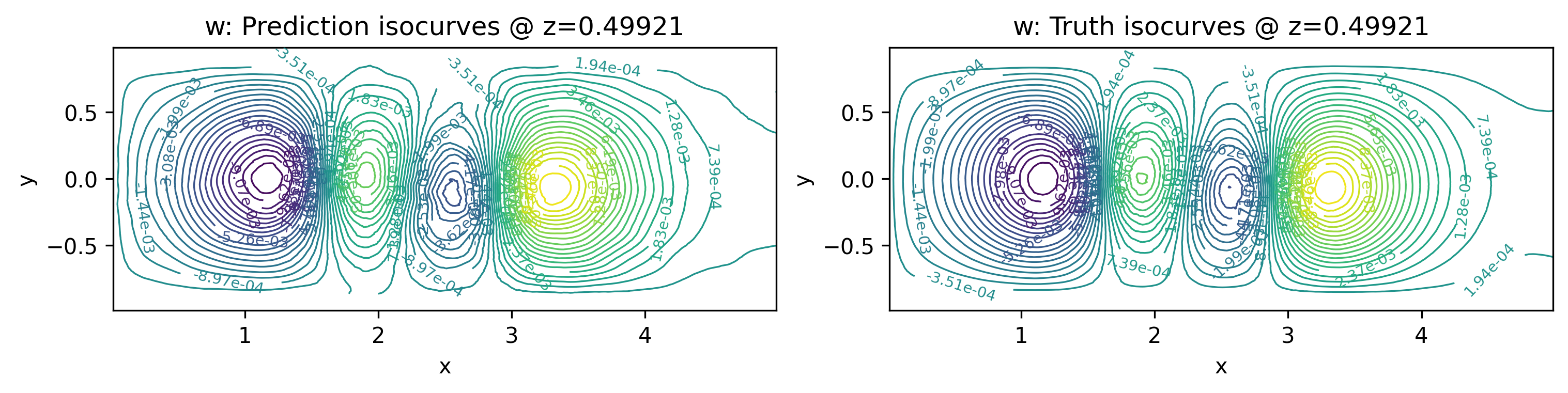}}
\caption{Prediction--reference isocurve comparisons for the rotated two-bump
test case $(\Delta\theta=5^\circ)$, on the same planes as
\Cref{fig:2br_cross_sectional_flow_comparison}: $p$ and $u$ on $y=0$, through
both bumps, $v$ on $z=0$, and $w$ on $z=0.50$, just below the bump crests.}
\label{fig:2br_iso_cross_sectional_flow_comparison}
\end{figure}

\begin{figure}[htbp]
\centering
\subfloat{\includegraphics[width=0.92\linewidth]{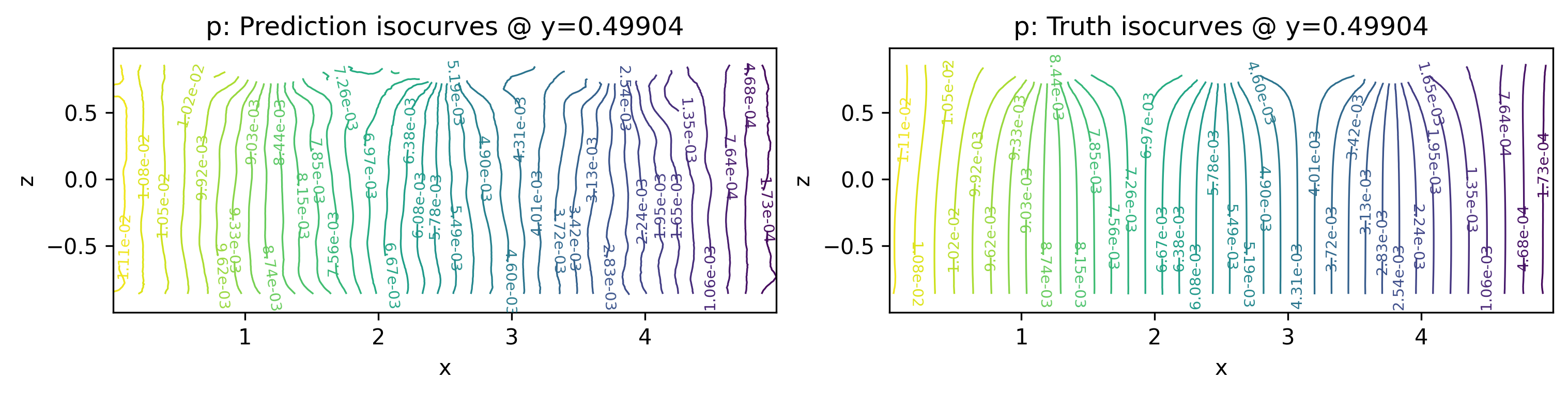}}
\par\medskip
\subfloat{\includegraphics[width=0.92\linewidth]{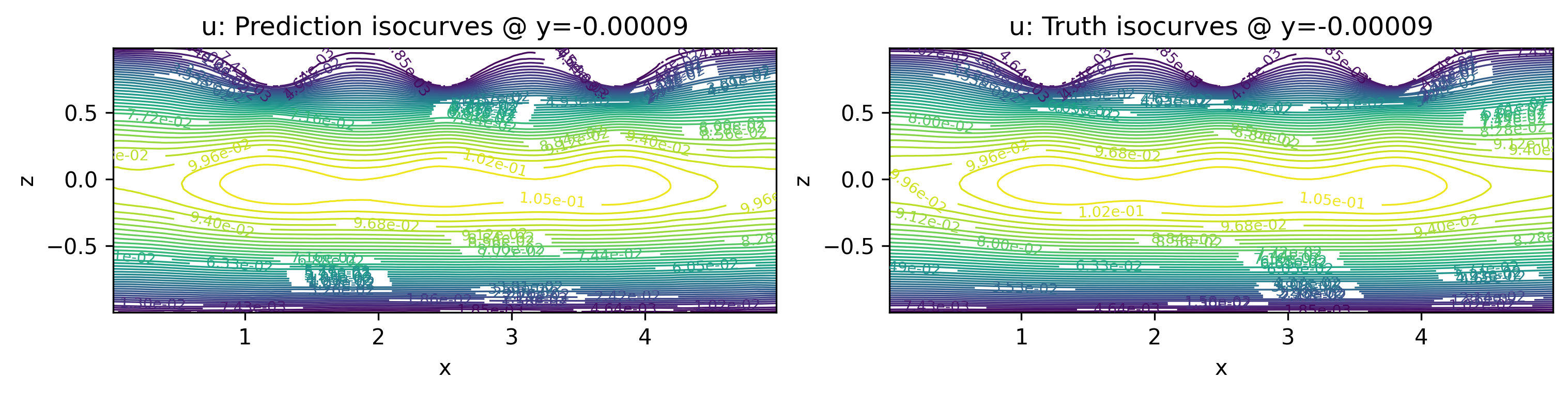}}
\par\medskip
\subfloat{\includegraphics[width=0.92\linewidth]{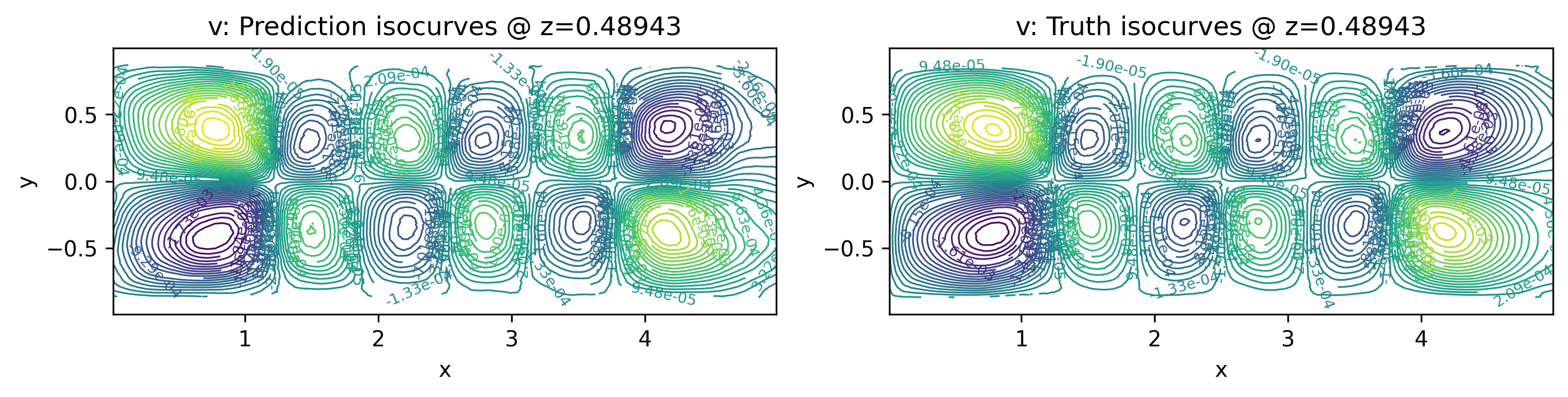}}
\par\medskip
\subfloat{\includegraphics[width=0.92\linewidth]{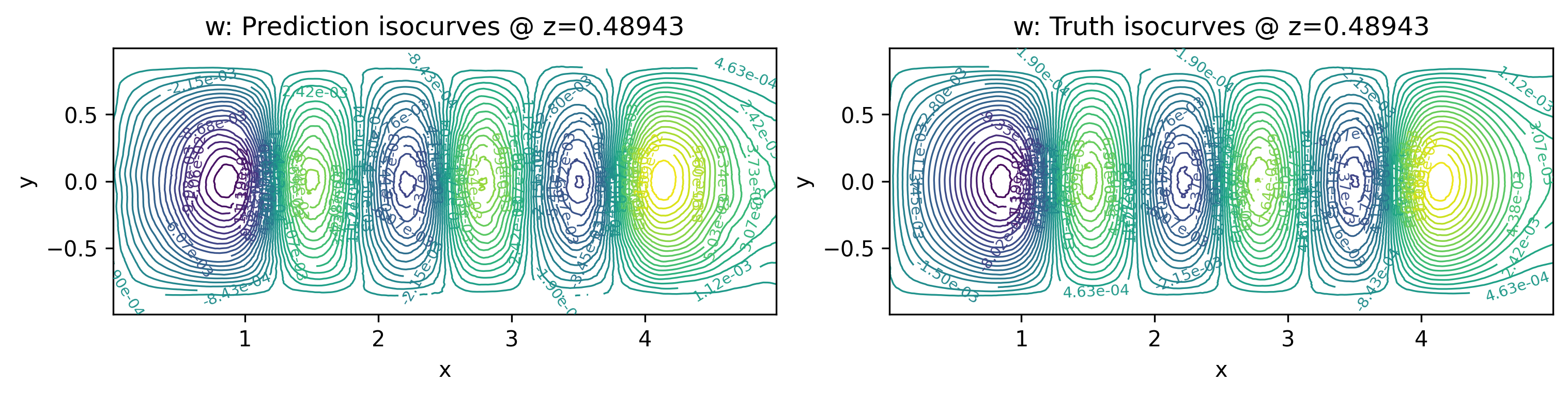}}
\caption{Prediction--reference isocurve comparisons for the three-bump test
case, on the same planes as \Cref{fig:3bcross_sectional_flow_comparison}: $p$
on $y=0.50$, $u$ on $y=0$, through all three bumps, and $v$ and $w$ on
$z=0.49$, between the pipe axis and the bump crests.}
\label{fig:3biso_cross_sectional_flow_comparison}
\end{figure}

\FloatBarrier

\subsection{One-Dimensional Line Profiles}
\hypertarget{target:line-profiles}{}
\label{subsec:line_profiles}

One-dimensional profiles provide a more localized comparison of LF2, NNLCI,
and the high-fidelity solution. Representative random and bump-targeted lines
are oriented along the $x$-, $y$-, or $z$-direction, the random lines crossing at least half of the pipe diameter, with the remaining coordinates fixed as reported in the captions. A point-in-tetrahedron test
retains only sampling locations inside the physical mesh. All three solutions
are then reconstructed at the same locations using a common Delaunay
triangulation and piecewise-linear barycentric interpolation \cite{barber1996quickhull,virtanen2020scipy}, preventing
differences in sampling or interpolation from biasing the comparison.

In all four configurations, NNLCI follows the high-fidelity profiles of $u$, $v$, and $w$ closely, apart from local over- or undershoots at some extrema, whereas LF2 deviates strongly, in particular in the transverse components. The pressure varies little across the pipe, so the vertical axes of its profiles are strongly magnified; there, NNLCI follows the shape of the reference but deviates by a few percent in places, most for one bump, consistent with the larger pressure error of this case in \Cref{tab:per_component_all}.

\begin{figure}[!htbp]
\centering
\subfloat{\includegraphics[width=0.48\linewidth]{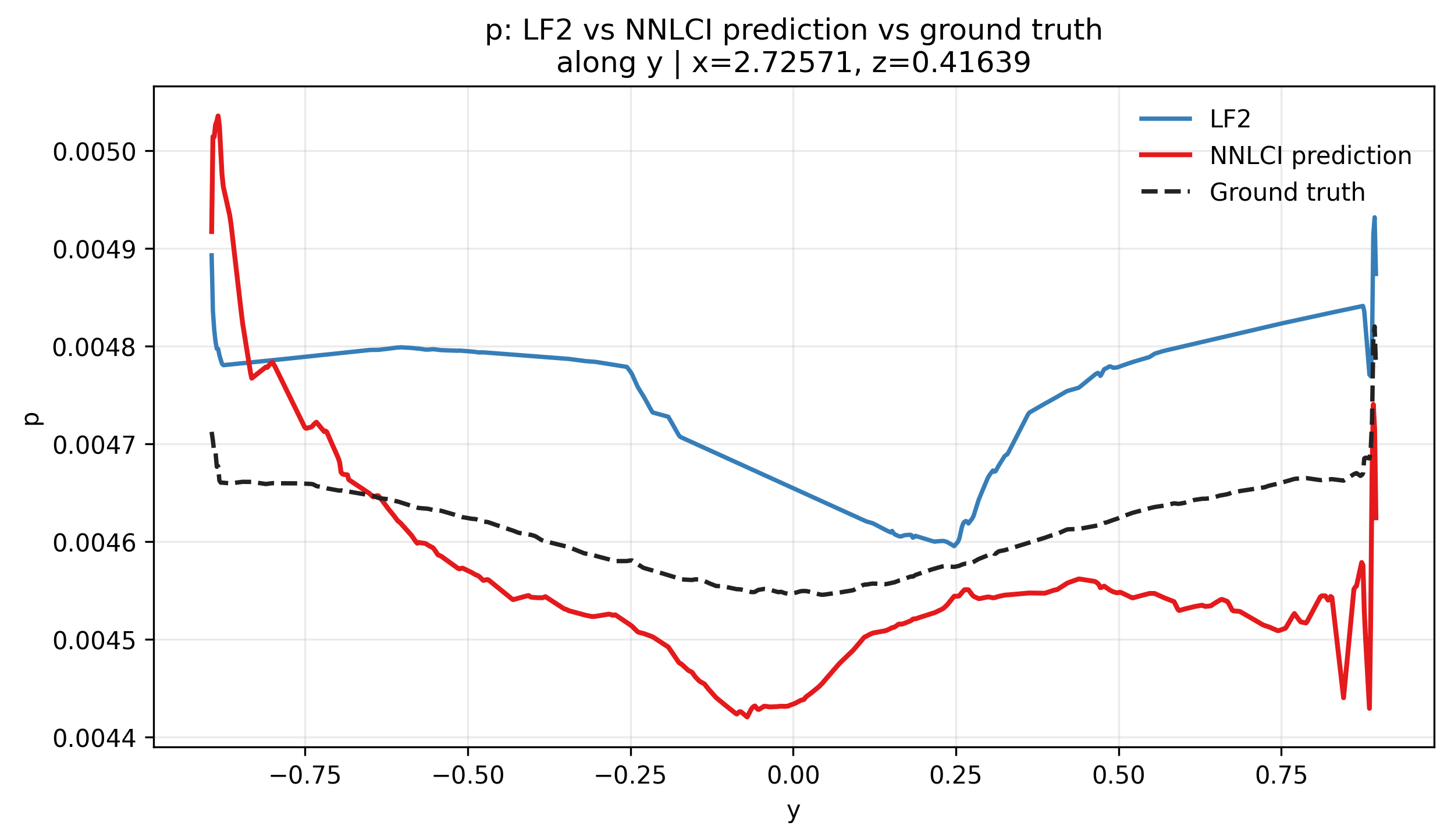}}
\hfill
\subfloat{\includegraphics[width=0.48\linewidth]{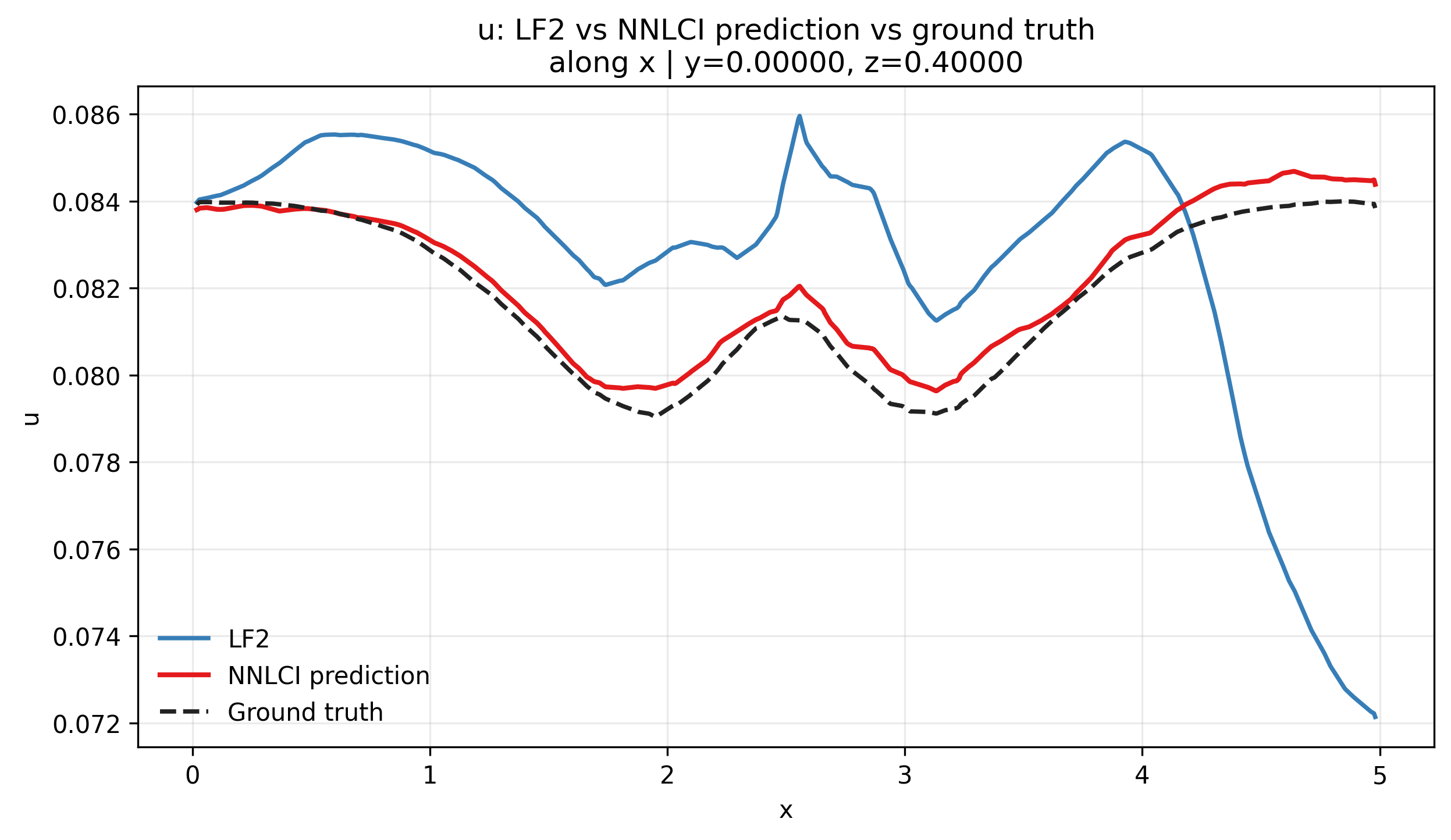}}
\par\medskip
\subfloat{\includegraphics[width=0.48\linewidth]{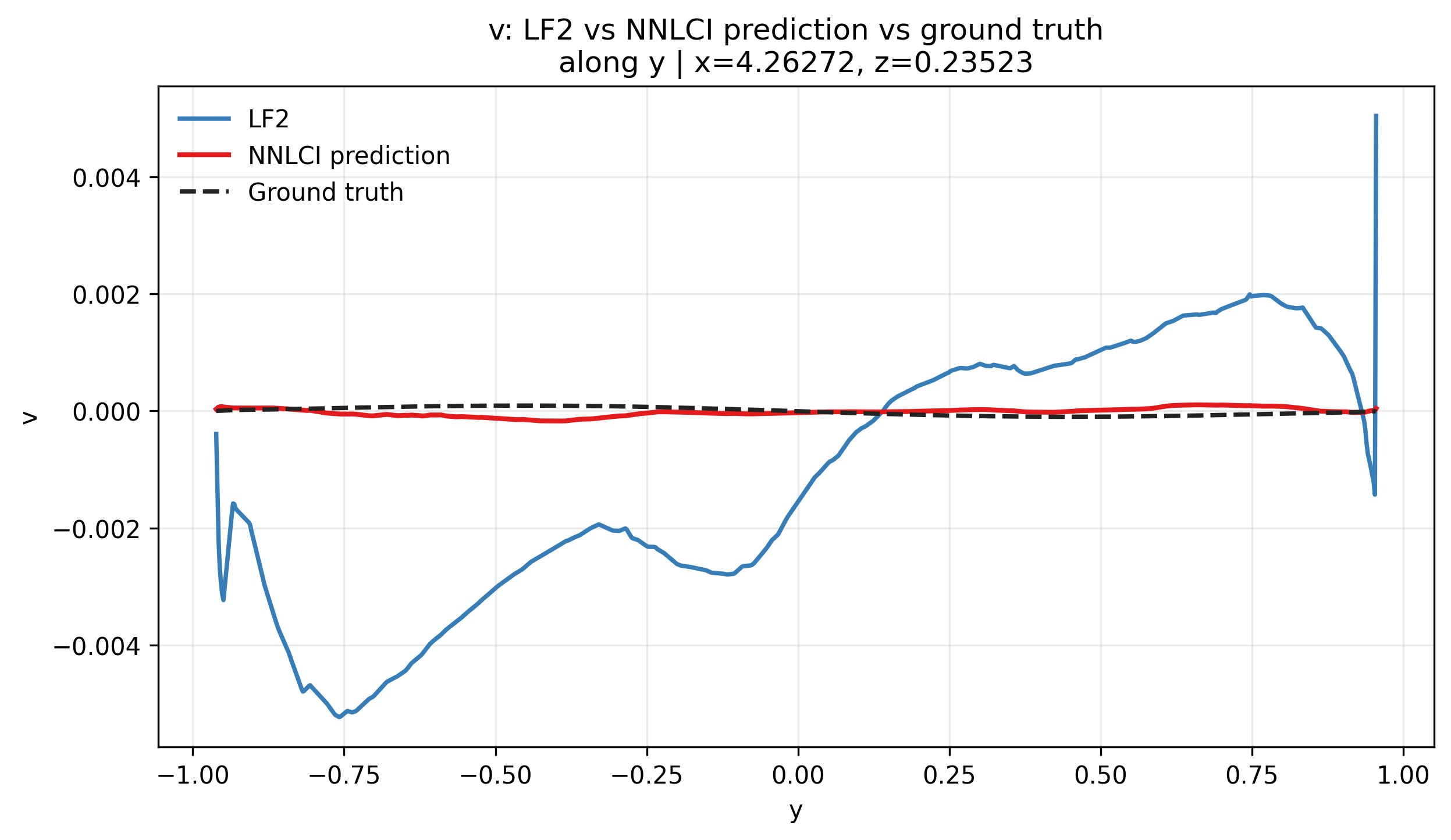}}
\hfill
\subfloat{\includegraphics[width=0.48\linewidth]{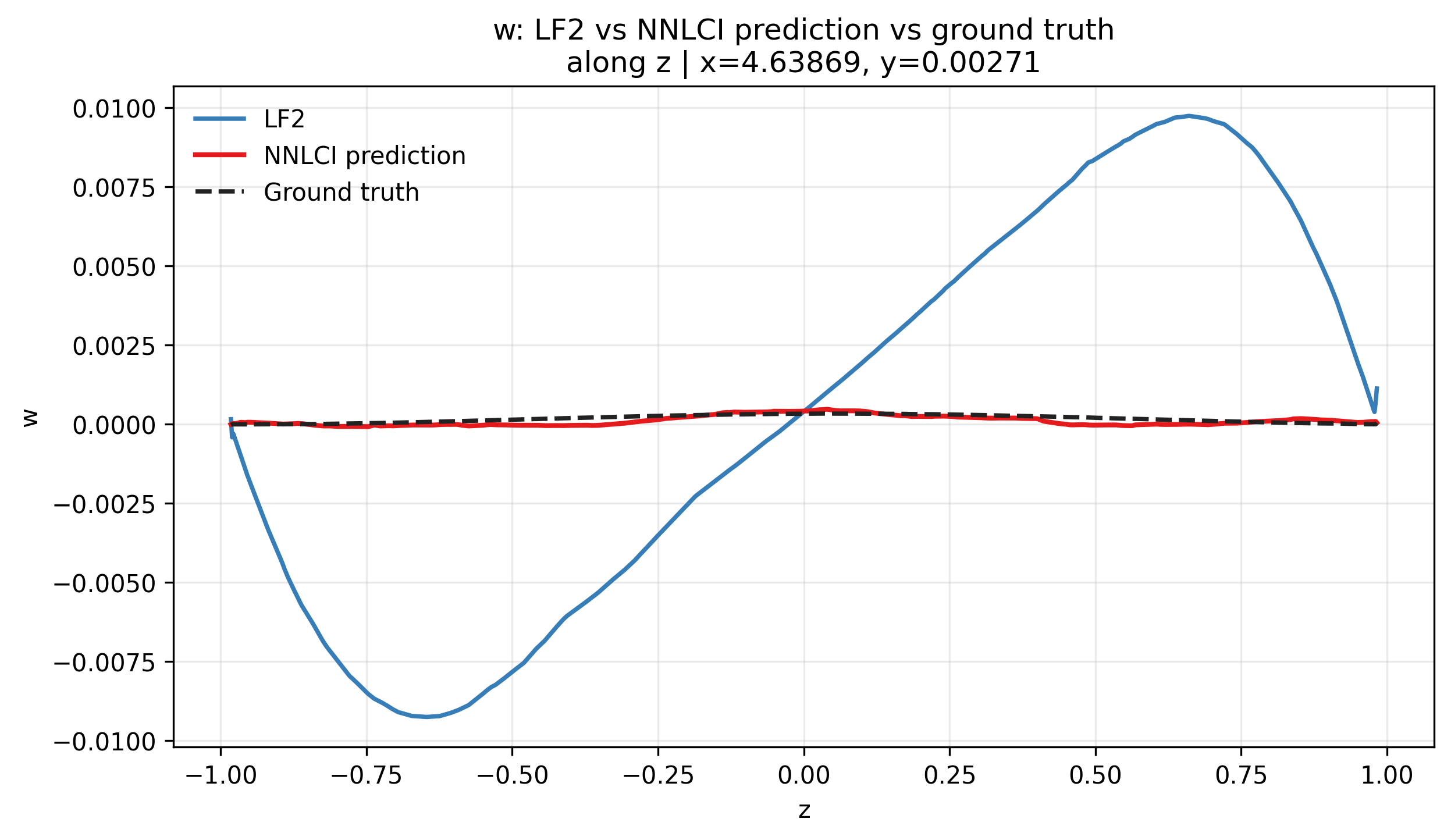}}
\caption{Representative one-dimensional profiles for the one-bump test case:
LF2, NNLCI, and the high-fidelity solution (ground truth). Top: $p$ along $y$ at
$(x,z)=(2.73,0.42)$ and $u$ along $x$ at $(y,z)=(0,0.40)$, through the bump.
Bottom: $v$ along $y$ at $(x,z)=(4.26,0.24)$ and $w$ along $z$ at
$(x,y)=(4.64,0)$.}
\label{fig:1b1d_flow_comparison}
\end{figure}

\begin{figure}[!htbp]
\centering
\subfloat{\includegraphics[width=0.48\linewidth]{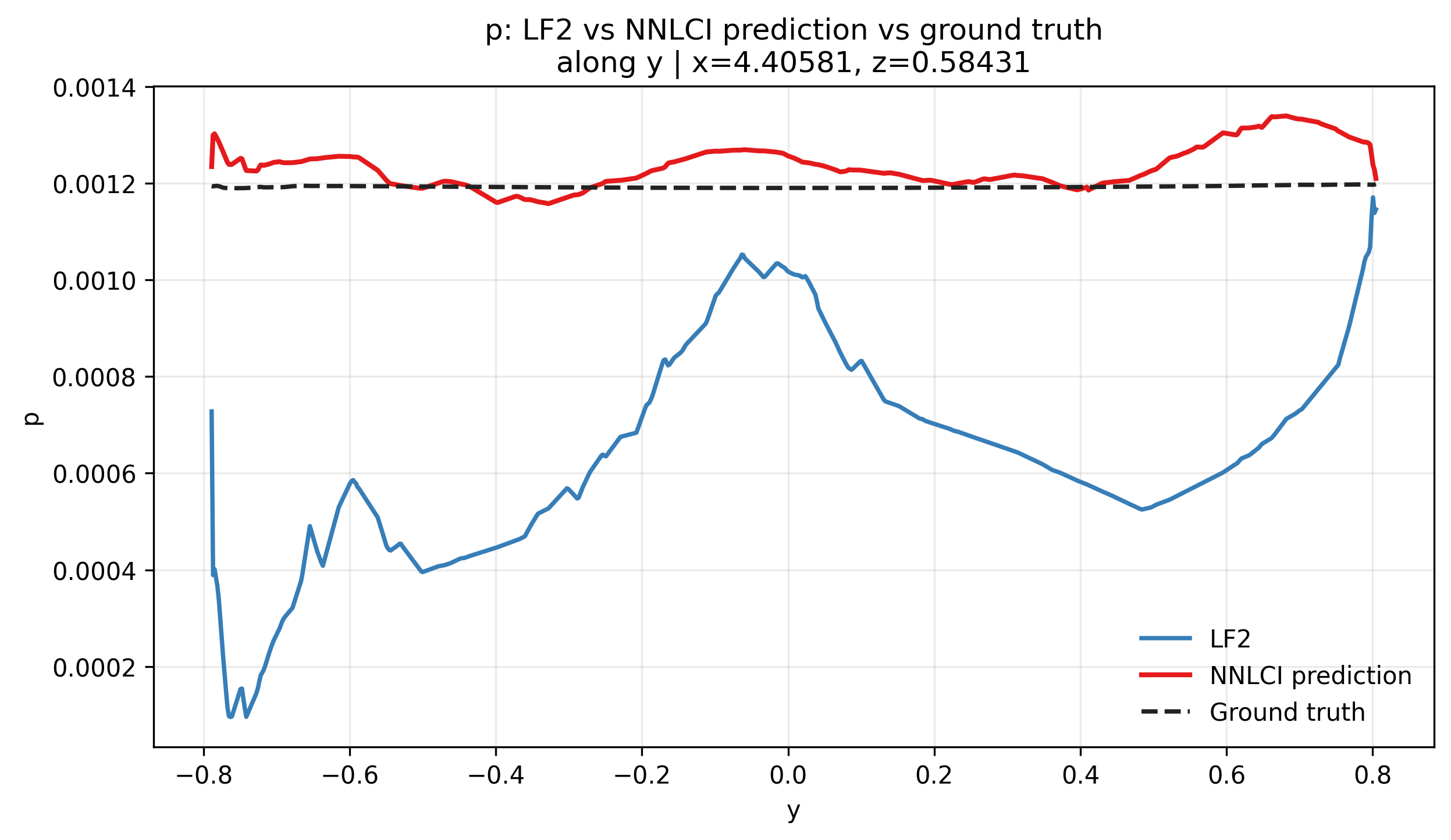}}
\hfill
\subfloat{\includegraphics[width=0.48\linewidth]{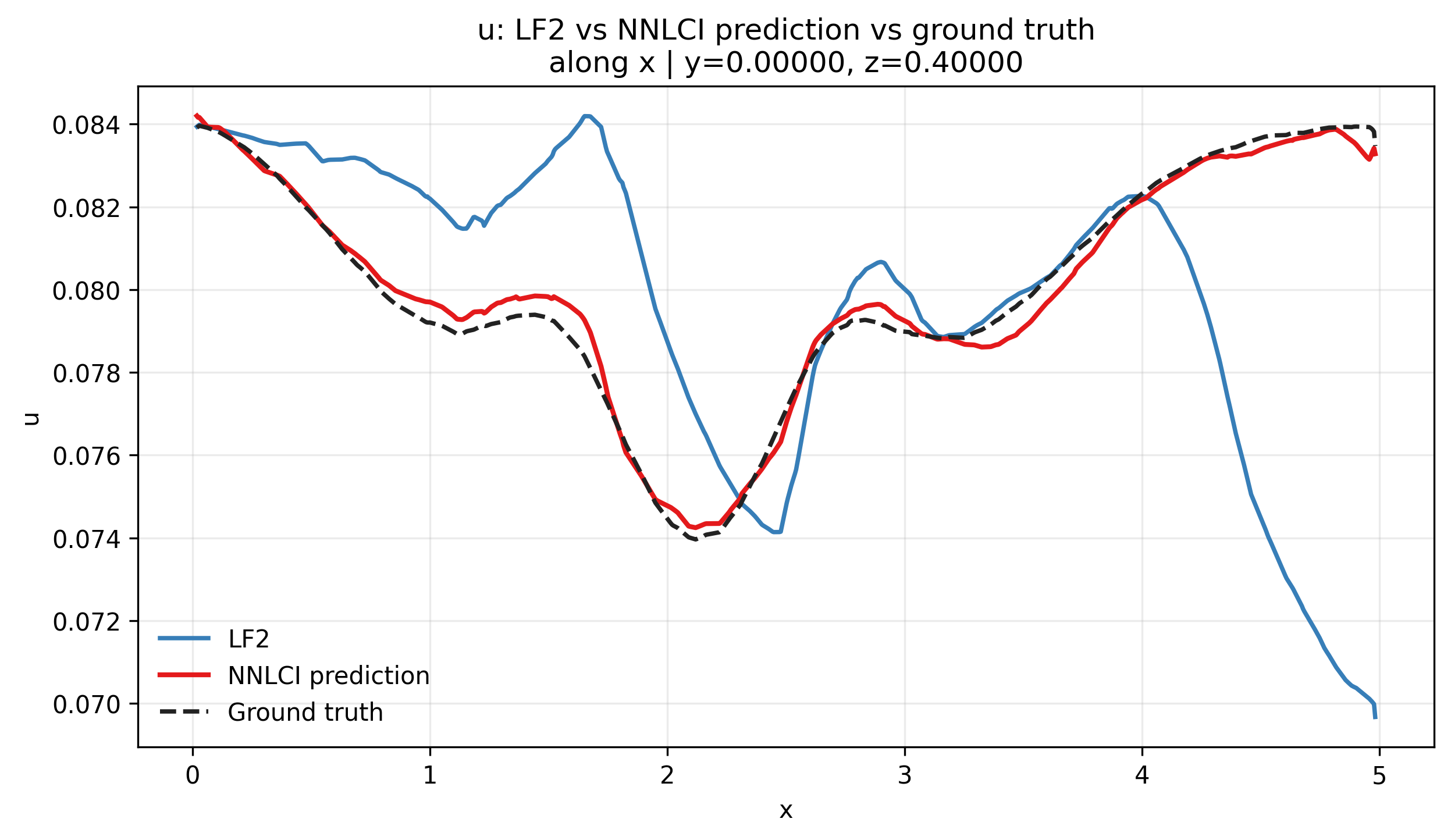}}
\par\medskip
\subfloat{\includegraphics[width=0.48\linewidth]{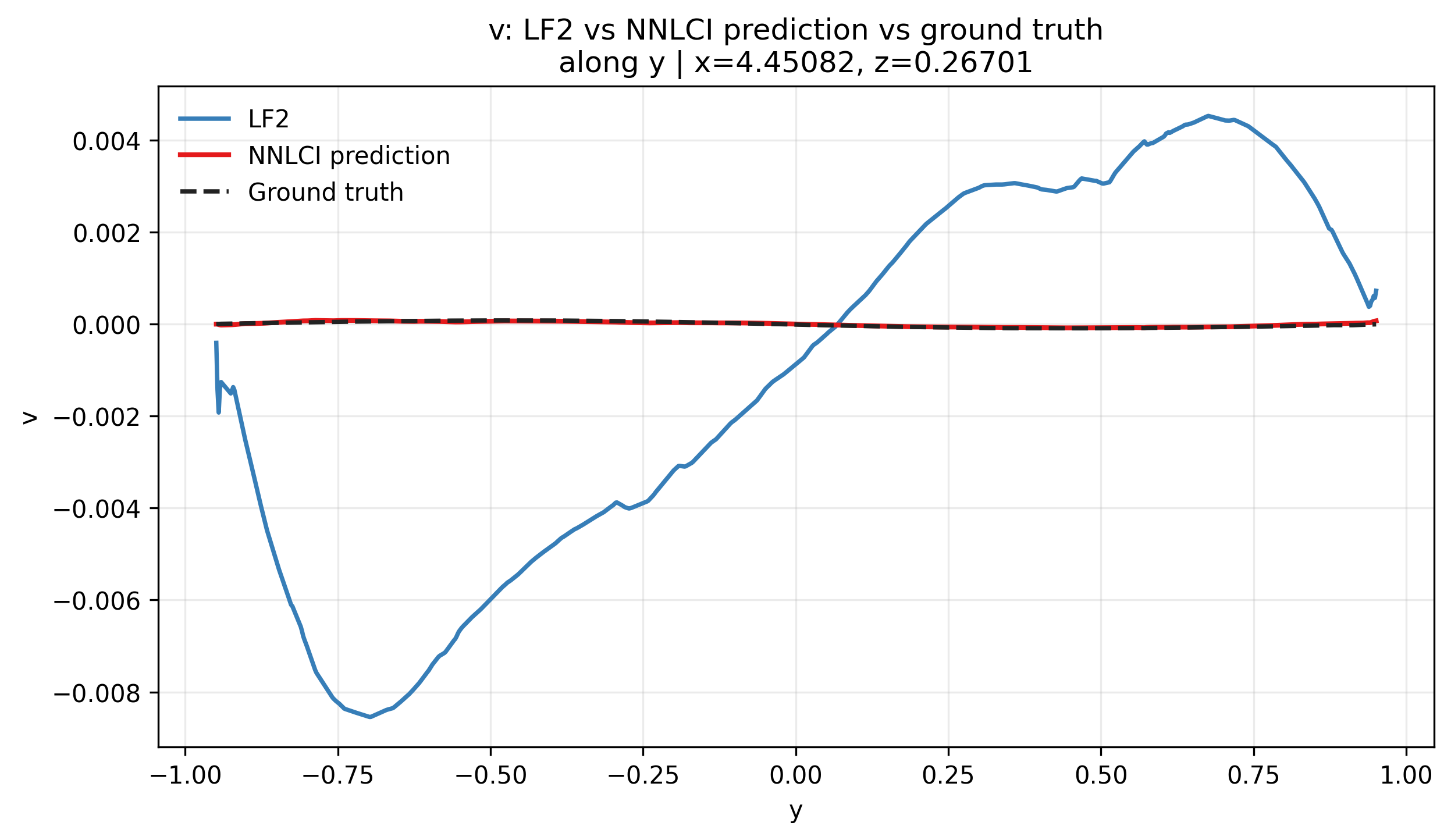}}
\hfill
\subfloat{\includegraphics[width=0.48\linewidth]{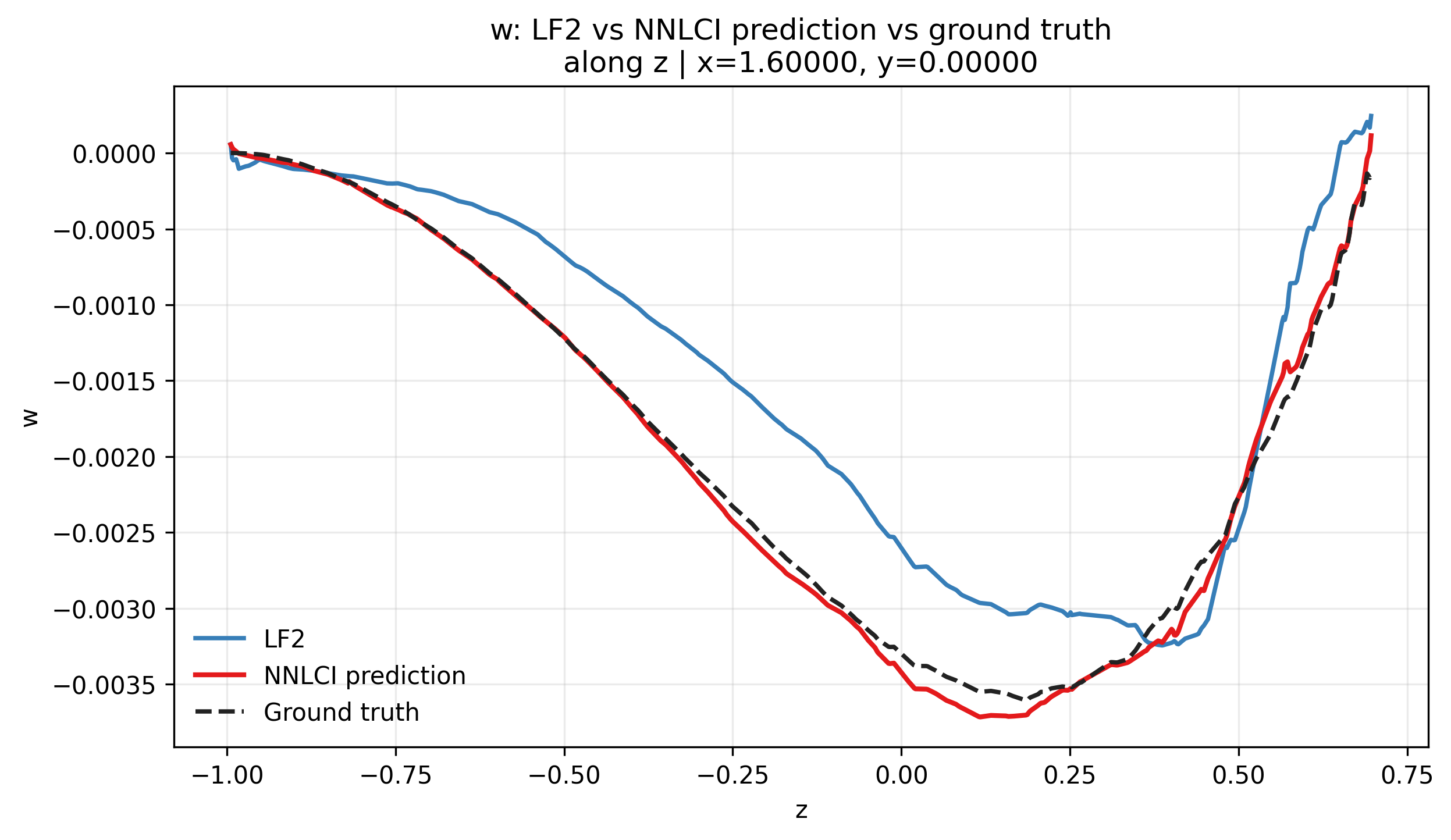}}
\caption{Representative one-dimensional profiles for the two-aligned-bump test
case: LF2, NNLCI, and the high-fidelity solution (ground truth). Top: $p$ along
$y$ at $(x,z)=(4.41,0.58)$ and $u$ along $x$ at $(y,z)=(0,0.40)$, through both
bumps. Bottom: $v$ along $y$ at $(x,z)=(4.45,0.27)$ and $w$ along $z$ at
$(x,y)=(1.60,0)$, through the first bump.}
\label{fig:2b1d_flow_comparison}
\end{figure}

\begin{figure}[!htbp]
\centering
\subfloat{\includegraphics[width=0.48\linewidth]{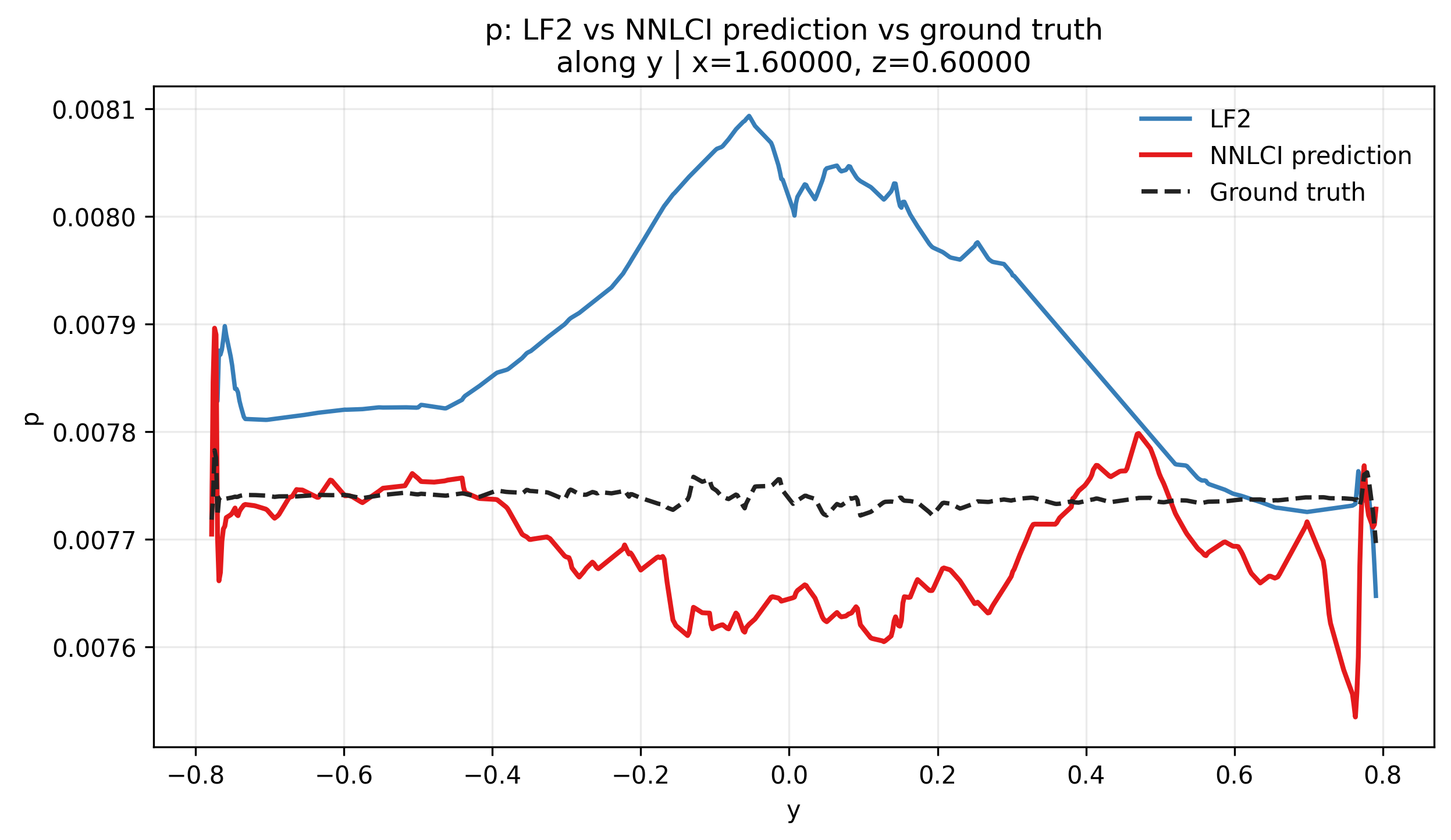}}
\hfill
\subfloat{\includegraphics[width=0.48\linewidth]{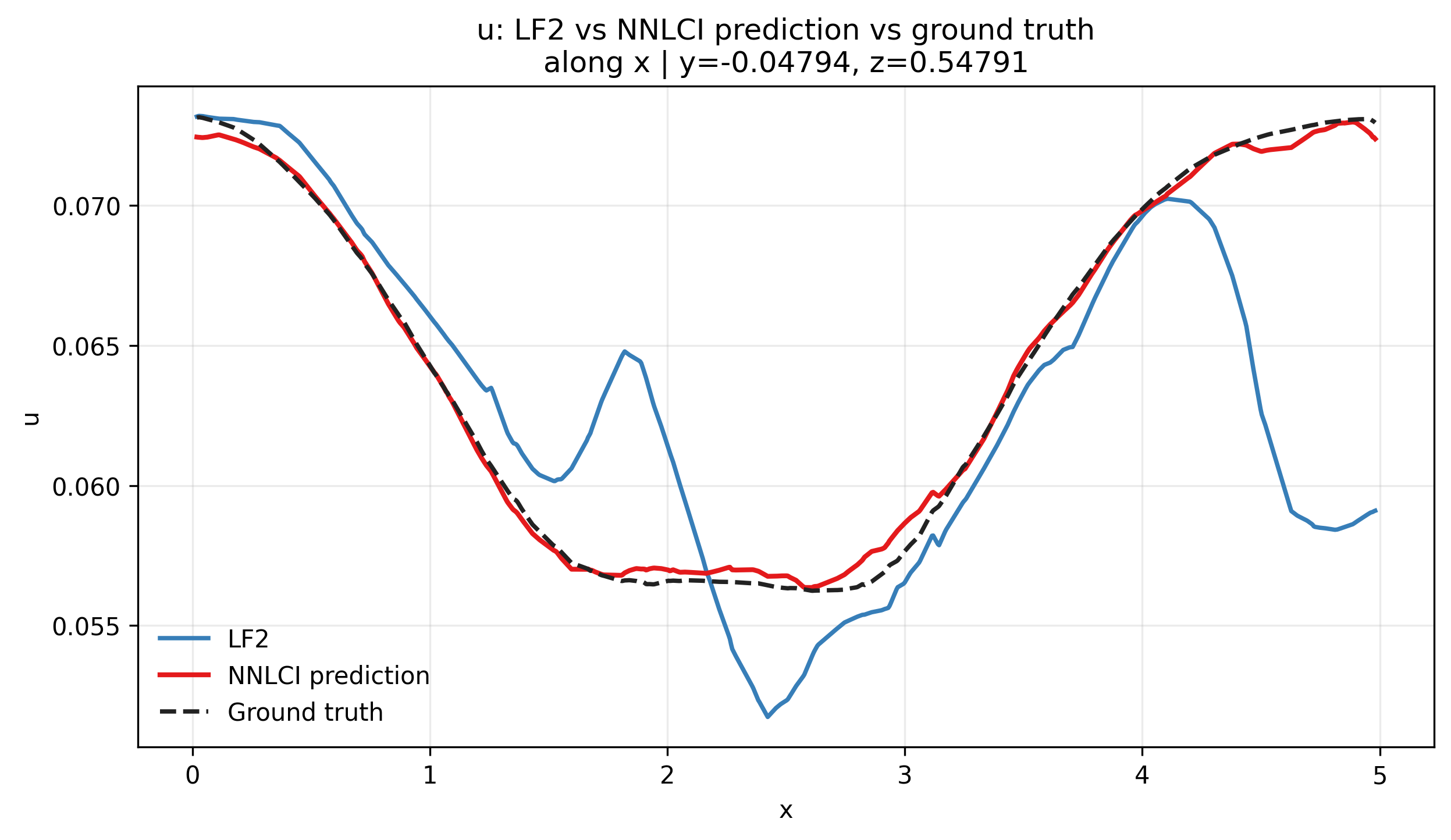}}
\par\medskip
\subfloat{\includegraphics[width=0.48\linewidth]{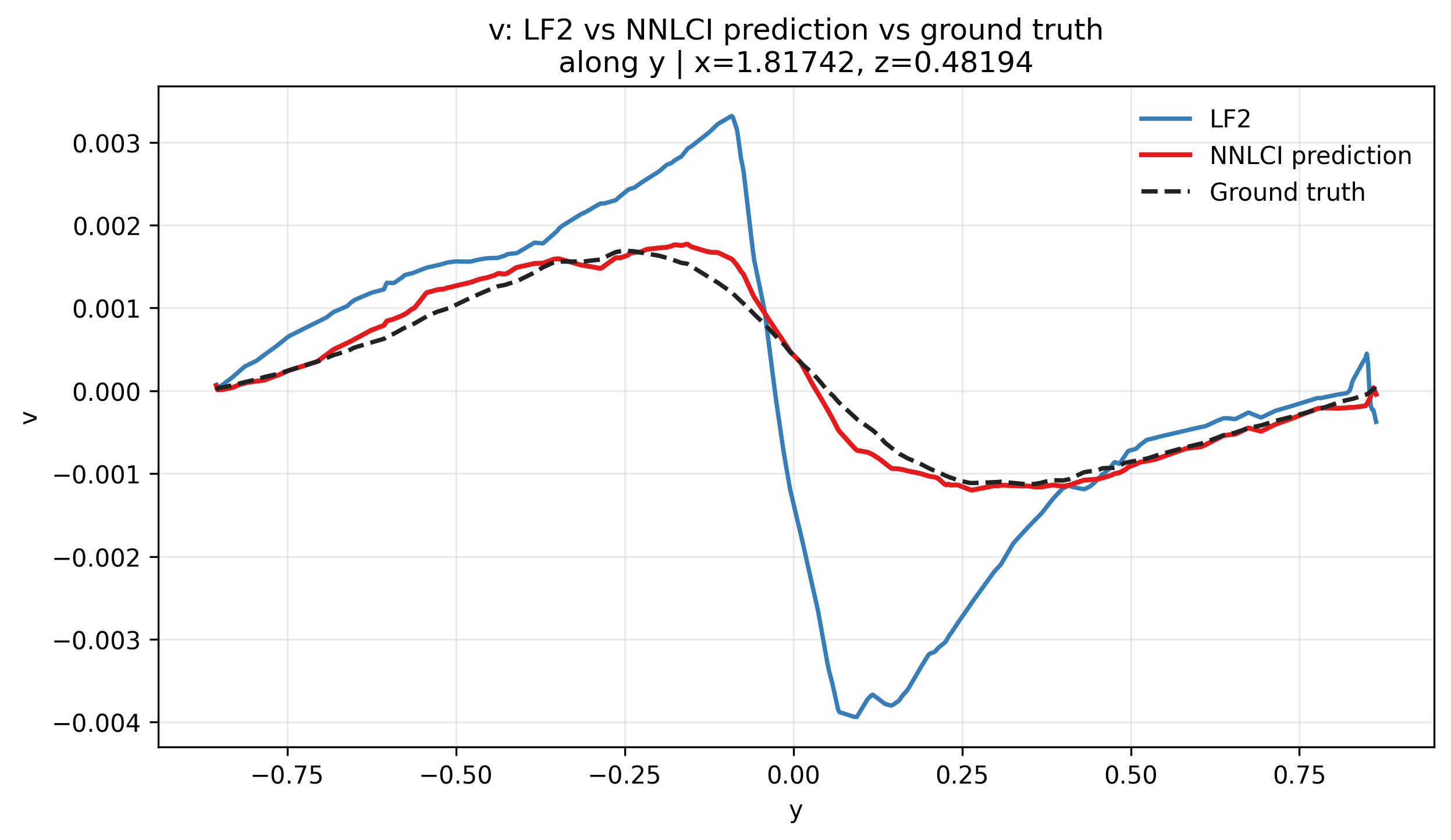}}
\hfill
\subfloat{\includegraphics[width=0.48\linewidth]{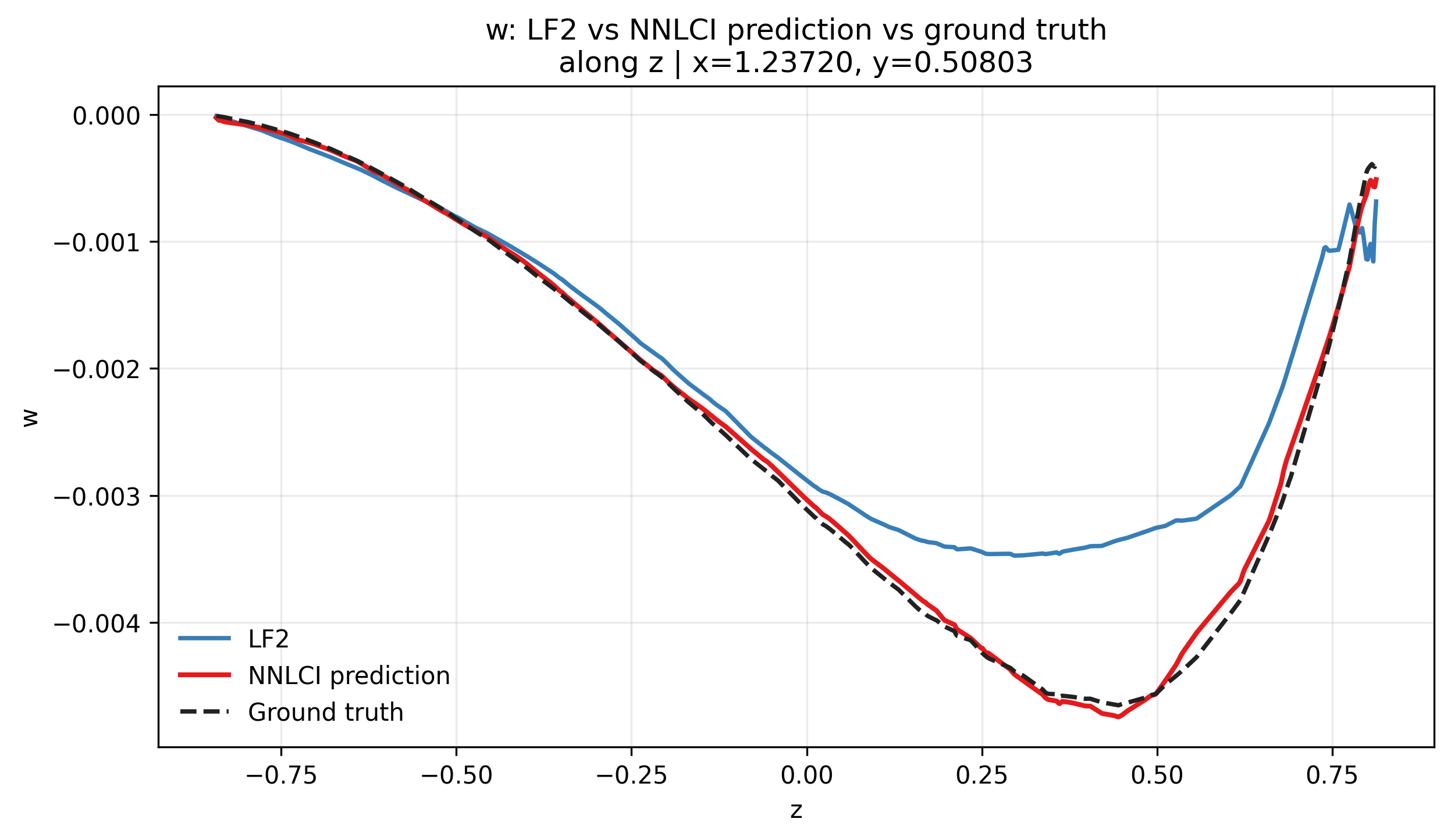}}
\caption{Representative one-dimensional profiles for the rotated two-bump test
case $(\Delta\theta=5^\circ)$: LF2, NNLCI, and the high-fidelity solution
(ground truth). Top: $p$ along $y$ at $(x,z)=(1.60,0.60)$, through the first
bump, and $u$ along $x$ at radius 0.55 in the direction of the rotated second
bump. Bottom: $v$ along $y$ at $(x,z)=(1.82,0.48)$ and $w$ along $z$ at
$(x,y)=(1.24,0.51)$.}
\label{fig:2b_rotated_1d_flow_comparison}
\end{figure}

\begin{figure}[!htbp]
\centering
\subfloat{\includegraphics[width=0.48\linewidth]{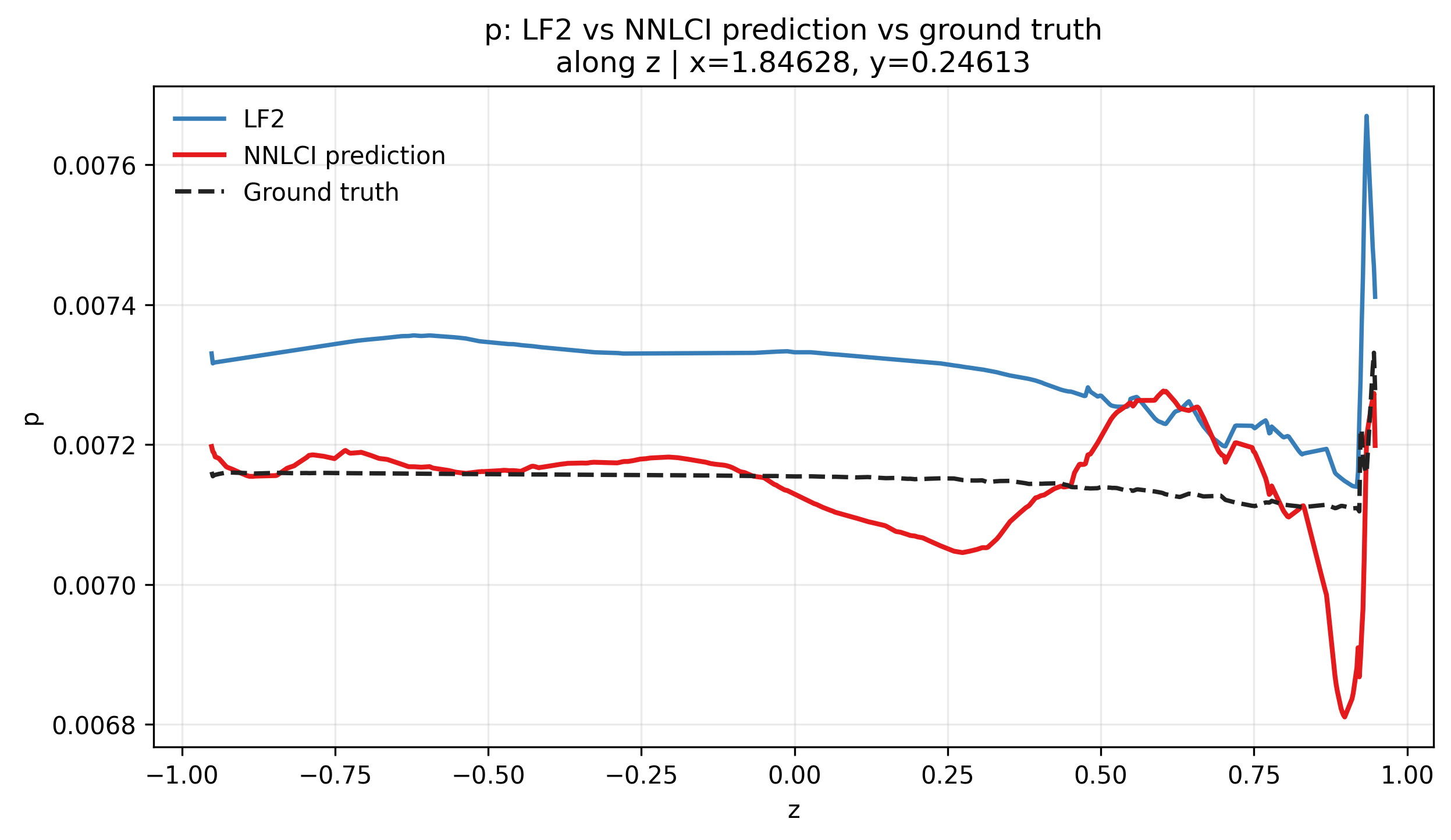}}
\hfill
\subfloat{\includegraphics[width=0.48\linewidth]{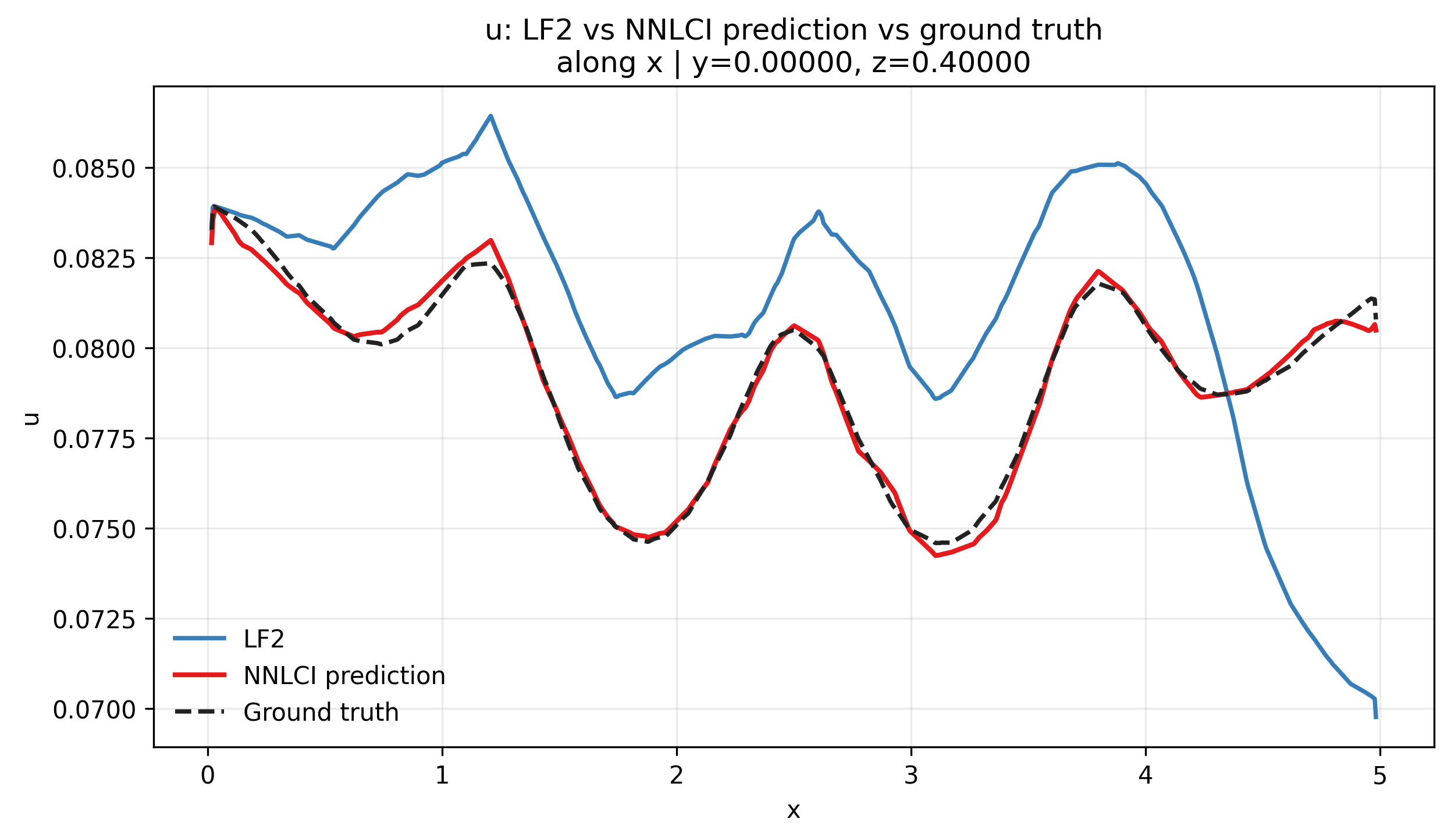}}
\par\medskip
\subfloat{\includegraphics[width=0.48\linewidth]{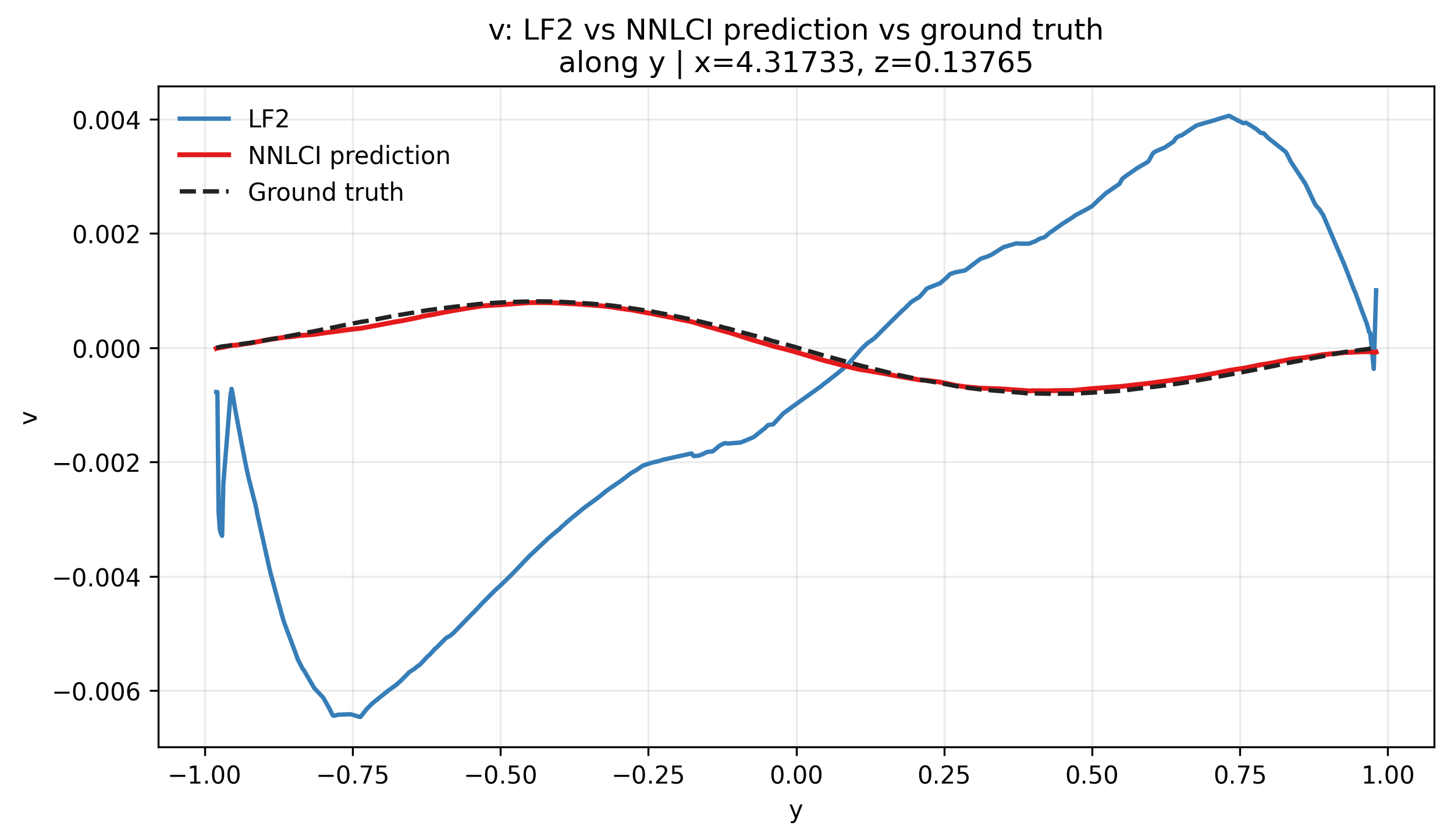}}
\hfill
\subfloat{\includegraphics[width=0.48\linewidth]{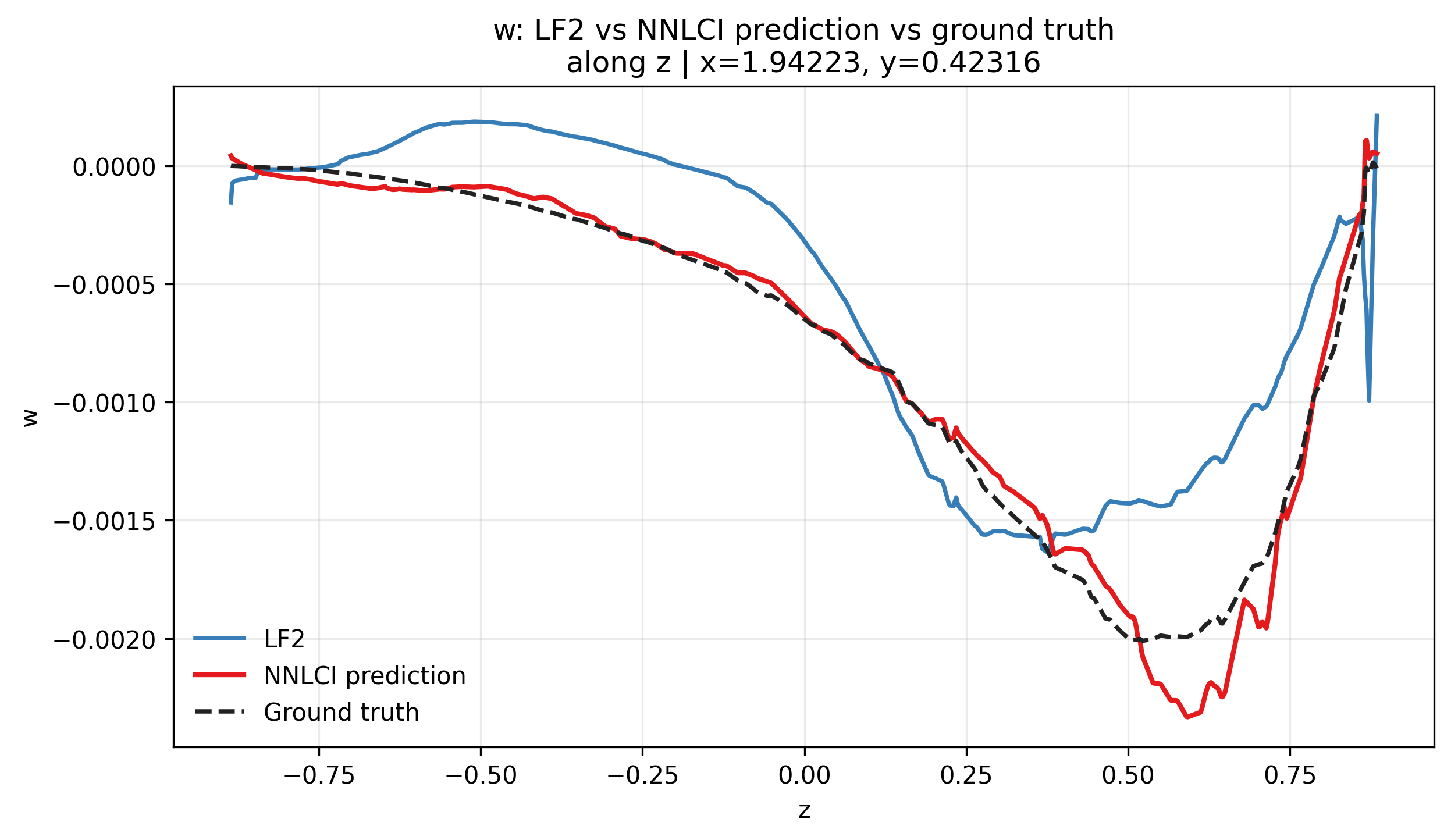}}
\caption{Representative one-dimensional profiles for the three-bump test case:
LF2, NNLCI, and the high-fidelity solution (ground truth). Top: $p$ along $z$
at $(x,y)=(1.85,0.25)$ and $u$ along $x$ at $(y,z)=(0,0.40)$, through all three
bumps. Bottom: $v$ along $y$ at $(x,z)=(4.32,0.14)$ and $w$ along $z$ at
$(x,y)=(1.94,0.42)$.}
\label{fig:3b1d_flow_comparison}
\end{figure}

\subsection{Accuracy Near the No-Slip Wall}
\hypertarget{target:near-wall}{}
\label{subsec:near_wall}

The remaining subsections examine the two-aligned-bump configuration of
Section~\ref{subsec:two_bump_base_test} in more detail, with three training seeds.
Next to the no-slip wall, the velocity varies most rapidly, and the stencils
of the high-fidelity centers extend beyond the polygonal low-fidelity
boundaries, so that their patches rely partly on the least-squares extension
of Section~\ref{subsec:matrix_generation}. Since the test matrices contain all
candidate centers, this layer is part of the test set. To quantify the
accuracy there, the errors are resolved by the distance from the wall,
estimated for each center as the difference between the largest radial
coordinate \(r=\sqrt{y^2+z^2}\) of the high-fidelity centers in the same axial
and circumferential bin, of width 0.1 and \(7.5^\circ\), and the radial
coordinate of the center (\Cref{tab:near_wall}).

\begin{table}[htbp]
\centering
\caption{Two-aligned-bump test case: composite relative \(L^2\) error by
approximate distance from the wall. NNLCI values are the mean and standard
deviation over three training seeds; \(T\) is the ratio of the LF2 error to
the mean NNLCI error.}
\label{tab:near_wall}
\small
\begin{tabular}{@{}lcccc@{}}
\toprule
Distance from the wall & Test points & NNLCI (\%) & LF2 (\%) & \(T\) \\
\midrule
Below 0.05     & 22{,}740  & \(7.5\pm0.8\)   & 36.4 &  4.9 \\
0.05 to 0.15   & 25{,}876  & \(2.5\pm0.2\)   & 16.1 &  6.4 \\
Above 0.15     & 104{,}938 & \(0.58\pm0.05\) &  7.1 & 12.3 \\
\midrule
All            & 153{,}554 & \(0.87\pm0.08\) &  7.9 &  9.1 \\
\bottomrule
\end{tabular}
\end{table}

Both errors grow toward the wall, consistent with the steep velocity gradients
there and with the greater reliance of the input patches on the extension. The
NNLCI error rises from 0.58\% at distances above 0.15 to 7.5\% within 0.05 of
the wall, and its advantage over LF2 decreases from a factor of 12.3 to 4.9.
The 830 test centers (0.5\%) that the patch-jump criterion would reject, most
of them next to the wall, are included as well: although their inputs contain
the large variations that the criterion detects, NNLCI is more accurate than LF2 at
94\% of them, with an error of \(11\pm4\%\) compared with 37\% for LF2. NNLCI
remains several times more accurate than LF2 in the near-wall layer,
where the errors of both are largest.

\subsection{Prediction Outside the Training Range}
\hypertarget{target:extrapolation}{}
\label{subsec:extrapolation}

The test cases of Section~\ref{subsec:quantitative_results} lie inside the training
ranges. To examine extrapolation, the three networks trained on the
two-aligned-bump configuration of Section~\ref{subsec:two_bump_base_test}, with
\(x_{b,2}\in\{2.56,2.77\}\) and \(U_{\max}\in\{0.09,0.11\}\), are applied
without retraining to a case outside both ranges,
\((x_{b,2},U_{\max})=(2.45,0.08)\), whose test matrix is constructed as in
Section~\ref{subsec:matrix_generation}.

\begin{table}[htbp]
\centering
\caption{Two-aligned-bump configuration: test inside and outside the training
ranges \(x_{b,2}\in[2.56,2.77]\) and \(U_{\max}\in[0.09,0.11]\), with the same
three networks. NNLCI values are the mean and standard deviation over the
three training seeds; the last column gives the percentage of test points at
which NNLCI is more accurate than LF2.}
\label{tab:extrapolation}
\small
\setlength{\tabcolsep}{4pt}
\begin{tabular}{@{}lcccccc@{}}
\toprule
 & & & \multicolumn{2}{c}{\(E_{\mathrm{rel}}\)} & & NNLCI \\
\cmidrule(lr){4-5}
Test case & \(x_{b,2}\) & \(U_{\max}\) & NNLCI & LF2 & \(T\) & better (\%) \\
\midrule
Interpolation & 2.66 & 0.10 & \((8.70\pm0.83)\times10^{-3}\) & \(7.90\times10^{-2}\) & 9.1 & 96.6 \\
Extrapolation & 2.45 & 0.08 & \((1.51\pm0.11)\times10^{-2}\) & \(7.42\times10^{-2}\) & 4.9 & 87.4 \\
\bottomrule
\end{tabular}
\end{table}

Outside the training range, the error of each network increases by a factor
of about 1.7, much more than the variation between the networks, and the
advantage over LF2 decreases from a factor of 9.1 to 4.9
(\Cref{tab:extrapolation}). All components degrade, the transverse velocities
most strongly: the error of \(v\) increases from 18.0\% to 26.7\% and that of
\(w\) from 5.6\% to 11.8\%, compared with 0.82\% to 1.37\% for \(u\) and
1.25\% to 1.74\% for \(p\). This is consistent with inputs that differ from
those seen in training: the bumps are closer together than in any training
case, and the network is not constructed to be invariant to the rescaling of
the Stokes solution by the inflow velocity. NNLCI remains about five
times more accurate than LF2 just outside the training range, but its accuracy
degrades, so the training data should cover the parameter range of interest.

\subsection{Ablation: Degree of the Local Least-Squares Extension}
\hypertarget{target:ablation-deg}{}
\label{subsec:ablation_degree}

Stencil points outside the low-fidelity mesh, either beyond the physical
boundary or between the polygonal low-fidelity boundary and the curved
physical boundary, need values where the finite-element solution is not
defined. This is analogous to ghost cells in finite-volume and
immersed-boundary methods, whose values are obtained by polynomial
extrapolation or reconstruction from interior data
\cite{leveque2002finite,tseng2003ghost,mittal2005immersed}; the extension of
Section~\ref{subsec:matrix_generation} is a weighted least-squares fit similar to moving least squares \cite{lancaster1981surfaces,levin1998mls}. To assess
the role of its polynomial degree, the two-aligned-bump case is repeated with
a linear fit, \(q=c_0+c_1\,\delta x+c_2\,\delta y+c_3\,\delta z\), with all
other settings unchanged, and each variant is trained with three seeds and
evaluated on all 153{,}554 test points.

\begin{table}[htbp]
\centering
\caption{Relative \(L^2\) errors of NNLCI for the two-aligned-bump test case
with a quadratic or a linear least-squares extension, on all 153{,}554 test
points: mean and standard deviation over three training seeds.}
\label{tab:ablation_degree}
\small
\setlength{\tabcolsep}{4pt}
\begin{tabular}{@{}lccccc@{}}
\toprule
 & Composite error
 & \multicolumn{4}{c}{Error of each component \(E_{\mathrm{rel},c}\) (\%)} \\
\cmidrule(lr){3-6}
Extension & \(E_{\mathrm{rel}}\) (\%) & \(p\) & \(u\) & \(v\) & \(w\) \\
\midrule
Quadratic & \(0.87\pm0.08\) & \(1.25\pm0.13\) & \(0.82\pm0.07\) & \(18.0\pm2.7\) & \(5.57\pm0.76\) \\
Linear    & \(1.01\pm0.03\) & \(0.94\pm0.10\) & \(0.99\pm0.03\) & \(14.7\pm0.9\) & \(4.28\pm0.04\) \\
\bottomrule
\end{tabular}
\end{table}

The quadratic extension gives the lower composite error in every seed
(\Cref{tab:ablation_degree}): its largest error, 0.96\%, is below the
smallest linear error, 0.99\%, so it remains below 1\% in every seed, whereas
the linear extension exceeds 1\% in two of the three. The advantage comes from
the axial velocity \(u\), which dominates the composite error; for \(p\),
\(v\), and \(w\), the linear extension is slightly more accurate.

The difference concentrates where the axial-velocity profile is curved. Within
0.05 of the wall (Section~\ref{subsec:near_wall}), the composite error is
\(7.5\pm0.8\%\) for the quadratic and \(9.7\pm0.4\%\) for the linear
extension, and near the inlet it is \(1.21\pm0.06\%\) and \(2.04\pm0.17\%\). A
quadratic continuation follows the curvature of the velocity profile beyond
the low-fidelity wall, whereas a linear one cannot; the quadratic fit also
matches the degree of the \(P_2\) velocity, whereas the \(P_1\) pressure is
captured slightly better by the linear fit. Near the outlet, where the
developed flow varies little along the pipe, the linear extension is slightly
more accurate in every seed, with \(0.90\pm0.02\%\) compared with
\(1.13\pm0.24\%\). The linear extension also stays within about 1.3 times the
field scale beyond the wall, so that it needs no rejection of training
samples, but it yields the larger composite error in every seed.

A constant fill is a simpler alternative. Zero padding is standard in 
convolutional networks and informs them about the location of the boundary
\cite{islam2020much}, and for convolutional surrogates of partial differential
equations the best padding is the one compatible with the boundary condition
of the data \cite{alguacil2021effects}. A zero value coincides with the no-slip
boundary value of the velocity, and in preliminary tests on this configuration
it gave lower errors than the quadratic extension. It is not adopted because
it relies on the solution vanishing at the boundary, which does not hold for
the pressure at the wall, the velocity at outflow boundaries, or moving walls
and nonzero Dirichlet data, whereas the least-squares extension uses only the
local low-fidelity solution. The remaining configurations use the quadratic
extension.
\FloatBarrier

\subsection{Computational Cost}
\hypertarget{target:comp-cost}{}
\label{subsec:computational_cost}

The finite-element solves were run on a CPU-only Intel Cascade Lake node of
a university computing cluster, with four cores and
32~GB of memory; the network was trained and evaluated on  
a node of the same cluster
with an
NVIDIA Tesla V100 GPU (16~GB), and the matrices were constructed on Google
Colab runtimes using the CPU only. For the two-aligned-bump configuration,
the offline cost, consisting of the four high-fidelity training solves, the
construction of the training matrices, and the training, is about 3.6~h and
is incurred once. A prediction for a new parameter value requires the two
low-fidelity solves (8~s, about 90 times cheaper than the 12-minute
high-fidelity solve), the construction of the input patches, and the network
evaluation (about 1~\(\mu\)s per point). The construction dominates: as a
serial loop at about 10~ms per point, it makes a full-field prediction take
about 27~min. Because the points are processed independently, it can be
parallelized, and its cost is proportional to the number of target points,
whereas a high-fidelity solve always covers the whole domain.

\section{Conclusions}
\hypertarget{target:conclusions}{}
\label{sec:conclusions}

This work extends the NNLCI framework to the three-dimensional steady
incompressible Stokes equations on unstructured tetrahedral meshes. The method
combines two converging low-fidelity finite-element solutions on a shared
local stencil with local mesh-size information to predict the high-fidelity
pressure and velocity. The proposed mesh-spacing estimator and hybrid
finite-element/local least-squares reconstruction provide consistent inputs on
irregular meshes, including stencils that cross the physical boundary, so that
predictions are obtained at all high-fidelity centers, including those next to
the no-slip wall.

The method was evaluated on four full-length pipe configurations: one bump,
two aligned bumps, a rotated two-bump arrangement, and three bumps. For the
held-out test cases, evaluated at all high-fidelity centers, NNLCI reduced the
composite volume-weighted relative \(L^2\) error by \(86\%\) to \(92\%\)
compared with the LF2-center baseline and reduced the error of every solution
component by at least a factor of \(3\); the largest reductions, by factors
of \(10\) to \(34\), occurred for the transverse velocity components, which
the coarse solutions resolve poorly. The cross-sectional, isocurve, and line-profile comparisons show close agreement for the velocity, with pressure deviations of a few percent in places. Within the sampled ranges of geometry and inflow, NNLCI interpolates accurately.

For the two-aligned-bump configuration, additional studies with three training
seeds show that NNLCI remains about five times more accurate than LF2 within
0.05 of the no-slip wall, where the errors of both are largest; that the
quadratic least-squares extension gives a lower error than a linear one in
every seed; and that outside the training range the error increases by a
factor of about 1.7, while NNLCI remains about five times more accurate than
LF2. In the present serial implementation, the construction of the inputs
dominates the cost of a prediction, but it is local and can be parallelized.
Future work will address a systematic study of extrapolation, a parallel
implementation of the input construction, and the extension of NNLCI to the
three-dimensional incompressible Navier--Stokes equations, which adds nonlinear advection, time dependence, and a dependence on the Reynolds number, and raises the question of long-time stability.

\section*{Data availability}
We are not aware of a public data set that provides the paired low- and
high-fidelity finite element solutions on matching parameterized geometries
that NNLCI requires. All data were generated with the finite element
simulations described in Section~\ref{sec:stokes_fem}. The training and test
matrices of the two-aligned-bump configuration are available at
[link withheld for double-anonymized review; provided to the editor]. The code and the
remaining data are available from the corresponding author on reasonable
request.
\section*{Declaration of generative AI and AI-assisted technologies in the manuscript preparation process}
During the preparation of this work the authors used ChatGPT (OpenAI,
GPT-6 Astra Pro) and Claude (Anthropic, Claude Opus 5.5) to assist with
writing and editing the text and with preparing analysis scripts. The authors
determined all content and performed all simulations and computations. After
using these tools, the authors reviewed and edited the content as needed and
take full responsibility for the content of the published article.

\bibliographystyle{siamplain}
\bibliography{references}

@book{kim1991microhydrodynamics,
  author    = {Kim, S. and Karrila, S. J.},
  title     = {Microhydrodynamics: Principles and Selected Applications},
  publisher = {Butterworth-Heinemann},
  year      = {1991},
  doi       = {10.1016/C2013-0-04644-0}
}

@book{boffi2013mixed,
  author    = {Boffi, D. and Brezzi, F. and Fortin, M.},
  title     = {Mixed Finite Element Methods and Applications},
  series    = {Springer Series in Computational Mathematics},
  publisher = {Springer},
  year      = {2013},
  doi       = {10.1007/978-3-642-36519-5}
}

@article{Forrester2007,
  author  = {Forrester, A. I. and S{\'{o}}bester, A. and Keane, A. J.},
  title   = {{Multi-fidelity optimization via surrogate modelling}},
  journal = {Proceedings of the Royal Society A: Mathematical, Physical and Engineering Sciences},
  volume  = {463},
  number  = {2088},
  pages   = {3251--3269},
  year    = {2007},
  doi     = {10.1098/rspa.2007.1900}
}

@article{Benner2015,
  author  = {Benner, Peter and Gugercin, Serkan and Willcox, Karen},
  title   = {{A survey of projection-based model reduction methods for parametric dynamical systems}},
  journal = {SIAM Review},
  volume  = {57},
  number  = {4},
  pages   = {483--531},
  year    = {2015},
  doi     = {10.1137/130932715}
}

@article{Rowley2004,
  author  = {Rowley, Clarence W. and Colonius, Tim and Murray, Richard M.},
  title   = {{Model reduction for compressible flows using POD and Galerkin projection}},
  journal = {Physica D: Nonlinear Phenomena},
  volume  = {189},
  number  = {1-2},
  pages   = {115--129},
  year    = {2004},
  doi     = {10.1016/j.physd.2003.03.001}
}

@article{Taira2020,
  author  = {Taira, K. and Hemati, M. S. and Brunton, S. L. and Sun, Y. and Duraisamy, K. and Bagheri, S. and Dawson, S. T. M. and Yeh, C.-A.},
  title   = {{Modal analysis of fluid flows: Applications and outlook}},
  journal = {AIAA Journal},
  volume  = {58},
  number  = {3},
  pages   = {998--1022},
  year    = {2020},
  doi     = {10.2514/1.j058462}
}

@article{Yano2014,
  author  = {Yano, M.},
  title   = {{A space-time Petrov--Galerkin certified reduced basis method: Application to the Boussinesq equations}},
  journal = {SIAM Journal on Scientific Computing},
  volume  = {36},
  number  = {1},
  pages   = {A232--A266},
  year    = {2014},
  doi     = {10.1137/120903300}
}

@article{LeeCarlberg2020,
  author  = {Lee, Kookjin and Carlberg, Kevin T.},
  title   = {Model reduction of dynamical systems on nonlinear manifolds using deep convolutional autoencoders},
  journal = {Journal of Computational Physics},
  volume  = {404},
  pages   = {108973},
  year    = {2020},
  doi     = {10.1016/j.jcp.2019.108973}
}

@article{Swischuk2019,
  author  = {Swischuk, R. and Mainini, L. and Peherstorfer, B. and Willcox, K.},
  title   = {{Projection-based model reduction: Formulations for physics-based machine learning}},
  journal = {Computers \& Fluids},
  volume  = {179},
  pages   = {704--717},
  year    = {2019},
  doi     = {10.1016/j.compfluid.2018.07.021}
}

@article{Yeh2018,
  author  = {Yeh, Shiang-Ting and Wang, Xingjian and Sung, Chih-Li and Mak, Simon and Chang, Yu-Hung and Zhang, Liwei and Wu, C. F. Jeff and Yang, Vigor},
  title   = {{Common proper orthogonal decomposition-based spatiotemporal emulator for design exploration}},
  journal = {AIAA Journal},
  volume  = {56},
  number  = {6},
  pages   = {2429--2442},
  year    = {2018},
  doi     = {10.2514/1.j056640}
}

@article{Kennedy2000,
  author  = {Kennedy, Marc C. and O'Hagan, Anthony},
  title   = {{Predicting the output from a complex computer code when fast approximations are available}},
  journal = {Biometrika},
  volume  = {87},
  number  = {1},
  pages   = {1--13},
  year    = {2000},
  doi     = {10.1093/biomet/87.1.1}
}

@article{Peherstorfer2018,
  author  = {Peherstorfer, B. and Willcox, K. and Gunzburger, M.},
  title   = {{Survey of multifidelity methods in uncertainty propagation, inference, and optimization}},
  journal = {SIAM Review},
  volume  = {60},
  number  = {3},
  pages   = {550--591},
  year    = {2018},
  doi     = {10.1137/16m1082469}
}

@article{Gano2005,
  author  = {Gano, S. E. and Renaud, J. E. and Sanders, B.},
  title   = {{Hybrid variable fidelity optimization by using a kriging-based scaling function}},
  journal = {AIAA Journal},
  volume  = {43},
  number  = {11},
  pages   = {2422--2433},
  year    = {2005},
  doi     = {10.2514/1.12466}
}

@article{Jiang2018,
  author  = {Jiang, P. and Xie, T. and Zhou, Q. and Shao, X. and Hu, J. and Cao, L.},
  title   = {{A space mapping method based on Gaussian process model for variable fidelity metamodeling}},
  journal = {Simulation Modelling Practice and Theory},
  volume  = {81},
  pages   = {64--84},
  year    = {2018},
  doi     = {10.1016/j.simpat.2017.11.010}
}

@article{Song2019,
  author  = {Song, X. and Lv, L. and Sun, W. and Zhang, J.},
  title   = {{A radial basis function-based multi-fidelity surrogate model: Exploring correlation between high-fidelity and low-fidelity models}},
  journal = {Structural and Multidisciplinary Optimization},
  volume  = {60},
  pages   = {965--981},
  year    = {2019},
  doi     = {10.1007/s00158-019-02248-0}
}

@article{Zhou2017,
  author  = {Zhou, Q. and Jiang, P. and Shao, X. and Hu, J. and Cao, L. and Wan, L.},
  title   = {{A variable fidelity information fusion method based on radial basis function}},
  journal = {Advanced Engineering Informatics},
  volume  = {32},
  pages   = {26--39},
  year    = {2017},
  doi     = {10.1016/j.aei.2016.12.005}
}

@article{brunton2020machine,
  author  = {Brunton, S. L. and Noack, B. R. and Koumoutsakos, P.},
  title   = {Machine learning for fluid mechanics},
  journal = {Annual Review of Fluid Mechanics},
  volume  = {52},
  pages   = {477--508},
  year    = {2020},
  doi     = {10.1146/annurev-fluid-010719-060214}
}

@article{vinuesa2022enhancing,
  author  = {Vinuesa, R. and Brunton, S. L.},
  title   = {Enhancing computational fluid dynamics with machine learning},
  journal = {Nature Computational Science},
  volume  = {2},
  pages   = {358--366},
  year    = {2022},
  doi     = {10.1038/s43588-022-00264-7}
}

@inproceedings{guo2016cnn,
  author    = {Guo, X. and Li, W. and Iorio, F.},
  title     = {Convolutional neural networks for steady flow approximation},
  booktitle = {Proceedings of the 22nd ACM SIGKDD International Conference on Knowledge Discovery and Data Mining},
  pages     = {481--490},
  year      = {2016},
  doi       = {10.1145/2939672.2939738}
}

@article{bhatnagar2019prediction,
  author  = {Bhatnagar, S. and Afshar, Y. and Pan, S. and Duraisamy, K. and Kaushik, S.},
  title   = {Prediction of aerodynamic flow fields using convolutional neural networks},
  journal = {Computational Mechanics},
  volume  = {64},
  pages   = {525--545},
  year    = {2019},
  doi     = {10.1007/s00466-019-01740-0}
}

@article{thuerey2020deep,
  author  = {Thuerey, N. and Weissenow, K. and Prantl, L. and Hu, X.},
  title   = {Deep learning methods for {R}eynolds-averaged {N}avier--{S}tokes simulations of airfoil flows},
  journal = {AIAA Journal},
  volume  = {58},
  number  = {1},
  pages   = {25--36},
  year    = {2020},
  doi     = {10.2514/1.J058291}
}

@article{Xu2020,
  author  = {Xu, Jiayang and Duraisamy, Karthik},
  title   = {{Multi-level convolutional autoencoder networks for parametric prediction of spatio-temporal dynamics}},
  journal = {Computer Methods in Applied Mechanics and Engineering},
  volume  = {372},
  pages   = {113379},
  year    = {2020},
  doi     = {10.1016/j.cma.2020.113379}
}

@article{Fukami19,
  author  = {Fukami, K. and Fukagata, K. and Taira, K.},
  title   = {Super-resolution reconstruction of turbulent flows with machine learning},
  journal = {Journal of Fluid Mechanics},
  volume  = {870},
  pages   = {106--120},
  year    = {2019},
  doi     = {10.1017/jfm.2019.238}
}

@article{Liu-etal-20,
  author  = {Liu, B. and Tang, J. and Huang, H. and Lu, X.-Y.},
  title   = {Deep learning methods for super-resolution reconstruction of turbulent flows},
  journal = {Physics of Fluids},
  volume  = {32},
  number  = {2},
  pages   = {025105},
  year    = {2020},
  doi     = {10.1063/1.5140772}
}

@article{Fukami2021,
  author  = {Fukami, Kai and Fukagata, Koji and Taira, Kunihiko},
  title   = {{Machine-learning-based spatio-temporal super resolution reconstruction of turbulent flows}},
  journal = {Journal of Fluid Mechanics},
  volume  = {909},
  pages   = {A9},
  year    = {2021},
  doi     = {10.1017/jfm.2020.948}
}

@article{Dissanayake1994,
  author  = {Dissanayake, M. W. M. G. and Phan-Thien, N.},
  title   = {{Neural-network-based approximations for solving partial differential equations}},
  journal = {Communications in Numerical Methods in Engineering},
  volume  = {10},
  number  = {3},
  pages   = {195--201},
  year    = {1994},
  doi     = {10.1002/cnm.1640100303}
}

@article{E2018,
  author  = {E, W. and Yu, B.},
  title   = {{The deep {R}itz method: A deep learning-based numerical algorithm for solving variational problems}},
  journal = {Communications in Mathematics and Statistics},
  volume  = {6},
  number  = {1},
  pages   = {1--12},
  year    = {2018},
  doi     = {10.1007/s40304-018-0127-z}
}

@article{raissi2019pinn,
  author  = {Raissi, M. and Perdikaris, P. and Karniadakis, G. E.},
  title   = {Physics-Informed Neural Networks: A Deep Learning Framework for Solving Forward and Inverse Problems Involving Nonlinear Partial Differential Equations},
  journal = {Journal of Computational Physics},
  volume  = {378},
  pages   = {686--707},
  year    = {2019},
  doi     = {10.1016/j.jcp.2018.10.045}
}

@article{Sirignano2018,
  author  = {Sirignano, Justin and Spiliopoulos, Konstantinos},
  title   = {{DGM: A deep learning algorithm for solving partial differential equations}},
  journal = {Journal of Computational Physics},
  volume  = {375},
  pages   = {1339--1364},
  year    = {2018},
  doi     = {10.1016/j.jcp.2018.08.029}
}

@article{jin2021nsfnets,
  author  = {Jin, X. and Cai, S. and Li, H. and Karniadakis, G. E.},
  title   = {{NSFnets} ({Navier--Stokes} Flow Nets): Physics-Informed Neural Networks for the Incompressible {Navier--Stokes} Equations},
  journal = {Journal of Computational Physics},
  volume  = {426},
  pages   = {109951},
  year    = {2021},
  doi     = {10.1016/j.jcp.2020.109951}
}

@article{kashefi2022pipn,
  author  = {Kashefi, A. and Mukerji, T.},
  title   = {Physics-Informed {PointNet}: A Deep Learning Solver for Steady-State Incompressible Flows and Thermal Fields on Multiple Sets of Irregular Geometries},
  journal = {Journal of Computational Physics},
  volume  = {468},
  pages   = {111510},
  year    = {2022},
  doi     = {10.1016/j.jcp.2022.111510}
}

@article{kashefi2023porous,
  author  = {Kashefi, A. and Mukerji, T.},
  title   = {Prediction of Fluid Flow in Porous Media by Sparse Observations and Physics-Informed {PointNet}},
  journal = {Neural Networks},
  volume  = {167},
  pages   = {80--91},
  year    = {2023},
  doi     = {10.1016/j.neunet.2023.08.006}
}

@article{ChenChen95,
  author  = {Chen, T. and Chen, H.},
  title   = {Universal Approximation to Nonlinear Operators by Neural Networks with Arbitrary Activation Functions and Its Application to Dynamical Systems},
  journal = {IEEE Transactions on Neural Networks},
  volume  = {6},
  number  = {4},
  pages   = {911--917},
  year    = {1995},
  doi     = {10.1109/72.392253}
}

@article{lu2021deeponet,
  author  = {Lu, L. and Jin, P. and Pang, G. and Zhang, Z. and Karniadakis, G. E.},
  title   = {Learning Nonlinear Operators via {DeepONet} Based on the Universal Approximation Theorem of Operators},
  journal = {Nature Machine Intelligence},
  volume  = {3},
  pages   = {218--229},
  year    = {2021},
  doi     = {10.1038/s42256-021-00302-5}
}

@inproceedings{Li21FNO,
  author    = {Li, Z. and Kovachki, N. B. and Azizzadenesheli, K. and Liu, B. and Bhattacharya, K. and Stuart, A. M. and Anandkumar, A.},
  title     = {{Fourier neural operator for parametric partial differential equations}},
  booktitle = {International Conference on Learning Representations},
  year      = {2021},
  doi       = {10.48550/arXiv.2010.08895}
}

@inproceedings{pfaff2021meshgraphnets,
  author    = {Pfaff, T. and Fortunato, M. and Sanchez-Gonzalez, A. and Battaglia, P. W.},
  title     = {Learning Mesh-Based Simulation with Graph Networks},
  booktitle = {International Conference on Learning Representations},
  year      = {2021},
  doi       = {10.48550/arXiv.2010.03409}
}

@article{Huang23_1,
  author  = {Huang, H. and Yang, V. and Liu, Y.},
  title   = {Neural Networks with Local Converging Inputs ({NNLCI}) for Solving Conservation Laws, Part {I}: {1D} Problems},
  journal = {Communications in Computational Physics},
  volume  = {34},
  number  = {2},
  pages   = {290--317},
  year    = {2023},
  doi     = {10.4208/cicp.OA-2022-0285}
}

@article{Huang23_2,
  author  = {Huang, H. and Yang, V. and Liu, Y.},
  title   = {Neural Networks with Local Converging Inputs ({NNLCI}) for Solving Conservation Laws, Part {II}: {2D} Problems},
  journal = {Communications in Computational Physics},
  volume  = {34},
  number  = {4},
  pages   = {907--933},
  year    = {2023},
  doi     = {10.4208/cicp.OA-2023-0026}
}

@misc{Cobb23,
  author       = {Cobb, H. and Lee, H. and Liu, Y.},
  title        = {Solving {M}axwell's equation in {2D} with neural networks with local converging inputs},
  howpublished = {arXiv preprint arXiv:2302.02860},
  year         = {2023},
  doi          = {10.48550/arXiv.2302.02860}
}

@article{Ding24,
  author  = {Ding, W. and Huang, H. and Lee, T. J. and Liu, Y. and Yang, V.},
  title   = {Neural Network with Local Converging Input for Unstructured-Grid Computational Fluid Dynamics},
  journal = {AIAA Journal},
  volume  = {62},
  number  = {8},
  pages   = {3155--3166},
  year    = {2024},
  doi     = {10.2514/1.J063885}
}

@article{Lee24,
  author  = {Lee, H. and Chao, Z. and Cobb, H. and Liu, Y. and Xie, D.},
  title   = {{A {PNP} ion channel deep learning solver with local neural network and finite element input data}},
  journal = {Machine Learning: Science and Technology},
  volume  = {5},
  number  = {4},
  pages   = {045001},
  year    = {2024},
  doi     = {10.1088/2632-2153/ad7e7a}
}

@article{POD-NNLCI-26,
  author  = {Lu, Q. and Zhang, L. and Ding, S. and Weng, Z. and Ding, W. and Wang, X.},
  title   = {Local multi-fidelity surrogates for data-efficient parametric studies of combustion problems},
  journal = {Proceedings of the Combustion Institute},
  volume  = {42},
  pages   = {106513},
  year    = {2026},
  doi     = {10.1016/j.proci.2026.106513}
}

@phdthesis{huang2025dissertation,
  author = {Huang, H.},
  title  = {Neural Networks for Solving Hyperbolic Conservation Laws},
  school = {Georgia Institute of Technology},
  year   = {2025}
}

@article{Fortin93,
  author  = {Fortin, M.},
  title   = {Finite element solution of the {Navier--Stokes} equations},
  journal = {Acta Numerica},
  volume  = {2},
  pages   = {239--284},
  year    = {1993},
  doi     = {10.1017/S0962492900002373}
}

@article{alnaes2015fenics,
  author  = {Aln{\ae}s, M. S. and Blechta, J. and Hake, J. and Johansson, A. and Kehlet, B. and Logg, A. and Richardson, C. and Ring, J. and Rognes, M. E. and Wells, G. N.},
  title   = {The {FEniCS} Project Version 1.5},
  journal = {Archive of Numerical Software},
  volume  = {3},
  number  = {100},
  pages   = {9--23},
  year    = {2015},
  doi     = {10.11588/ans.2015.100.20553}
}

@book{logg2012fenics,
  author    = {Logg, A. and Mardal, K.-A. and {Wells, eds.}, G. N.},
  title     = {Automated Solution of Differential Equations by the Finite Element Method: The {FEniCS} Book},
  series    = {Lecture Notes in Computational Science and Engineering},
  volume    = {84},
  publisher = {Springer},
  address   = {Berlin},
  year      = {2012},
  doi       = {10.1007/978-3-642-23099-8}
}

@article{donea1982ale,
  author  = {Donea, J. and Giuliani, S. and Halleux, J. P.},
  title   = {An Arbitrary {L}agrangian--{E}ulerian Finite Element Method for Transient Dynamic Fluid--Structure Interactions},
  journal = {Computer Methods in Applied Mechanics and Engineering},
  volume  = {33},
  number  = {1--3},
  pages   = {689--723},
  year    = {1982},
  doi     = {10.1016/0045-7825(82)90128-1}
}

@article{amestoy2001mumps,
  author  = {Amestoy, P. R. and Duff, I. S. and L'Excellent, J.-Y. and Koster, J.},
  title   = {A fully asynchronous multifrontal solver using distributed dynamic scheduling},
  journal = {SIAM Journal on Matrix Analysis and Applications},
  volume  = {23},
  number  = {1},
  pages   = {15--41},
  year    = {2001},
  doi     = {10.1137/S0895479899358194}
}

@article{levin1998mls,
  author  = {Levin, D.},
  title   = {The Approximation Power of Moving Least-Squares},
  journal = {Mathematics of Computation},
  volume  = {67},
  number  = {224},
  pages   = {1517--1531},
  year    = {1998},
  doi     = {10.1090/S0025-5718-98-00974-0}
}

@misc{hendrycks2016gelu,
  author       = {Hendrycks, D. and Gimpel, K.},
  title        = {Gaussian error linear units ({GELUs})},
  howpublished = {arXiv preprint arXiv:1606.08415},
  year         = {2016},
  doi          = {10.48550/arXiv.1606.08415}
}

@inproceedings{he2015delving,
  author    = {He, K. and Zhang, X. and Ren, S. and Sun, J.},
  title     = {Delving deep into rectifiers: Surpassing human-level performance on {ImageNet} classification},
  booktitle = {Proceedings of the IEEE International Conference on Computer Vision (ICCV)},
  pages     = {1026--1034},
  year      = {2015},
  doi       = {10.1109/ICCV.2015.123}
}

@inproceedings{loshchilov2019adamw,
  author    = {Loshchilov, I. and Hutter, F.},
  title     = {Decoupled Weight Decay Regularization},
  booktitle = {International Conference on Learning Representations},
  year      = {2019},
  doi       = {10.48550/arXiv.1711.05101}
}

@article{barber1996quickhull,
  author  = {Barber, C. B. and Dobkin, D. P. and Huhdanpaa, H.},
  title   = {The Quickhull Algorithm for Convex Hulls},
  journal = {ACM Transactions on Mathematical Software},
  volume  = {22},
  number  = {4},
  pages   = {469--483},
  year    = {1996},
  doi     = {10.1145/235815.235821}
}

@article{virtanen2020scipy,
  author  = {Virtanen, P. and Gommers, R. and Oliphant, T. E. and others},
  title   = {{SciPy} 1.0: Fundamental Algorithms for Scientific Computing in {P}ython},
  journal = {Nature Methods},
  volume  = {17},
  pages   = {261--272},
  year    = {2020},
  doi     = {10.1038/s41592-019-0686-2}
}

@book{leveque2002finite,
  author    = {LeVeque, R. J.},
  title     = {Finite Volume Methods for Hyperbolic Problems},
  publisher = {Cambridge University Press},
  address   = {Cambridge},
  year      = {2002},
  doi       = {10.1017/CBO9780511791253}
}

@article{tseng2003ghost,
  author  = {Tseng, Y. H. and Ferziger, J. H.},
  title   = {A ghost-cell immersed boundary method for flow in complex geometry},
  journal = {Journal of Computational Physics},
  volume  = {192},
  number  = {2},
  pages   = {593--623},
  year    = {2003},
  doi     = {10.1016/j.jcp.2003.07.024}
}

@article{mittal2005immersed,
  author  = {Mittal, R. and Iaccarino, G.},
  title   = {Immersed boundary methods},
  journal = {Annual Review of Fluid Mechanics},
  volume  = {37},
  pages   = {239--261},
  year    = {2005},
  doi     = {10.1146/annurev.fluid.37.061903.175743}
}

@article{lancaster1981surfaces,
  author  = {Lancaster, Peter and {\v{S}}alkauskas, Kestutis},
  title   = {Surfaces generated by moving least squares methods},
  journal = {Mathematics of Computation},
  volume  = {37},
  number  = {155},
  pages   = {141--158},
  year    = {1981},
  doi     = {10.1090/S0025-5718-1981-0616367-1}
}

@inproceedings{islam2020much,
  author    = {Islam, M. A. and Jia, S. and Bruce, N. D. B.},
  title     = {How much position information do convolutional neural networks encode?},
  booktitle = {International Conference on Learning Representations},
  year      = {2020},
  doi       = {10.48550/arXiv.2001.08248}
}

@inproceedings{alguacil2021effects,
  author    = {Alguacil, A. and Pinto, W. G. and Bauerheim, M. and Jacob, M. C. and Moreau, S.},
  title     = {Effects of boundary conditions in fully convolutional networks for learning spatio-temporal dynamics},
  booktitle = {Machine Learning and Knowledge Discovery in Databases: Applied Data Science Track (ECML PKDD 2021)},
  publisher = {Springer},
  year      = {2021},
  doi       = {10.1007/978-3-030-86517-7_7}
}

@article{squires2005microfluidics,
  author  = {Squires, T. M. and Quake, S. R.},
  title   = {Microfluidics: Fluid physics at the nanoliter scale},
  journal = {Reviews of Modern Physics},
  volume  = {77},
  number  = {3},
  pages   = {977--1026},
  year    = {2005},
  doi     = {10.1103/RevModPhys.77.977}
}

@article{benzi2005saddle,
  author  = {Benzi, M. and Golub, G. H. and Liesen, J.},
  title   = {Numerical solution of saddle point problems},
  journal = {Acta Numerica},
  volume  = {14},
  pages   = {1--137},
  year    = {2005},
  doi     = {10.1017/S0962492904000212}
}

@article{taylor1973numerical,
  author  = {Taylor, C. and Hood, P.},
  title   = {A numerical solution of the {N}avier--{S}tokes equations using the finite element technique},
  journal = {Computers \& Fluids},
  volume  = {1},
  number  = {1},
  pages   = {73--100},
  year    = {1973},
  doi     = {10.1016/0045-7930(73)90027-3}
}

@inproceedings{paszke2019pytorch,
  author    = {Paszke, A. and Gross, S. and Massa, F. and Lerer, A. and others},
  title     = {{PyTorch}: An imperative style, high-performance deep learning library},
  booktitle = {Advances in Neural Information Processing Systems},
  volume    = {32},
  pages     = {8024--8035},
  year      = {2019},
  doi       = {10.48550/arXiv.1912.01703}
}

@article{richardson1911approximate,
  author  = {Richardson, L. F.},
  title   = {The approximate arithmetical solution by finite differences of physical problems involving differential equations, with an application to the stresses in a masonry dam},
  journal = {Philosophical Transactions of the Royal Society of London, Series A},
  volume  = {210},
  pages   = {307--357},
  year    = {1911},
  doi     = {10.1098/rsta.1911.0009}
}

@book{roache1998verification,
  author    = {Roache, P. J.},
  title     = {Verification and Validation in Computational Science and Engineering},
  publisher = {Hermosa Publishers},
  address   = {Albuquerque, NM},
  year      = {1998}
}
\end{document}